%% file: 2025_ZhRy_transport.tex
\documentclass[11pt,oneside,fleqn,reqno]{article}
\usepackage{amsmath}
\usepackage{amsthm}
\usepackage{amssymb}
\usepackage[pdftex]{graphicx,color}
\usepackage{mathtools}
\usepackage{wrapfig}
\usepackage{natbib} \citeindextrue
\usepackage{subfigure}
\usepackage{dsfont}
\usepackage{multirow}
\usepackage{multicol}
\usepackage{algorithm}
\usepackage{algorithmic}
\usepackage{bm}
\usepackage{comment}
\usepackage[
  colorlinks=true,
  linkcolor=blue,
  citecolor=blue,
  urlcolor=blue
]{hyperref}

\DeclareMathOperator{\tr}{tr}

\allowdisplaybreaks


\input{notation_ams}

\input{macro_ams}

\newtheorem{assum}{Assumption}[section]

\begin{document}

\title{Semidiscrete optimal transport with unknown costs}
\author{Yinchu Zhu \and Ilya O. Ryzhov}

\date{\today}

\maketitle

\begin{abstract}
Semidiscrete optimal transport is a challenging generalization of the classical transportation problem in linear programming. The goal is to design a joint distribution for two random variables (one continuous, one discrete) with fixed marginals, in a way that minimizes expected cost. We formulate a novel variant of this problem in which the cost functions are unknown, but can be learned through noisy observations; however, only one function can be sampled at a time. We develop a semi-myopic algorithm that couples online learning with stochastic approximation, and prove that it achieves optimal convergence rates, despite the non-smoothness of the stochastic gradient and the lack of strong concavity in the objective function.
\end{abstract}

\doublespace

\section{Introduction}\label{sec:intro}

In the semidiscrete optimal transport problem \citep{PeCu19}, we are given a continuous random vector $X$ with known density $f$, and a discrete random variable $Y$ with known probability mass function $p$. Our task is to choose a \textit{joint} distribution for $\left(X,Y\right)$ to minimize an expected cost. Formally, we write
\begin{equation}\label{eq:obj}
\inf_{h} \sum^{K}_{k=1} \int_{\mathcal{X}} c\left(x,k\right)h\left(x,k\right)dx
\end{equation}
subject to
\begin{eqnarray}
\int_{\mathcal{X}} h\left(x,k\right)dx &=& p_k, \qquad k = 1,...,K,\label{eq:target}\\
\sum^K_{k=1} h\left(x,k\right) &=& f\left(x\right), \qquad x\in\mathcal{X},\label{eq:density}\\
h\left(x,k\right) &\geq& 0, \qquad x\in\mathcal{X}, \; k = 1,...,K.\label{eq:nonneg}
\end{eqnarray}
where $\mathcal{X}\subseteq \mathds{R}^d$ is the support of $X$ and $\left\{1,...,K\right\}$ is the support of $Y$. The marginal distributions are fixed by the problem inputs $p_k > 0$ and $f$. The objective function can also be written as $\mathbb{E}\left[c\left(X,Y\right)\right]$ under the assumption that $P\left(X\in dx,Y=k\right) = h\left(x,k\right)dx$. In other words, $h$ is the mixed joint likelihood of $\left(X,Y\right)$.

The name ``optimal transport'' has the same origin as the well-known ``transportation problem'' in linear programming \citep{FoFu56}, which can be interpreted as a special case of (\ref{eq:obj}) where both $X$ and $Y$ have finite support. In that case, the joint distribution of $\left(X,Y\right)$ is described by a finite set of probabilities $h_{jk}$, whose cost coefficients are given by $c_{jk}$ for all possible values $\left(j,k\right)$ of $\left(X,Y\right)$. The optimization problem then reduces to a linear program, whose solution can be interpreted as a minimum-cost plan for matching supply from one set of locations (or facilities) indexed by $j$ with demand at a different set of locations indexed by $k$. Such plans have numerous applications in logistics. In semidiscrete optimal transport, $Y$ remains discrete but $X$ becomes continuous, which makes the problem more difficult and generally not solvable using LP methods. The semidiscrete version also has applications in logistics, specifically in geographical partitioning problems \citep{CaCaDe16,HaSc20} where demand can arise anywhere on the map, and the goal is to design service zones that assign spatial regions to supply facilities.


The optimal transport literature universally assumes that the cost function $c$ is known: in fact, most papers only work with some particular form for $c$, most commonly a Euclidean distance or $p$-norm \citep{Sa15}. Our paper is the first to focus on a situation where $c$ is \textit{unknown}, but can be estimated in a sequential manner from data. Our motivation for this setting is a class of applications where $k\in\left\{1,...,K\right\}$ represents a ``choice'' or ``alternative'' available to a decision-maker. One observes a sample of $X$ from the density $f$, then selects an alternative $k$ and incurs a cost $c\left(X,k\right)$. The goal is to assign $X$ values to alternatives in a way that minimizes expected cost while obeying the constraints (\ref{eq:target}) on how often each alternative can be chosen.


Consider the following application to online advertising. An advertising platform has $K$ clients. A guaranteed contract \citep{Bharadwaj10} ensures that client $k$ receives a certain number of impressions through the platform, which can be expressed as a proportion $p_k$, agreed upon ahead of time, of the platform's total user base. Users have heterogeneous attributes which influence how they respond to advertisements. The platform observes the attributes $X$ of each randomly arriving user and selects a client, whose advertisement is then shown to the user in real time. Only one client can be selected per user, and the platform has the freedom to make individual assignments, provided that the long-term proportion of users assigned to client $k$ is equal to $p_k$. An assignment of a user with attributes $x$ to the $k$th client can be evaluated using a performance metric such as the clickthrough rate, which can be represented using, e.g., the linear model $c\left(x,k\right) = \beta^\top_k x$ \citep{LiChLaSc10} or the logistic model $c\left(x,k\right) = \frac{1}{1+e^{-\beta^\top_k x}}$ \citep{ChCa10}. We can maximize clickthrough rates by minimizing the negative of these functions in (\ref{eq:obj}). The coefficients $\beta_k \in \mathbb{R}^d$ are unknown (note that each client has a different set of coefficients), reflecting the fact that the platform cannot predict clickthrough rates with perfect accuracy. However, they can be inferred over time by observing how users behave. The goal is to learn $\beta_k$ efficiently while optimizing the expected performance.

More generally, our paper considers parametric cost functions of the form $c\left(x,k\right) = \chi\left(\beta^\top_k x\right)$, where $\chi$ can be any link function associated with a generalized linear statistical model \citep{McNe89}. We do not know $\beta_k$, but when an observed $x$ is assigned to the $k$th alternative, we collect an observation of the form $c\left(x,k\right)+\varepsilon$, where $\varepsilon$ is an additive noise. We then use an appropriate statistical procedure (e.g., linear or logistic regression) to estimate $\beta_k$ from the observations for that $k$ value. Only one $k$ can be observed at a time, creating a tradeoff: one more observation of the $k$th cost function means one less for the others.

This is essentially the same tradeoff that arises in classical optimal learning problems, such as ranking and selection \citep{ChChLePu15} and multi-armed bandits \citep{GiGlWe11}, where one similarly observes noisy samples of the values of various ``alternatives'' with the goal of efficiently identifying the best. Our problem belongs to the class of \textit{contextual} learning, where the value of an alternative additionally depends on a random vector $X$; see, e.g., \cite{HaRyDe16}, \cite{AlChZo21} and \cite{ShHoZh21} from the ranking and selection literature, or \cite{AbPaSz11} and \cite{BaBaKh21} from the bandit community. However, the presence of the target constraint (\ref{eq:target}) in our problem is a significant departure from existing models. In optimal learning, there is a clear notion of ``correct selection,'' when the alternative with the smallest true value also has the smallest estimated value, and of ``regret'' when it does not. The proportional assignments $p_k$ in those problems vary between algorithms \citep{GlJu04}, and many papers aim to learn proportions that maximize the probability of correct selection \citep{Ru20,ChRy22}. However, under (\ref{eq:target}), we sometimes have to make choices that do \textit{not} achieve the lowest cost, even if the costs are known. Thus, the standard definitions of correct selection or regret do not apply in this setting.

In this paper, we bridge optimal transport theory and optimal learning to formalize a notion of correct selection, and we create a sampling algorithm that makes correct selections at an optimal rate. As a byproduct, we also obtain new results for the more standard setting of semidiscrete optimal transport with \textit{known} costs. Although the known-cost setting is not the main focus of this research, we use it as a starting point to develop the analytical technique that we eventually use to handle unknown costs. We first explain the essential elements of this technique in Section \ref{sec:intknown} below, and then discuss the main results in Section \ref{sec:intunknown}.


\subsection{Stochastic approximation and local strong concavity}\label{sec:intknown}

There is a vast literature on optimal transport with known $c$. \cite{Vi21} gives an introduction to the deep theory of this field. Computational approaches include methods based on discretization, such as the Sinkhorn \citep{Cu13} and Greenkhorn \citep{LiHoJo22} algorithms or the proximal splitting procedure of \cite{PaPeOu14}, as well as methods based on PDEs \citep{BeFrOb14} and difference of convex programming \citep{BoPe24}.

The specific setting of \textit{semidiscrete} optimal transport has distinct structure that greatly aids computation. Problem (\ref{eq:obj})-(\ref{eq:nonneg}) admits an optimal policy $\pi^*$, a decision rule that assigns an observed $X$ to some $\pi^*\left(X\right)\in\left\{1,...,K\right\}$, in the presence of probabilistic targets. This characterization gives us a notion of correct selection that we will later use in the unknown-cost setting: given $X$, we want to choose the same $k$ as the optimal policy. The crucial distinguishing feature of the semidiscrete problem is that $\pi^*$ is completely characterized by a finite-dimensional vector of parameters, namely, the shadow prices of (\ref{eq:target}) that optimally solve the Kantorovich dual of (\ref{eq:obj})-(\ref{eq:nonneg}). As long as we know $c$ and are able to observe samples from $f$, we can learn these parameters using stochastic approximation (SA) techniques. SA is very simple to implement, performs well empirically \citep{GeCuPeBa16}, and is guaranteed to converge to the optimal parameters due to the concavity of the Kantorovich dual problem. An important detail is that SA does not require discretization, unlike many general-purpose optimal transport methods.

We build on this approach and use SA under both known and unknown costs. Our main focus is on theoretical performance guarantees. In the literature on SA, the strongest possible guarantee \citep{BaMo11} is that the sequence of iterates converges in $L^2$ at the canonical rate $\mathcal{O}\left(\frac{1}{n}\right)$, where $n$ is the number of samples drawn from $f$. In general, however, this rate is only achievable with Polyak averaging \citep{PoJu92} included into the SA procedure and with the assumption of strong convexity (concavity) on the SA objective function. Even in later work (e.g., \citealp{BaMo13}), which has yielded sharper rates, strong convexity is still needed to bound the $L^2$ error of the SA iterates. The main technical issue in our case is that the objective of the Kantorovich dual is concave, but not strongly concave. The optimal transport literature is aware of this difficulty, and has developed a workaround (see, e.g., \citealp{GeCuPeBa16} or \citealp{TaShKu22}) that replaces the objective by a smoothed approximation. This approach recovers the canonical rate at the expense of one of the most attractive features of the algorithm, namely, the ability to solve the original problem exactly.

In marked contrast, we do not modify the problem. Instead, we adopt a perspective from the recent bandit literature (e.g., \citealp{GoZe13}, \citealp{BaBa20}, and \citealp{BaBaKh21}), which has found that convergence rates improve when the data-generating distribution exhibits a sufficient degree of random variation (``richness''). With comparable assumptions on the distribution of $X$, the objective is still not strongly concave, but it has a kind of ``local'' strong concavity. Our paper is the first to identify this property and show that it enables recovery of the canonical rate without smoothing or averaging. This part of our analysis holds generally, without assuming any particular structure on $c$, as long as it is known. However, our analysis of unknown costs also draws upon this result.

\subsection{Main contribution: semi-myopic learning of unknown costs}\label{sec:intunknown}

When $c$ is unknown, we still use SA to learn the dual variables, but the situation becomes much more complicated because the gradient estimates are no longer unbiased (they now have to be computed using estimated cost coefficients $\beta_k$). The problem couples optimal learning together with SA in such a way that both aspects create difficulties for each other: statistical error in the estimated costs creates bias in the gradient estimator, but the gradient estimator drives the decision rule and thus determines which alternatives are sampled.


We solve this problem using a simple, computationally efficient online algorithm with a ``semi-myopic'' structure. Most of the time, the algorithm approximates the optimal policy using plug-in estimators of both the costs and the dual variables, but occasionally it is forced to sample each individual cost function. These periods of forced exploration become less frequent as time goes on, to reduce their impact on performance. Semi-myopic structure has a long history in optimal learning, dating back to the work of \cite{BrRu12} on dynamic pricing, and followed by extensions to ranking and selection \citep{GaKa16} and contextual bandits \citep{GoZe13}. Virtually all of these papers require assumptions on the richness of the data-generating distribution in order to obtain optimal rates. Semi-myopic algorithms also vary in how often they need to conduct forced exploration. The gap between exploration periods grows linearly in \cite{BrRu12}; quadratically in \cite{KeZe14}, \cite{GaKa16}, and \cite{BaKe21}; and exponentially in \cite{GoZe13} and \cite{BaBa20}. The gap for our procedure, in the more complex setting of semidiscrete optimal transport, grows sub-exponentially.

Our paper is the first to couple this type of exploration with stochastic approximation. As in SA, we update the iterates using (biased) gradient estimates; as in optimal learning, information is collected sequentially in a semi-myopic fashion. Building on our results for the known-cost setting, we prove that SA still achieves the canonical rate under this setup. As a corollary, the expected number of incorrect decisions made by the semi-myopic policy (the most relevant analog of ``regret'' in our setting) grows at a rate $\mathcal{O}\left(\sqrt{n}\right)$, the best possible. This performance is achieved in an online manner, without requiring the sample size to be known in advance. Numerical experiments show that, in some situations, cost uncertainty produces only minor performance loss relative to an ``ideal'' SA procedure with unbiased gradients.

In summary, our work bridges and contributes to three distinct streams of literature. We contribute to \textit{optimal transport} by being the first to formulate and study the semidiscrete problem with unknown costs. This problem couples \textit{optimal learning} and \textit{stochastic approximation} in a novel and challenging way. We contribute to optimal learning by studying a novel contextual problem with probabilistic targets, and by proving that semi-myopic exploration can learn the optimal policy efficiently. We contribute to SA by showing that optimal convergence rates can be achievable without strong concavity, as long as the data-generating distribution is sufficiently rich.

\subsection{Organization of the paper}

Section \ref{sec:preliminaries} presents preliminaries on the semidiscrete optimal transport problem, most importantly the structure of the optimal policy. We then state the algorithm and key theoretical results: Section \ref{sec:algorithms} presents stochastic approximation (Section \ref{sec:mainresultsknown}) and semi-myopic learning (Section \ref{sec:mainresultsunknown}) together with our main assumptions and rate results. Then, Section \ref{sec:known} presents the main steps of our technical analysis for SA (local strong concavity under known costs), while Section \ref{sec:unknown} does the same for semi-myopic learning of unknown costs. Section \ref{sec:exp} presents numerical examples and insights, and Section \ref{sec:conc} concludes.

\section{Semidiscrete optimal transport: preliminaries}\label{sec:preliminaries}

We begin by reviewing the structure of the semidiscrete optimal transport problem. The literature has characterized the structure of the optimal solution to (\ref{eq:obj})-(\ref{eq:nonneg}) when $c$ is known; these results can be found in, e.g., \cite{CaCaDe16} or \cite{GeCuPeBa16}, but we summarize them here to provide background, sketching the key steps in the derivation. In brief, the problem (\ref{eq:obj})-(\ref{eq:nonneg}) is solved by a \textit{policy}, or decision rule for assigning any observed $X$ to a $k$ value. In other words, $Y$ should be set equal to a certain deterministic function of $X$. It is necessary to understand this function before we can formally define a performance metric for the problem that we will eventually tackle.


For notational convenience, we write $c_k\left(x\right) = c\left(x,k\right)$, to visually distinguish between costs $c_k$ corresponding to different ``alternatives,'' from which we will eventually make a selection. We do not impose any parametric structure on $c_k$. By Kantorovich duality (Theorem 1.3 in \citealp{Vi21}), the problem (\ref{eq:obj})-(\ref{eq:nonneg}) has the same optimal value as the functional optimization problem
\begin{equation}\label{eq:dualobj}
\sup_{\left(\phi,\psi\right)} \mathbb{E}\left(\phi\left(X\right)+\psi\left(Y\right)\right),
\end{equation}
subject to the constraint
\begin{equation}\label{eq:dualcons}
\phi\left(X\right) + \psi\left(Y\right) \leq \sum_k c_k\left(X\right) 1_{\left\{Y=k\right\}} \quad \text{a.s.}
\end{equation}
In (\ref{eq:dualobj}), the marginal density of $X$ is $f$, and the marginal pmf of $Y$ is $p$, as in the primal problem. To handle (\ref{eq:dualcons}), we take
\begin{equation*}
\phi\left(x\right) = \min_{j\in\left\{1,...,K\right\}} c_j\left(x\right) - \psi\left(j\right).
\end{equation*}
Letting $g = \left(\psi\left(1\right),...,\psi\left(K\right)\right)^\top$, we can rewrite (\ref{eq:dualobj}) as
\begin{equation}\label{eq:newobj}
\sup_{g \in \mathbb{R}^K} \mathbb{E}\left(\min_j c_j\left(X\right) - g_j\right) + p^\top g.
\end{equation}
The minimum of a finite number of linear functions is concave, and expectations preserve concavity, so the objective in (\ref{eq:newobj}) is concave in $g$. An important distinguishing characteristic of semidiscrete optimal transport is that the dual is both concave and finite-dimensional, and thus it can be solved without the use of discretization or nonconvex programming (see, e.g., \citealp{BoPe24}).

In fact, there are no constraints on $g$, so we can simply take the gradient and set it equal to zero. Under some mild conditions, the gradient can be interchanged with the expectation, an instance of ``infinitesimal perturbation analysis,'' a well-known technique in the simulation community \citep{Fu06,Ki06}. We show this formally for completeness, as \cite{GeCuPeBa16} did not do so. The proof is deferred to the Appendix.

\begin{prop}\label{prop:ipa}
Let $F\left(g,x\right) = \min_j c_j\left(x\right) - g_j$. Fix $g\in\mathbb{R}^k$ and suppose that, for any $j \neq k$, the pairwise difference $\left(c_j\left(X\right)-g_j\right) - \left(c_k\left(X\right)-g_k\right)$ has a density in a neighborhood of zero. Then, $\nabla_g \mathbb{E}\left(F\left(g,X\right)\right) = \mathbb{E}\left(\nabla_g F\left(g,X\right)\right)$, where
\begin{equation}\label{eq:ipagradient}
\left(\nabla_g F\left(g,X\left(\omega\right)\right)\right)_k = -1_{\left\{k = \arg\min_j c_j\left(X\left(\omega\right)\right)-g_j\right\}}
\end{equation}
is the $k$th element of the stochastic gradient $\nabla_g F\left(g,X\right)$.
\end{prop}

Under the conditions of Proposition \ref{prop:ipa}, we combine (\ref{eq:newobj}) with (\ref{eq:ipagradient}) and find that any optimal $g^*$ solves the system of equations
\begin{equation}\label{eq:insight}
P\left(c_k\left(X\right) - g^*_k = \min_j c_j\left(X\right) - g^*_j\right) = p_k, \qquad k = 1,...,K.
\end{equation}
In words, $g^*$ is a vector of bonuses and penalties (since its elements can be either positive or negative) which are subtracted from the costs $c_j\left(X\right)$ to ensure that each element is the smallest with precisely the target probability.

These same bonuses and penalties can be used to construct an optimal solution for the primal problem (\ref{eq:obj})-(\ref{eq:nonneg}). By weak duality, we have
\begin{equation}\label{eq:weakduality}
\mathbb{E}\left(\min_j c_j\left(X\right) -g^*_j\right) + p^\top g^* \leq \inf_h \mathbb{E}\left(\sum_k c_k\left(X\right)1_{\left\{Y=k\right\}}\right),
\end{equation}
where $h$ is the joint likelihood of $\left(X,Y\right)$ satisfying (\ref{eq:target})-(\ref{eq:nonneg}), as before. Now define $Y^* = \arg\min_j c_j\left(X\right) - g^*_j$. By (\ref{eq:insight}), we have $P\left(Y^* = k\right) = p_k$, so the joint likelihood of $\left(X,Y^*\right)$ satisfies (\ref{eq:target})-(\ref{eq:nonneg}). We then write
\begin{eqnarray*}
\mathbb{E}\sum_k c_k\left(X\right)1_{\left\{Y^*=k\right\}} &=& \mathbb{E}\sum_k c_k\left(X\right) 1_{\left\{c_k\left(X\right) - g^*_k = \min_j c_j\left(X\right) - g^*_j\right\}}\\
&=& \mathbb{E}\sum_k \left(c_k\left(X\right)-g^*_k\right) 1_{\left\{c_k\left(X\right) - g^*_k = \min_j c_j\left(X\right) - g^*_j\right\}} + \mathbb{E}\sum_k g^*_k 1_{\left\{c_k\left(X\right) - g^*_k = \min_j c_j\left(X\right) - g^*_j\right\}}\\
&=& \mathbb{E}\left(\min_j c_j\left(X\right) - g^*_j\right) + p^\top g^*\\
&\leq & \inf_{h} \mathbb{E}\left(\sum_k c_k\left(X\right)1_{\left\{Y=k\right\}}\right)
\end{eqnarray*}
where the last line follows by (\ref{eq:weakduality}). Then, letting
\begin{equation}\label{eq:truepolicy}
\pi^*\left(X\right) = \arg\min_j c_j\left(X\right) - g^*_j,
\end{equation}
it follows that the joint distribution of $\left(X,\pi^*\left(X\right)\right)$ optimally solves (\ref{eq:obj})-(\ref{eq:nonneg}).

The policy $\pi^*$ is a natural benchmark against which other policies may be compared. Suppose that $\left\{X^n\right\}^{\infty}_{n=1}$ is a sequence of i.i.d. samples from $f$, and $\left\{\pi^n\right\}^{\infty}_{n=1}$ is a sequence of random variables taking values in $\left\{1,...,K\right\}$. We say that a \textit{correct selection} is made at time $n$ if $\pi^n = \pi^*\left(X^n\right)$. We may use the expected total number $\mathbb{E}\left(\sum^n_{m=1} 1_{\left\{\pi^m\neq\pi^*\left(X^m\right)\right\}}\right)$ of incorrect selections to evaluate the performance of the sequence up to time $n$. The probability of incorrect selection (PICS), written as $P\left(\pi^n\neq \pi^*\left(X^n\right)\right)$, plays an important role in the analysis of this performance metric.

Many papers on optimal learning present their results in terms of the regret, which in our context would be the difference in cost achieved by $\pi^n$ vs. $\pi^*\left(X^n\right)$. However, because of (\ref{eq:target}), regret relative to $\pi^*$ can be negative, unlike in traditional bandit problems. We therefore focus on the expected number of incorrect selections, which has a very clear, straightforward interpretation.

\section{Algorithm and main results}\label{sec:algorithms}

Our overarching goal is to learn $\pi^*$ iteratively based on a sequence $\left\{X^n\right\}^{\infty}_{n=1}$ of i.i.d. samples from the data-generating density $f$. Ultimately, we wish to do this in a setting where the cost functions $c_k$ are unknown and have to be estimated. Our algorithmic approach combines two elements: stochastic approximation (for learning the optimal dual variables $g^*$) and semi-myopic exploration (for learning the costs).

Integrating these elements is quite challenging. Therefore, we begin by examining stochastic approximation separately, in a setting where the costs are known. Section \ref{sec:mainresultsknown} states the assumptions and main results needed to establish convergence rates for SA in semidiscrete optimal transport. These results, particularly the local strong concavity property, will continue to play a vital role when we switch to the unknown-cost setting. Section \ref{sec:mainresultsunknown} presents our algorithm and main results for that case.

\subsection{Stochastic approximation: known costs}\label{sec:mainresultsknown}

It was observed by \cite{GeCuPeBa16} that, when the cost functions $c_k$ are known, (\ref{eq:insight}) can be approached as a stochastic root-finding problem \citep{PaKi11}. We can straightforwardly apply the stochastic approximation (SA) algorithm
\begin{equation}\label{eq:basicsa}
g^{n+1}_k = g^n_k + \alpha_{n}\left(-1_{\left\{k = \arg\min_j c_j\left(X^{n+1}\right)-g^n_j\right\}}+p_k\right), \quad k = 1,...,K,
\end{equation}
where $\left\{\alpha_{n}\right\}^{\infty}_{n=0}$ is a stepsize sequence satisfying the usual conditions $\sum_n \alpha_n = \infty$, $\sum_n \alpha^2_n < \infty$. The gradient is bounded, so the convergence $g^n\rightarrow g^*$ easily follows from classical SA theory \citep{KuYi03}. The only input to (\ref{eq:basicsa}) is the stepsize $\alpha_n$. The algorithm requires knowledge of the targets $p_k$ and the cost functions $c_j$, but does \textit{not} require us to know the density $f$ from which each $X^{n+1}$ is sampled. This is an attractive feature of SA, as data-generating distributions are extremely difficult to estimate. If the goal is to maximize expected costs, as in the motivating example from Section \ref{sec:intro}, we can use (\ref{eq:basicsa}) with a trivial sign change.

Although our main focus is on unknown costs, the convergence rate of (\ref{eq:basicsa}) is an important prerequisite for that analysis. As mentioned in Section \ref{sec:intknown}, the main technical issue is the fact that the objective function $g \mapsto \mathbb{E}\left(F\left(g,X\right)\right)$ is concave, but not strongly concave. We will explain in Section \ref{sec:known} how this issue is resolved; for the moment, let us state the assumptions and results. It is important to note that these assumptions do \textit{not} require any specific parametric structure on $c_k$. Such structure will only be imposed in Section \ref{sec:mainresultsunknown} when we consider unknown costs.

From (\ref{eq:truepolicy}), we see that adding a constant to every component of $g^*$ does not change the optimal decision. Therefore, we only need to learn the differences $g^*_k-g^*_1$ for $k > 1$, so we may assume without loss of generality that $g^*_1 = 0$. Our main assumption concerns the ``richness'' of the data-generating distribution. It is based on the recent optimal learning literature, which has found, in various settings, that desirable convergence rates require the observed values of $X$ to exhibit a sufficient degree of random variation. This is made more precise as follows.



\begin{assum}\label{a1new}
Let $Z$ be the random vector whose $j$th component is equal to $c_j\left(X\right) - g^*_j$. Suppose that there exist constants $\kappa_1,...,\kappa_4>0$ (which may depend on $g^*$) such that:
\begin{enumerate}
\item[i)] The density of $Z$ exists and is bounded below by $\kappa_2$ on $\left[-\kappa_1,\kappa_1\right]^K$, and bounded above by $\kappa_3$ everywhere on its support.
\item[ii)] Both $X$ and $Z$ have bounded support, i.e., $\|X\|_2 \leq \kappa_4$ and $\|Z\|_2 \leq \kappa_4$.
\end{enumerate}
\end{assum}

Assumption \ref{a1new}(i) is a so-called ``margin condition,'' requiring sufficient random variation near the decision boundary (where multiple components of $Z$ are close to zero). Conditions of this type in the existing literature include Assumption A2 of \cite{GoZe13}, Assumption 2 of \cite{BaBa20}, and Assumption 2 of \cite{BaBaKh21}, all of which also assume boundedness as in Assumption \ref{a1new}(ii). The main difference between these papers and our work is that they impose margin conditions for pairs of choices (in our setting, these would be the pairwise differences $Z_j - Z_k$), while we make the slightly stronger assumption that the margin condition holds on the entire multivariate density of $Z$. The distinctions between assumptions are due to the fact that, in the above-cited papers, the penalties $g^*_j$ are not present at all, and the ultimate goal is only to learn the cost functions, which is done in a ``marginal'' way, by observing one $k$ at a time. In our setting, even when the costs are known, we still have the problem of learning $g^*_j$, which is done ``jointly'' by updating our estimates of all the penalties in each iteration of (\ref{eq:basicsa}). It therefore becomes necessary to consider the decision boundary in $K$ dimensions.

When Assumption \ref{a1new} is imposed, one then obtains certain useful technical properties for the pairwise differences. These properties will be useful for our analysis later. We state them below; the proof is deferred to the Appendix. Note that the second property considers the pairwise differences jointly by minimizing over them, which is one reason why Assumption \ref{a1new} holds on the multivariate density of $Z$.

\begin{lem}\label{a1}
Let Assumption \ref{a1new} hold. There exist constants $\kappa'_1,...,\kappa'_4>0$ (which depend on $\kappa_1,...,\kappa_4$) such that:
\begin{enumerate}
\item[i)] For any $1 \leq j,k \leq K$ with $j\neq k$, the random variable
\begin{equation*}
Z_{j,k} = \left(c_j\left(X\right) - g^*_j\right) - \left(c_k\left(X\right) - g^*_k\right)
\end{equation*}
has a density that is bounded below by $\kappa'_2$ on the interval $\left[-\kappa'_1,\kappa'_1\right]$, and bounded above by $\kappa'_3$ everywhere.
\item[ii)] For any $j\neq k$ and any $0 < \kappa < \kappa'_1$, we have
\begin{equation*}
P\left(\min_{\ell \neq j} Z_{\ell,k} \geq \kappa'_1 \mid 0 < Z_{j,k} \leq \kappa\right) \geq \kappa'_4.
\end{equation*}
\end{enumerate}
\end{lem}


Under known costs, our algorithm for learning $g^*$ is given by (\ref{eq:basicsa}) using a simple stepsize $\alpha_{n} = \frac{\alpha}{n+1}$ with $\alpha>0$ being a pre-specified constant. Our main result is a rate bound on the convergence in $L^2$ of the sequence $\left\{\delta^n\right\}^{\infty}_{n=0}$, with $\delta^n = g^n - g^*$.

\begin{thm}\label{thm:known}
Let Assumption \ref{a1new} hold. There exists some $\bar{\alpha} > 0$, which depends only on $\kappa_1,\kappa_2,\kappa_3,\kappa_4$, such that $\mathbb{E}\left(\|\delta^n\|^2_2\right) \leq \frac{D}{n}$ if $\alpha > \bar{\alpha}$.
\end{thm}

Thus, under Assumption \ref{a1new}, classical SA achieves the canonical $\mathcal{O}\left(\frac{1}{n}\right)$ rate. It is interesting to note that this does not happen in general: one usually resorts to Polyak averaging \citep{PoJu92} to recover the optimal convergence rate, which is indeed done in \cite{BaMo11}. However, in the specific setting of semidiscrete optimal transport with known costs, neither averaging nor strong concavity are necessary as long as the data-generating process is rich in the sense of Assumption \ref{a1new}. This result is completely new to our paper, and is valuable for preserving one of the most attractive characteristics of the SA approach to semidiscrete optimal transport, namely its ability to learn (in the limit) the exact optimal policy without resorting to approximation or discretization of the problem.

The result of Theorem \ref{thm:known} depends on two parameters, the threshold $\bar{\alpha}$ used by the stepsize, and the constant $D$ in the bound. We briefly clarify their dependence on the constants $\kappa_1,...,\kappa_4$ obtained from Assumption \ref{a1new}. The threshold $\bar{\alpha}$ is the main source of this dependence, because $D$ is essentially an increasing function of $\alpha > \bar{\alpha}$. Our proofs indicate that $\bar{\alpha}$ is decreasing in $\kappa_1$ and $\kappa_2$ (that is, $\bar{\alpha}$ is larger when these constants are smaller) and increasing in $\kappa_3$ and $\kappa_4$. It is not possible to precisely quantify each relationship because, for example, $\bar{\alpha}$ depends on the support of $Z$ (which is controlled by $\kappa_4$) indirectly, through the support of the pairwise differences $Z_{j,k}$. However, roughly speaking, expanding the support of $X$ generally increases the support of $Z_{j,k}$ and thus the threshold.

\subsection{Semi-myopic learning: unknown costs}\label{sec:mainresultsunknown}

Now, let us turn to the case where the cost functions are unknown. In each iteration, we still observe $X^{n+1}$, but we do not know $c_k\left(X^{n+1}\right)$. We only have access to approximate cost functions $\hat{c}^n_k$, which now have to be used to update $g^n$. We can, however, select one particular $k$ and collect a noisy observation of the form $W^{n+1}_k = c_k\left(X^{n+1}\right) + \varepsilon^{n+1}_k$, where $\left\{\varepsilon^n_k\right\}$ are i.i.d. noise terms with mean zero, assumed to be independent of $\left\{X^n\right\}$. Additionally, we can only observe one function at a time: if we choose $k$ at iteration $n$, we cannot see $W^{n+1}_j$ for $j\neq k$.

The techniques developed in our paper apply to cost functions of the form $c_k\left(X\right) = \chi\left(\beta^\top_k X\right)$, where $\chi$ is a link function associated with some generalized linear model \citep{McNe89} used in statistics. The simplest link function is the identity, in which case the cost functions are linear. Other possible $\chi$ include the logistic link $\chi\left(t\right) = \frac{1}{1+e^{-t}}$ and the Poisson link $\chi\left(t\right) = e^t$. In all of these cases, $\chi$ is known; the choice of link function depends on the application at hand, with, e.g., the logistic link modeling bounded costs, the Poisson link modeling integer-valued costs, and so on. However, in all cases, the parameters $\beta_k$ are unknown, and the approximate cost functions are given by $\hat{c}^n_k\left(X\right) = \chi\left(\left(\hat{\beta}^n_k\right)^\top X\right)$, where $\hat{\beta}^n_k$ is a maximum likelihood estimator of $\beta_k$ computed using all past observations (up to the $n$th time stage) of the $k$th cost function.

\begin{algorithm}[t]
 \mbox{}\hrulefill\mbox{}
 \begin{description}
    \item[Step 0:] Initialize $n =0$, $g^0$ and $\hat{\beta}^0_k$ for $k=1,...,K$. Choose a deterministic stepsize sequence $\left\{\alpha_n\right\}^{\infty}_{n=0}$.
    \item[Step 1:] Observe $X^{n+1}$ and select $\pi^{n+1} \in \left\{1,...,K\right\}$.
    \item[Step 2:] Update
    \begin{eqnarray}
    g^{n+1}_k &=& g^n_k + \alpha_{n}\left(-1_{\left\{k = \arg\min_j \hat{c}^n_j\left(X^{n+1}\right) - g^n_j\right\}}+p_k\right).\label{eq:approxsa}
    \end{eqnarray}
    \item[Step 3:] Observe
    \begin{equation*}
    W^{n+1}_{\pi^{n+1}} = c_j\left(X^{n+1}\right) + \varepsilon^{n+1}_{\pi^{n+1}}
    \end{equation*}
    and use this quantity to calculate an updated $\hat{\beta}^{n+1}_{\pi^{n+1}}$. For all $k\neq\pi^{n+1}$, let $\hat{\beta}^{n+1}_k = \hat{\beta}^n_k$.
    \item[Step 4:] Increment $n$ by $1$ and return to Step 1.
 \end{description}
 \mbox{}\hrulefill\mbox{}
 \vspace{0.1in}
 \caption{Algorithmic framework for semidiscrete optimal transport with unknown (estimated) costs.}\label{fig:sa}
\end{algorithm}

The cost function to be observed, formally denoted as $\pi^{n+1}$, is selected in an online manner: it depends on the available estimators $\hat{\beta}^n_k$ and $g^n$, but also on the new data $X^{n+1}$. Thus, $\pi^{n+1}$ affects the approximate costs in the next iteration, and through these approximations, the next gradient update as well. This algorithmic framework, laid out in Algorithm \ref{fig:sa}, can be viewed as a coupling of the SA procedure (\ref{eq:basicsa}) with optimal learning (e.g., bandit problems). As in optimal learning, we can only sample one ``alternative'' (cost function) at a time, but the policy used to do this is linked to the bias of the SA update.

As discussed in Section \ref{sec:intunknown}, our analysis uses a particular construction of $\pi^{n+1}$ known as a semi-myopic policy. This approach approximates (\ref{eq:truepolicy}) using plug-in estimates of $g^*$ and $\beta_k$. Most decisions are made based on this approximation, but we occasionally force ourselves to sample different $k$ to prevent the estimates from stalling. Forced exploration occurs less frequently as time goes on.

Formally, we first define $\hat{\pi}^{n+1} = \arg\min_j \hat{c}^n_j\left(X^{n+1}\right) - g^n_j$ to be the plug-in estimate of the optimal decision, based on estimates $\hat{\beta}^n$ and $g^n$. We also define, for each $k = 1,...,K$, the set
\begin{equation*}
\mathcal{T}_k = \left\{\lceil\exp\left(a\cdot m^{\frac{1}{9}}\right)\rceil : \frac{m}{k+1}\in\mathbb{N}\right\}
\end{equation*}
of time periods where we will force ourselves to explore $k$. Any $a>0$ works for the theory; in practice, this value is tunable (for instance, $a \geq 5$ produces fairly infrequent forced exploration). Then, the $\left(n+1\right)$st decision is made according to the policy
\begin{equation}\label{eq:ourpolicy}
\pi^{n+1} = \left\{
\begin{array}{c l}
k & \mbox{ if } n\in\mathcal{T}_k,\\
\hat{\pi}^{n+1} & \mbox{otherwise.}
\end{array}
\right.
\end{equation}
The estimate $g^n_j$ is updated using (\ref{eq:approxsa}), and the regression coefficients are computed using a suitable generalized linear model. Thus, in order to implement the algorithm, we require only an exploration parameter $a > 0$ and a stepsize sequence; as in Section \ref{sec:mainresultsknown}, we will use $\alpha_n = \frac{\alpha}{n+1}$, so only a single parameter $\alpha$ needs to be specified.


We now proceed to the assumptions and main results. Here and in Section \ref{sec:unknown}, we present our main results for the special case of \textit{linear} costs $c_k\left(X\right) = \beta^\top_k X$, for which we use the ridge-like estimator
\begin{equation}\label{eq:ridge}
\hat{\beta}^{n+1}_k = \left(\rho^{n+1} \cdot I + \sum^{n+1}_{m=1} X^m\left(X^m\right)^\top 1_{\left\{\pi^m=k\right\}}\right)^{-1}\sum^{n+1}_{m=1} W^m_k X^m 1_{\left\{\pi^m=k\right\}},
\end{equation}
where $\rho^n = 1 + \left(\log n\right)^3$. A rigorous extension to more general $\chi$ is provided in the Appendix. The results and analysis for that case are very similar, with the estimator $\hat{\beta}^{n+1}_k$ being a modification of maximum likelihood, just as (\ref{eq:ridge}) is a modification of ordinary least squares. The reason to focus on linear costs here is for ease of presentation: the analysis becomes somewhat more streamlined, avoiding technical nuisances and allowing the reader to understand the main concepts. We emphasize, however, that the results hold for general $\chi$ and the extension does not add much difficulty to what we present here.

As before, we let $\delta^n = g^n - g^*$. We also define $\Delta^n_k = \hat{\beta}^n_k - \beta_k$. Recall the distinction between $\hat{\pi}^n$, which is the plug-in estimate of the optimal decision, and $\pi^n$, which is the semi-myopic policy combining that estimate with forced exploration as defined in (\ref{eq:ourpolicy}).

We require two additional assumptions on the data-generating distribution. The first is a standard assumption on the residual noise terms, used in virtually all proofs in contemporary learning theory. The second is very similar to Assumption A3 in \cite{GoZe13}, and is weaker than Assumption 3 of \cite{BaBaKh21}.

\begin{assum}\label{a2}
Suppose that there exist constants $\kappa_5,\kappa_6>0$ such that:
\begin{enumerate}
\item[i)] The residual noise is conditionally sub-Gaussian: for any $k$, $\mathbb{E}\left(\exp\left(t\varepsilon^n_k\right)\mid X^n\right) \leq \exp\left(\frac{1}{2}\kappa_5 t^2\right)$.
\item[ii)] Let $\lambda_{\min}\left(\cdot\right)$ be the function that returns the smallest eigenvalue of the matrix argument. For any $k$,
\begin{equation*}
\lambda_{\min}\left(\mathbb{E}\left(X \left(X\right)^\top1_{\left\{\pi^*\left(X\right)=k\right\}}\right)\right) \geq \kappa_6.
\end{equation*}
\end{enumerate}
\end{assum}

The bulk of the analysis focuses on bounding the convergence rate of the estimation error for both $g^n$ and $\hat{\beta}^n_k$. These results are stated in Theorem \ref{thm:errorrates}. Then, Theorem \ref{thm:decision} obtains a rate for the accuracy of the plug-in estimate $\hat{\pi}^{n}$, in the sense of the PICS metric defined in Section \ref{sec:preliminaries}.

\begin{thm}\label{thm:errorrates}
Let Assumptions \ref{a1new} and \ref{a2} hold, and suppose that the semi-myopic policy $\pi^n$ is used to sample cost functions. There exists $\bar{\alpha}$ depending only on $\kappa_1,...,\kappa_6$, such that, when $\alpha > \bar{\alpha}$, we have
\begin{equation*}
\mathbb{E}\left(\max_j \|\Delta^n_j\|^2_2\right) \leq \frac{C}{n}, \qquad \mathbb{E}\left(\|\delta^n\|^2_2\right) \leq \frac{C'}{n}
\end{equation*}
for some $C,C'>0$ depending on $\alpha,\kappa_1,...,\kappa_6$.
\end{thm}

As in the case of Theorem \ref{thm:known}, we may briefly discuss the dependence of the threshold $\bar{\alpha}$ on the constants $\kappa_1,...,\kappa_6$. As before, the threshold is decreasing in $\kappa_1,\kappa_2$ and increasing in $\kappa_3$. The dependence on $\kappa_4$ is more complex because the support of $X$ indirectly influences various constants in the proof, but it is possible to say that the threshold becomes increasing in $\kappa_4$ for large enough values of this constant. The new constants $\kappa_5,\kappa_6$ that are specific to the unknown-cost setting do not influence $\bar{\alpha}$, only the rate constants $C,C'$. Unfortunately, the dependence of $C,C'$ on $\kappa_1,...,\kappa_6$ is non-monotonic and very difficult to characterize.

\begin{thm}\label{thm:decision}
Suppose that we are in the situation of Theorem \ref{thm:errorrates}.  Then, there exists $D > 0$, which depends only on $\kappa_1,...,\kappa_6$, such that the plug-in estimate $\hat{\pi}^n$ satisfies
\begin{equation}\label{eq:decision}
P\left(\hat{\pi}^{n} \neq \pi^*\left(X^n\right)\right) \leq \frac{D}{\sqrt{n}}.
\end{equation}
\end{thm}

The square-root rate in (\ref{eq:decision}) is due to the fact that some $X$ values can make it arbitrarily difficult to distinguish between choices. Faster rates (e.g., exponential rates, as in \citealp{GaDuCh19}) can only be obtained if one assumes some minimal amount of separation between objective values, for example by making the support of $X$ finite. If this is not the case, it will be impossible to improve on (\ref{eq:decision}).

While we focus on PICS rather than regret in this work, our result is comparable to an optimal regret bound. For example, \cite{GoZe13} obtains a single-period regret rate of $\mathcal{O}\left(\frac{1}{n}\right)$ (a known universal lower bound) from a bound that multiplies together two terms, one involving the estimation error $\|\Delta^n_j\|_2$ of the regression coefficients, and one that looks similar to PICS. Individually, both terms converge at a rate of $\mathcal{O}\left(\frac{1}{\sqrt{n}}\right)$, as they do in our paper. One can view that problem as a special case of ours with known $g^*\equiv 0$, which implies that it is not possible to guarantee faster rates than these.

A straightforward consequence of Theorem \ref{thm:decision} is that the expected number of incorrect selections made by $\hat{\pi}^{n+1}$ (essentially adding up (\ref{eq:decision}) over $n$) grows according to $\mathcal{O}\left(\sqrt{n}\right)$. This rate does not change if we count incorrect selections made by the semi-myopic policy (incorporating the forced exploration periods) rather than simply the plug-in estimate, because forced exploration occurs only in $\mathcal{O}\left(\left(\log n\right)^9\right)$ time periods.

\begin{corol}
Suppose that we are in the situation of Theorem \ref{thm:errorrates}.  Then, there exists $D' > 0$, which depends only on $\kappa_1,...,\kappa_6$, such that
\begin{equation*}
\mathbb{E}\left(\sum^n_{m=1} 1_{\left\{\pi^{m} \neq \pi^*\left(X^m\right)\right\}}\right) \leq D' \cdot \sqrt{n}.
\end{equation*}
\end{corol}

\section{Stochastic approximation: technical analysis}\label{sec:known}

Earlier, in Section \ref{sec:algorithms}, we separated the presentation of the main results according to the core components of our algorithm: first, stochastic approximation under known costs (Theorem \ref{thm:known}), and then semi-myopic learning under unknown costs (Theorems \ref{thm:errorrates}-\ref{thm:decision}). We make the same distinction when presenting the proofs. In this section, we confine ourselves to the setting of Section \ref{sec:mainresultsunknown} and prove Theorem \ref{thm:known}. The unknown-cost case will be treated later in Section \ref{sec:unknown}.

As was discussed previously, the main technical challenge in establishing Theorem \ref{thm:known} is the lack of strong concavity in the objective function. A crucial first step is the derivation (in Section \ref{sec:concavity}) of a kind of ``local strong concavity'' property that is weaker than strong concavity, but still sufficient to prove (in Section \ref{sec:rateknown}) the main rate result. We will later use this property in our analysis of unknown costs, but it holds more generally and does not require $c_k$ to have any parametric structure. For that reason, the results in this section are of some stand-alone interest, because we are the first to show that semidiscrete optimal transport can be solved at an optimal rate without the need for discretization or smoothing, as long as the data-generating distribution is sufficiently rich.


\subsection{Properties of the objective}\label{sec:concavity}

We establish two technical results concerning the smoothness of the objective $\mathbb{E}\left(F\left(g,X\right)\right)$. The proof techniques will be used again later on, in Section \ref{sec:unknown}, where these properties will be generalized to the setting of unknown cost functions. We will assume Assumption \ref{a1new} throughout, denoting by $\kappa'_1,...,\kappa'_4$ the constants obtained from Lemma \ref{a1}.

\begin{lem}\label{lem:technical1}
Let Assumption \ref{a1new} hold, and take any $\delta\in\mathds{R}^K$ satisfying $\delta_1=0$. Then,
\begin{equation*}
\mathbb{E}\left(F\left(g^*,X\right)\right)-\mathbb{E}\left(F\left(g^*+\delta,X\right)\right) - \delta^\top p \geq \frac{\kappa'_2\kappa'_4}{4K}\left(\sum^K_{k=1} \min\left\{\delta^2_k,\left(\kappa'_1\right)^2\right\}\right).
\end{equation*}
\end{lem}

\noindent\textbf{Proof:} To avoid notational clutter, we write $g^*$ as simply $g$ in this proof, since no other $g$ will be considered. By definition of $F$, we can write
\begin{eqnarray*}
&\,& F\left(g,X\right)-F\left(g+\delta,X\right)\\
 &=&  \sum^K_{k=1} \left(c_k\left(X\right)-g_k\right)1_{\left\{c_k\left(X\right)-g_k = F\left(g,X\right)\right\}}-\sum^K_{k=1} \left(c_k\left(X\right)-g_k - \delta_k\right)1_{\left\{c_k\left(X\right)-g_k-\delta_k = F\left(g+\delta,X\right)\right\}}\\
 &=& T_1 + T_2,
\end{eqnarray*}
where
\begin{eqnarray}
T_1 &=& \sum^K_{k=1} \delta_k 1_{\left\{c_k\left(X\right)-g_k = F\left(g,X\right)\right\}},\nonumber\\
T_2 &=& \sum^K_{k=1} \left(c_k\left(X\right)-g_k-\delta_k\right)\left(1_{\left\{c_k\left(X\right)-g_k = F\left(g,X\right)\right\}}-1_{\left\{c_k\left(X\right)-g_k-\delta_k = F\left(g+\delta,X\right)\right\}}\right).\label{eq:T2}
\end{eqnarray}
Clearly, $\mathbb{E}\left(T_1\right) = \delta^\top p$. We will work toward a lower bound on $\mathbb{E}\left(T_2\right)$, which will require some algebra on (\ref{eq:T2}). Because the indices minimizing $F\left(g+\delta,X\right)$ and $F\left(g,X\right)$ are unique by Assumption \ref{a1new}, we can write
\begin{eqnarray*}
T_2 &=& \sum^K_{k=1} \left[\left(c_k\left(X\right)-g_k-\delta_k\right)-F\left(g+\delta,X\right)\right]\cdot \left(1_{\left\{c_k\left(X\right)-g_k = F\left(g,X\right)\right\}}-1_{\left\{c_k\left(X\right)-g_k-\delta_k = F\left(g+\delta,X\right)\right\}}\right)\\
&=& \sum^K_{k=1}\left[\left(c_k\left(X\right)-g_k-\delta_k\right)-F\left(g+\delta,X\right)\right]\cdot 1_{\left\{c_k\left(X\right)-g_k = F\left(g,X\right),c_k\left(X\right) - g_k-\delta_k > F\left(g+\delta,X\right)\right\}}
\end{eqnarray*}
The mapping $z\mapsto \left[\left(c_k\left(X\right)-g_k-\delta_k\right)-z\right]\cdot 1_{\left\{c_k\left(X\right)-g_k = F\left(g,X\right),c_k\left(X\right) - g_k-\delta_k > z\right\}}$ is decreasing in $z$. Because $F\left(g+\delta,X\right)\leq c_j\left(X\right)-g_j-\delta_j$ for any $j$ by definition, we then have
\begin{eqnarray*}
&\,& \left[\left(c_k\left(X\right)-g_k-\delta_k\right)-F\left(g+\delta,X\right)\right]\cdot 1_{\left\{c_k\left(X\right)-g_k = F\left(g,X\right),c_k\left(X\right) - g_k-\delta_k > F\left(g+\delta,X\right)\right\}}\\
&\geq & \frac{1}{K}\sum^K_{j=1}\left[\left(c_k\left(X\right)-g_k-\delta_k\right)-\left(c_j\left(X\right)-g_j-\delta_j\right)\right]\cdot 1_{\left\{c_k\left(X\right)-g_k = F\left(g,X\right),c_k\left(X\right) - g_k-\delta_k > c_j\left(X\right)-g_j-\delta_j\right\}}.
\end{eqnarray*}
Consequently,
\begin{equation}\label{eq:T2-2}
T_2 \geq \frac{1}{K} \sum^K_{k=1}\sum^K_{j=1}\left[\left(c_k\left(X\right)-g_k-\delta_k\right)-\left(c_j\left(X\right)-g_j-\delta_j\right)\right]\cdot 1_{\left\{c_k\left(X\right)-g_k = F\left(g,X\right),c_k\left(X\right) - g_k-\delta_k > c_j\left(X\right)-g_j-\delta_j\right\}}.
\end{equation}

We now seek to bound the expectation of each term in (\ref{eq:T2-2}). For convenience, let $Z_{j,k} = \left(c_j\left(X\right)-g_j\right)-\left(c_k\left(X\right)-g_k\right)$ for any $j,k$. Similarly, let $\bar{\delta}_{j,k} = \delta_j-\delta_k$. We write
\begin{eqnarray*}
&\,& \left[\left(c_k\left(X\right)-g_k-\delta_k\right)-\left(c_j\left(X\right)-g_j-\delta_j\right)\right]\cdot 1_{\left\{c_k\left(X\right)-g_k = F\left(g,X\right),c_k\left(X\right) - g_k-\delta_k > c_j\left(X\right)-g_j-\delta_j\right\}}\\
&=& \left(\bar{\delta}_{j,k}-Z_{j,k}\right)\cdot 1_{\left\{\min_{\ell \neq j}Z_{\ell,k}\geq 0,\; 0\leq Z_{j,k}<\bar{\delta}_{j,k}\right\}}.
\end{eqnarray*}
Taking expectations, we obtain
\begin{eqnarray}
&\,& \mathbb{E}\left\{\left[\left(c_k\left(X\right)-g_k-\delta_k\right)-\left(c_j\left(X\right)-g_j-\delta_j\right)\right]\cdot 1_{\left\{c_k\left(X\right)-g_k = F\left(g,X\right),c_k\left(X\right) - g_k-\delta_k > c_j\left(X\right)-g_j-\delta_j\right\}}\right\}\nonumber\\
&=& \mathbb{E}\left\{ \left(\bar{\delta}_{j,k}-Z_{j,k}\right)\cdot 1_{\left\{0\leq Z_{j,k}<\bar{\delta}_{j,k}\right\}}P\left(\min_{\ell \neq j}Z_{\ell,k}\geq 0 \mid Z_{j,k}\right)\right\}\label{eq:towerproperty}\\
&=& \int^{\bar{\delta}_{j,k}}_0 \left(\bar{\delta}_{j,k}-z\right)P\left(\min_{\ell \neq j}Z_{\ell,k}\geq 0 \mid Z_{j,k}=z\right)P\left(Z_{j,k}\in dz\right)\nonumber\\
&\geq& \frac{1}{2}\bar{\delta}_{j,k} \int^{\frac{\bar{\delta}_{j,k}}{2}}_0 P\left(\min_{\ell \neq j}Z_{\ell,k}\geq 0 \mid Z_{j,k}=z\right)P\left(Z_{j,k}\in dz\right)\label{eq:Deltamanipulations}\\
&\geq &\frac{1}{2}\bar{\delta}_{j,k} \int^{\frac{\bar{\delta}_{j,k}}{2}}_0 P\left(\min_{\ell \neq j}Z_{\ell,k}\geq \kappa'_1 \mid Z_{j,k}=z\right)P\left(Z_{j,k}\in dz\right)\label{eq:Deltamanipulations2}
\end{eqnarray}
where (\ref{eq:towerproperty}) follows by the tower property, and (\ref{eq:Deltamanipulations}) follows because $\bar{\delta}_{j,k}-z\geq 0$ for $z \in \left[0,\bar{\delta}_{j,k}\right]$.

Using Lemma \ref{a1}(ii), and taking $0 < \kappa<\kappa'_1$, we write
\begin{eqnarray}
\kappa'_4 &\leq & P\left(\min_{\ell\neq j} Z_{\ell,k} \geq \kappa'_1 \mid 0\leq Z_{j,k} \leq \kappa\right)\nonumber\\
&=& \frac{\int^\kappa_0 P\left(\min_{\ell \neq j}Z_{\ell,k}\geq \kappa'_1 \mid Z_{j,k}=z\right)P\left(Z_{j,k}\in dz\right)}{\int^\kappa_0 P\left(Z_{j,k}\in dz\right)},\label{eq:usesecondpart}
\end{eqnarray}
and, since $P\left(Z_{j,k}\in dz\right) \geq \kappa'_2$ on $0 \leq z \leq \kappa'_1$, (\ref{eq:usesecondpart}) yields
\begin{equation}\label{eq:kappa2squared}
\int^\kappa_0 P\left(\min_{\ell \neq j}Z_{\ell,k}\geq \kappa'_1 \mid Z_{j,k}=z\right)P\left(Z_{j,k}\in dz\right) \geq \kappa\cdot \kappa'_2\kappa'_4.
\end{equation}
Then, returning to (\ref{eq:Deltamanipulations}), we consider two cases. First, suppose that $0< \bar{\delta}_{j,k} < \kappa'_1$. Then, (\ref{eq:kappa2squared}) can be applied directly, yielding
\begin{eqnarray}
&\,& \mathbb{E}\left\{\left[\left(c_k\left(X\right)-g_k-\delta_k\right)-\left(c_j\left(X\right)-g_j-\delta_j\right)\right]\cdot 1_{\left\{c_k\left(X\right)-g_k = F\left(g,X\right),c_k\left(X\right) - g_k-\delta_k > c_j\left(X\right)-g_j-\delta_j\right\}}\right\}\nonumber\\
&\geq& \frac{1}{4}\kappa'_2\kappa'_4\bar{\delta}^2_{j,k}.\label{eq:boundbyDelta}
\end{eqnarray}
In the second case, we consider $\bar{\delta}_{j,k} \geq \kappa'_1$. The mapping
\begin{equation*}
x\mapsto \left(x-Z_{j,k}\right)\cdot 1_{\left\{0\leq Z_{j,k}<x\right\}}P\left(\min_{\ell \neq j}Z_{\ell,k}\geq 0 \mid Z_{j,k}\right)
\end{equation*}
is increasing on the interval $\left[\kappa'_1,\infty\right)$. Therefore, repeating the arguments used to derive (\ref{eq:Deltamanipulations2}), and applying (\ref{eq:kappa2squared}) again, we obtain
\begin{eqnarray}
&\,& \mathbb{E}\left\{\left[\left(c_k\left(X\right)-g_k-\delta_k\right)-\left(c_j\left(X\right)-g_j-\delta_j\right)\right]\cdot 1_{\left\{c_k\left(X\right)-g_k = F\left(g,X\right),c_k\left(X\right) - g_k-\delta_k > c_j\left(X\right)-g_j-\delta_j\right\}}\right\}\nonumber\\
&\geq& \frac{1}{2}\kappa'_1 \int^{\frac{\kappa'_1}{2}}_0 P\left(\min_{\ell \neq j}Z_{\ell,k}\geq \kappa'_1 \mid Z_{j,k}=z\right)P\left(Z_{j,k}\in dz\right)\nonumber\\
&\geq& \frac{1}{4}\left(\kappa'_1\right)^2\kappa'_2\kappa'_4.\label{eq:boundbykappa1}
\end{eqnarray}
Combining (\ref{eq:boundbyDelta}) and (\ref{eq:boundbykappa1}), we have
\begin{eqnarray}
&\,& \mathbb{E}\left\{\left[\left(c_k\left(X\right)-g_k-\delta_k\right)-\left(c_j\left(X\right)-g_j-\delta_j\right)\right]\cdot 1_{\left\{c_k\left(X\right)-g_k = F\left(g,X\right),c_k\left(X\right) - g_k-\delta_k > c_j\left(X\right)-g_j-\delta_j\right\}}\right\}\nonumber\\
&\geq & \frac{1}{4}\kappa'_2\kappa'_4\min\left\{\bar{\delta}^2_{j,k},\left(\kappa'_1\right)^2\right\}\label{eq:boundbyminimum}
\end{eqnarray}
for any $\bar{\delta}_{j,k}> 0$. Since the left-hand side of (\ref{eq:boundbyminimum}) is always positive, we have
\begin{eqnarray}
&\,& \mathbb{E}\left\{\left[\left(c_k\left(X\right)-g_k-\delta_k\right)-\left(c_j\left(X\right)-g_j-\delta_j\right)\right]\cdot 1_{\left\{c_k\left(X\right)-g_k = F\left(g,X\right),c_k\left(X\right) - g_k-\delta_k > c_j\left(X\right)-g_j-\delta_j\right\}}\right\}\nonumber\\
&\geq & \frac{1}{4}\kappa'_2\kappa'_4\min\left\{\bar{\delta}^2_{j,k},\left(\kappa'_1\right)^2\right\}\cdot 1_{\left\{\bar{\delta}_{j,k}\geq 0\right\}}\label{eq:boundbyminind}
\end{eqnarray}
for any value (positive or negative) of $\bar{\delta}_{j,k}$.

We may now derive the desired result. Combining (\ref{eq:T2-2}) and (\ref{eq:boundbyminind}), we have
\begin{equation}\label{eq:T2-3}
\mathbb{E}\left(T_2\right) \geq \frac{1}{4K}\kappa'_2\kappa'_4\sum^K_{k=1}\sum^K_{j=1} \min\left\{\bar{\delta}^2_{j,k},\left(\kappa'_1\right)^2\right\}\cdot 1_{\left\{\bar{\delta}_{j,k}\geq 0\right\}}.
\end{equation}
Because, in (\ref{eq:T2-3}), all of the terms in the double sum are positive, we may take another lower bound in which we only keep those terms in which either $j =1$ or $k=1$. However, for any $j,k$, either $\bar{\delta}_{j,k}\geq 0$ or $\bar{\delta}_{k,j}\geq 0$. Therefore, we have
\begin{equation*}
\mathbb{E}\left(T_2\right) \geq \frac{1}{4K}\kappa'_2\kappa'_4\sum^K_{k=1} \min\left\{\delta^2_k,\left(\kappa'_1\right)^2\right\},
\end{equation*}
whence
\begin{equation*}
\mathbb{E}\left(T_1+T_2\right) \geq \delta^\top p + \frac{1}{4K}\kappa'_2\kappa'_4\sum^K_{k=1} \min\left\{\delta^2_k,\left(\kappa'_1\right)^2\right\},
\end{equation*}
as required.\qed

The second property represents, again, a kind of local strong concavity. In ordinary concavity, the right-hand side of inequality (\ref{eq:localstrongconc}) below is simply zero. If we had strong concavity, the right-hand side would be $\|\delta\|^2_2$. Thus, we have that property only when $\|\delta\|_2$ is small, i.e., $g$ is in a neighborhood of $g^*$. For notational convenience, we let $p^g$ be the vector whose components are given by $p^g_k = P\left(c_k\left(X\right)-g_k = F\left(g,X\right)\right)$, noting that $p^{g^*} = p$.

\begin{lem}\label{lem:technical2}
Suppose that we are in the situation of Lemma \ref{lem:technical1}. Then,
\begin{equation}\label{eq:localstrongconc}
\delta^\top\left(p^{g^*+\delta}-p\right) \geq \frac{\kappa'_2\kappa'_4}{4K}\min\left\{\|\delta\|^2_2,\left(\kappa'_1\right)^2\right\}.
\end{equation}
\end{lem}

\noindent\textbf{Proof:} To avoid notational clutter, we write $g^*$ as simply $g$ in this proof, since no other $g$ will be considered. By the concavity of the function $g\mapsto F\left(g,X\right)$, we have
\begin{equation*}
\mathbb{E}\left(F\left(g+\delta,X\right)\right) - \mathbb{E}\left(F\left(g,X\right)\right) + \delta^\top p^{g+\delta} \geq 0.
\end{equation*}
We then apply Lemma \ref{lem:technical1}. It remains only to show that
\begin{equation*}
\sum^K_{k=1} \min\left\{\delta^2_k,\left(\kappa'_1\right)^2\right\} \geq \min\left\{\sum^K_{k=1}\delta^2_k,\left(\kappa'_1\right)^2\right\}.
\end{equation*}
To see this, consider two cases. First, suppose that there is some $j$ for which $\left|\delta_j\right| \geq \kappa'_1$. Then,
\begin{equation*}
\sum^K_{k=1} \min\left\{\delta^2_k,\left(\kappa'_1\right)^2\right\} \geq \min\left\{\delta^2_j,\left(\kappa'_1\right)^2\right\} = \left(\kappa'_1\right)^2 \geq \min\left\{\sum^K_{k=1}\delta^2_k,\left(\kappa'_1\right)^2\right\}.
\end{equation*}
Second, suppose that $\left|\delta_k\right| < \kappa'_1$ for all $k$. Then,
\begin{equation*}
\sum^K_{k=1} \min\left\{\delta^2_k,\left(\kappa'_1\right)^2\right\} = \sum^K_{k=1}\delta^2_k \geq \min\left\{\sum^K_{k=1}\delta^2_k,\left(\kappa'_1\right)^2\right\}.
\end{equation*}
This completes the proof.\qed

\subsection{Proof of Theorem \ref{thm:known}}\label{sec:rateknown}

We rewrite (\ref{eq:basicsa}) in vector form as $g^{n+1} = g^n - \alpha_{n}\left(\zeta^{n+1}-p\right)$, where $\zeta^{n+1}_k = 1_{\left\{k=\arg\min_j c_j\left(X^{n+1}\right)-g^n_j\right\}}$. We may also write a similar recursion
\begin{equation}\label{eq:basicdelta}
\delta^{n+1}=\delta^n-\frac{\alpha}{n+1}\left(\zeta^{n+1}-p\right)
\end{equation}
for the error sequence. Before proceeding to the convergence rate, we first prove two intermediate results: an almost sure (but non-vanishing) bound on $\|\delta_n\|_2$, and a concentration inequality showing that the tail probabilities of $\|\delta^n\|_2$ vanish quickly outside a certain bounded range. The proofs are deferred to the Appendix (Section \ref{sec:proofs}).

\begin{lem}\label{lem:technicalknown1}
Let $s_n = \|\delta^0\|_2 + 3\alpha \log\left(n+1\right)$. Then, $P\left(\|\delta^n\|_2\leq s_n\right) = 1$ for $n\geq 1$.
\end{lem}

\begin{lem}\label{lem:technicalknown2}
Let Assumption \ref{a1new} hold. There exists a constant $C >0$ such that $P\left(\|\delta^n\|_2 > 2\right) \leq \frac{C}{n^9}$.
\end{lem}

Finally, we prove Theorem \ref{thm:known}. Let $C_0 = 2$ and define $p^n = p^{g^*+\delta^n}$. Using (\ref{eq:basicdelta}), we may write
\begin{eqnarray}
\|\delta^{n+1}\|^2_2 &\leq& \|\delta^n\|^2_2 + \frac{\alpha^2}{\left(n+1\right)^2}\|\zeta^{n+1}-p\|^2_2 - \frac{2\alpha}{n+1}\left(\zeta^{n+1}-p\right)^\top \delta^n\nonumber\\
&\leq & \|\delta^n\|^2_2 + \frac{2\alpha^2}{\left(n+1\right)^2} - \frac{2\alpha}{n+1}\left(\zeta^{n+1}-p\right)^\top \delta^n\nonumber\\
&=& \|\delta^n\|^2_2 + \frac{2\alpha^2}{\left(n+1\right)^2} - \frac{2\alpha}{n+1}\left(\zeta^{n+1}-p^n\right)^\top \delta^n - \frac{2\alpha}{n+1}\left(p^n-p\right)^\top \delta^n\nonumber\\
&\leq& \|\delta^n\|^2_2 + \frac{2\alpha^2}{\left(n+1\right)^2} -\frac{2\alpha}{n+1}\left(\zeta^{n+1}-p^n\right)^\top \delta^n - \frac{2\alpha C_1}{n+1}\min\left\{\|\delta^n\|^2_2,\left(\kappa'_1\right)^2\right\},\label{eq:deltadecomposition}
\end{eqnarray}
where (\ref{eq:deltadecomposition}) uses Lemma \ref{lem:technical2} with an appropriate $C_1 > 0$. Letting $\mathcal{F}^n$ be the sigma-algebra generated by $X^1,...,X^n$, we have
\begin{equation*}
\mathbb{E}\left(\|\delta^{n+1}\|^2_2\mid \mathcal{F}^n\right) \leq \|\delta^n\|^2_2 + \frac{2\alpha^2}{\left(n+1\right)^2} - \frac{2\alpha C_1}{n+1}\min\left\{\|\delta\|^2_2,\left(\kappa'_1\right)^2\right\}.
\end{equation*}
On the event $\left\{\|\delta^n\|_2 \leq C_0\right\}$, we have
\begin{eqnarray*}
\mathbb{E}\left(\|\delta^{n+1}\|^2_2\mid \mathcal{F}^n\right) &\leq&  \|\delta^n\|^2_2 + \frac{2\alpha^2}{\left(n+1\right)^2} - \frac{2\alpha C_1}{n+1}\|\delta^n\|^2_2\min\left\{1,\frac{\left(\kappa'_1\right)^2}{\|\delta^n\|^2}\right\}\\
&\leq & \|\delta^n\|^2_2 + \frac{2\alpha^2}{\left(n+1\right)^2} - \frac{2\alpha C_1}{n+1}\|\delta^n\|^2_2\min\left\{1,\frac{\left(\kappa'_1\right)^2}{C^2_0}\right\}\\
&=& \left(1 - \frac{2\alpha C_1}{n+1}\min\left\{1,\frac{\left(\kappa'_1\right)^2}{C^2_0}\right\}\right)\|\delta^n\|^2_2 + \frac{2\alpha^2}{\left(n+1\right)^2}.
\end{eqnarray*}
Therefore,
\begin{eqnarray*}
\mathbb{E}\left(\|\delta^{n+1}\|^2_2\mid \mathcal{F}^n\right) &=& \mathbb{E}\left(\|\delta^{n+1}\|^2_2 1_{\left\{\|\delta^n\|_2\leq C_0\right\}}\mid \mathcal{F}^n\right) + \mathbb{E}\left(\|\delta^{n+1}\|^2_2 1_{\left\{\|\delta^n\|_2> C_0\right\}}\mid \mathcal{F}^n\right)\\
&\leq & \left(1 - \frac{2\alpha C_1}{n+1}\min\left\{1,\frac{\left(\kappa'_1\right)^2}{C^2_0}\right\}\right)\|\delta^n\|^2_2 + \frac{2\alpha^2}{\left(n+1\right)^2} + \mathbb{E}\left(\|\delta^{n+1}\|^2_2 1_{\left\{\|\delta^n\|_2> C_0\right\}}\mid \mathcal{F}^n\right)\\
&\leq & \left(1 - \frac{2\alpha C_1}{n+1}\min\left\{1,\frac{\left(\kappa'_1\right)^2}{C^2_0}\right\}\right)\|\delta^n\|^2_2 + \frac{2\alpha^2}{\left(n+1\right)^2} + \alpha C_2 \log\left(n+1\right)1_{\left\{\|\delta^n\|_2 > C_0\right\}},
\end{eqnarray*}
where $C_2 > 0$ is a constant obtained from Lemma \ref{lem:technicalknown1}. Taking unconditional expectations and applying Lemma \ref{lem:technicalknown2}, we obtain
\begin{equation*}
\mathbb{E}\left(\|\delta^{n+1}\|^2_2\right) \leq \left(1 - \frac{2\alpha C_1}{n+1}\min\left\{1,\frac{\left(\kappa'_1\right)^2}{C^2_0}\right\}\right)\mathbb{E}\left(\|\delta^n\|^2_2\right) + \frac{2\alpha^2}{\left(n+1\right)^2} + \alpha C_2 \log\left(n+1\right)\frac{C}{n^9}.
\end{equation*}
Therefore, for some suitable $C_3,C_4 > 0$, we have
\begin{equation*}
\mathbb{E}\left(\|\delta^{n+1}\|^2_2\right) \leq \left(1 - \frac{\alpha C_3}{n+1}\right)\mathbb{E}\left(\|\delta^n\|^2_2\right) + \frac{C_4}{\left(n+1\right)^2}.
\end{equation*}
Letting $a_n = n\mathbb{E}\left(\|\delta^{n+1}\|^2_2\right)$, we have
\begin{equation}\label{eq:weirda}
a_{n+1} \leq \left(1 - \frac{\alpha C_3 - 1}{n}\right)a_n + \frac{C_4}{n}.
\end{equation}
A technical lemma in the Appendix (Lemma \ref{lem:weirdineq}) proves that a non-negative sequence $\left\{a_n\right\}$ satisfying (\ref{eq:weirda}) is uniformly bounded if the coefficient of $a_n$ on the right-hand side is positive. Thus, we obtain the desired result as long as $\alpha > \frac{1}{C_3}$.

\section{Semi-myopic learning: technical analysis}\label{sec:unknown}

Finally, we may prove our main result (Theorems \ref{thm:errorrates}-\ref{thm:decision}). As in Section \ref{sec:mainresultsunknown}, we will focus on the case of linear costs $c_k\left(X\right) = \beta^\top_k X$ where (\ref{eq:ridge}) is used for estimation. The Appendix presents a fully rigorous extension to the case of more general link functions $\chi$. The analysis in that extension will be very similar, with some additional technical complications in several steps.

The order of the arguments is similar to Section \ref{sec:known}. Section \ref{sec:concavity2} proves analogous properties to those in Section \ref{sec:concavity} for the case of unknown costs. Section \ref{sec:sketch} outlines the proof of the main rate results, with some intermediate steps moved to the Appendix. The core of the proof is the derivation of new concentration inequalities for both the SA error and the estimation error. The rates of these inequalities allow us to place a sub-exponential gap between periods of forced exploration.

\subsection{Properties of the objective}\label{sec:concavity2}

We proceed largely the same way as in Section \ref{sec:concavity}. This time, however, we generalize the notation to allow for different choices of the regression coefficients. Let $F^b\left(g,x\right) = \min_j b^\top_j x - g_j$, and let $p^{b,g}$ be the vector whose components are given by $p^{b,g}_k = P\left(b^\top_k X - g_k = F^b\left(g,X\right)\right)$. Note that $F^{\beta} = F$ and $p^{\beta,g^*} = p$.

The first two results are extensions of Lemmas \ref{lem:technical1}-\ref{lem:technical2}. Their main purpose is to prove local strong concavity in the presence of estimation error for both $g^*$ and $\beta$. The overall structure of the proofs closely follows that of their counterparts in Section \ref{sec:concavity}, but some additional work is required to handle the error of the regression coefficients. The proofs are deferred to the Appendix.

\begin{lem}\label{lem:technicalunknown1}
Let Assumption \ref{a1new} hold, and take $\delta\in\mathds{R}_K$ satisfying $\delta_1=0$. Fix $b_1,...,b_K$ and let $\Delta_k = b_k-\beta_k$. Define $\Delta^{\max} = \max_k \|\Delta_k\|_2$. The following inequality holds:
\begin{eqnarray*}
&\,& \mathbb{E}\left(F^{b}\left(g^*,X\right)\right)-\mathbb{E}\left(F^{b}\left(g^*+\delta,X\right)\right) - \delta^\top p^{b,g^*}\\
&\geq& \frac{\kappa'_2\kappa'_4}{8K}\left(\sum^K_{k=1} \min\left\{\delta^2_k,\left(\kappa'_1\right)^2\right\}1_{\left\{\min\left\{\left|\delta_k\right|,\kappa'_1\right\} \geq 8\max\left\{1,\frac{\kappa'_3}{\kappa'_2\kappa'_4}\right\}\kappa_4\Delta^{\max}\right\}}\right)1_{\left\{\Delta^{\max}\leq \frac{\kappa'_1}{2\kappa_4}\right\}}.
\end{eqnarray*}
\end{lem}

\begin{lem}\label{lem:technicalunknown2}
Suppose that we are in the situation of Lemma \ref{lem:technicalunknown1}. Then, we have
\begin{equation*}
\delta^\top\left(p^{b,g^*+\delta}-p^{b,g^*}\right) \geq \max\left\{0,C\min\left\{\|\delta\|^2_2,\left(\kappa'_1\right)^2\right\}-C'\left(\Delta^{\max}\right)^2\right\}
\end{equation*}
where $C,C'>0$ are constants.
\end{lem}

The final property is completely new to the case of unknown costs, and bounds the estimation error of the vector $p$ of target probabilities in a setting where we know $g^*$, but not $\beta$.

\begin{lem}\label{lem:technicalunknown3}
Suppose that we are in the situation of Lemma \ref{lem:technicalunknown1}. Then,
\begin{equation*}
\|p^{b,g^*}-p\|_2 \leq D\cdot\Delta^{\max},
\end{equation*}
where $D > 0$ is a constant.
\end{lem}

\noindent\textbf{Proof:} To avoid notational clutter, we write $g^*$ as simply $g$ in this proof, since no other $g$ will be considered. Note, however, that we do distinguish between $b$ and $\beta$.

As in the proof of Lemma \ref{lem:technicalunknown1}, we define $Z^{\beta}_{j,k} = \beta^\top_j X-g_j-\left(\beta^\top_k X-g_k\right)$. For additional convenience, denote $d_k = \Delta^\top_k X$. Then, $b^\top_k X = \beta^\top_k X + d_k$. Observe that, for any $k$,
\begin{eqnarray}
\left|p^{b,g^*}_k - p_k\right| &=& \left|P\left(b^\top_k X - g_k = F^b\left(g,X\right)\right)-P\left(\beta^\top_k X - g_k = F^\beta\left(g,X\right)\right)\right|\nonumber\\
&\leq& P\left(b^\top_k X - g_k = F^b\left(g,X\right),\; \beta^\top_k X - g_k > F^\beta\left(g,X\right)\right)\nonumber\\
&\,& + P\left(b^\top_k X - g_k > F^b\left(g,X\right),\; \beta^\top_k X - g_k = F^\beta\left(g,X\right)\right).\label{eq:twoprobs}
\end{eqnarray}
By Assumption \ref{a1new}(ii), $\left|d_k\right| \leq \kappa_4\Delta^{\max}$. We focus on the first term of (\ref{eq:twoprobs}) and derive
\begin{eqnarray*}
&\,& P\left(b^\top_k X - g_k = F^b\left(g,X\right),\; \beta^\top_k X - g_k > F^\beta\left(g,X\right)\right)\\
&=& P\left(\beta^\top_k X - g_k + d_k = \min_\ell \beta^\top_\ell X - g_\ell + d_\ell,\; \beta^\top_k X - g_k > F^\beta\left(g,X\right)\right)\\
&\leq& \sum_j P\left(\beta^\top_k X - g_k + d_k = \min_\ell \beta^\top_\ell X - g_\ell + d_\ell,\; \beta^\top_k X - g_k > \beta^\top_j X - g_j\right)\\
&\leq & \sum_j P\left(\beta^\top_k X - g_k + d_k \leq \beta^\top_j X - g_j + d_j,\; \beta^\top_k X - g_k > \beta^\top_j X - g_j\right)\\
&=& \sum_j P\left(d_k-d_j \leq Z^{\beta}_{j,k} < 0\right)\\
&\leq& \sum_j P\left(-2\kappa_4\Delta^{\max} \leq Z^{\beta}_{j,k} < 0\right).
\end{eqnarray*}
Using very similar arguments, the same bound can be shown for the second term of (\ref{eq:twoprobs}). Thus, we have
\begin{equation}\label{eq:kappaboundonps}
\left|p^{b,g^*}_k - p_k\right| \leq \sum_j P\left(\left|Z^{\beta}_{j,k}\right|<2\kappa_4\Delta^{\max}\right).
\end{equation}
Consider the case $\Delta^{\max}\leq \frac{\kappa'_1}{2\kappa_4}$. By Lemma \ref{a1}(i), (\ref{eq:kappaboundonps}) yields
\begin{equation*}
P\left(\left|Z^{\beta}_{j,k}\right|<2\kappa_4\Delta^{\max}\right) \leq 4\kappa'_3\kappa_4\Delta^{\max}.
\end{equation*}
At the same time, since we always have $\left|p^{b,g^*}_k - p_k\right| \leq 2$, we may continue (\ref{eq:kappaboundonps}) as
\begin{equation}\label{eq:kappaboundonps2}
\left|p^{b,g^*}_k - p_k\right| \leq 4K\kappa'_3\kappa_4\Delta^{\max}\cdot 1_{\left\{\Delta^{\max}\leq\frac{\kappa'_1}{2\kappa_4}\right\}} + 2\cdot 1_{\left\{\Delta^{\max}>\frac{\kappa'_1}{2\kappa_4}\right\}}.
\end{equation}
We observe the elementary inequality that, for any $a_1,a_2,a_3 > 0$ and $\rho\geq 0$,
\begin{equation*}
a_1\rho 1_{\left\{\rho\leq a_3\right\}} + a_2 1_{\left\{\rho > a_3\right\}} \leq \rho\max\left\{a_1,\frac{a_2}{a_3}\right\}.
\end{equation*}
Applying this to (\ref{eq:kappaboundonps2}) yields the desired result.\qed

\subsection{Proof outline for main rate results}\label{sec:sketch}

Finally, we provide an outline of the steps made in proving Theorems \ref{thm:errorrates}-\ref{thm:decision}. The notation from Section \ref{sec:concavity2} is carried over. The notation $\|\cdot\|_{\text{sp}}$ will be used to denote the spectral norm (largest singular value).

The first result relates the probability of incorrect selection (for an arbitrary policy) to the estimation error of both $g^*$ and $\beta$. It is used in several places. Theorem \ref{thm:decision} will be a direct consequence of it once Theorem \ref{thm:errorrates} is established.

\begin{lem}\label{lem:probineq}
Let Assumption \ref{a1new} hold. For arbitrary $b,g$, let $\delta = g - g^*$ and $\Delta^{\max} = \max_j \|b_j - \beta_j\|_2$. Define $\pi^{b,g}\left(X\right) = \arg\min_j b^\top X_j - g_j$. Then, there exists $C_0 > 0$ such that
\begin{equation*}
P\left(\pi^{b,g} \neq \pi^{\beta,g^*}\right) \leq C_0 \left(\kappa_4 \Delta^{\max} + \|\delta\|_{\infty}\right).
\end{equation*}
\end{lem}

\noindent\textbf{Proof:} We observe that
\begin{equation*}
P\left(\pi^{b,g} \neq \pi^{\beta,g^*}\right) \leq \sum^K_{k=1} P\left(\pi^{b,g} = k,\; \pi^{\beta,g^*}\neq k\right),
\end{equation*}
so it is sufficient to consider a single $k$. Let $d_k = \left(b_k-\beta_k\right)^\top X$. By repeating some arguments in the proof of Lemma \ref{lem:technicalunknown3} (starting from eq. (\ref{eq:twoprobs}) onwards), we obtain
\begin{eqnarray}
&\,& P\left(\pi^{b,g} = k,\; \pi^{\beta,g^*}\neq k\right)\nonumber\\
&=& P\left(b^\top_k X - g_k = \min_j b^\top_j X - g_j,\; \beta^\top_k X - g^*_k > \min_j \beta^\top_j X - g^*_j\right)\nonumber\\
&\leq& \sum_j P\left(\beta^\top_k X - g^*_k + d_k - \delta_k \leq \beta^\top_j X - g^*_j + d_j - \delta_j,\; \beta^\top_k X - g^*_k > \beta^\top_j X - g^*_j\right)\nonumber\\
&\leq& \sum_j P\left(d_k - \delta_k - d_j + \delta_j \leq \beta^\top_j X - g^*_j - \left(\beta^\top_k X - g^*_k\right) < 0\right)\nonumber\\
&\leq& \sum_j P\left(-2\left(\kappa_4 \Delta^{\max} + \|\delta\|_{\infty}\right) \leq \beta^\top_j X - g^*_j - \left(\beta^\top_k X - g^*_k\right) < 0\right).\label{eq:repeatunknown3}
\end{eqnarray}
The arguments in the proof of Lemma \ref{lem:technicalunknown3} can then be repeated to yield the desired bound on (\ref{eq:repeatunknown3}).\qed

Next, we state three intermediate technical results related to the information matrix of the observations collected using the proposed semi-myopic policy. The proofs of these results are deferred to the Appendix.

\begin{lem}\label{lem:errorrates1}
Let Assumption \ref{a1new} and \ref{a2} hold. For arbitrary $b,g$, let $\delta = g - g^*$ and $\Delta^{\max} = \max_j \|b_j - \beta_j\|_2$. Define $\pi^{b,g}\left(X\right) = \arg\min_j b^\top X_j - g_j$, and let
\begin{equation}\label{eq:defineSk}
S_k\left(b,g\right) = \mathbb{E}\left(X X^\top 1_{\left\{\pi^{b,g}\left(X\right)=k\right\}}\right).
\end{equation}
Then, there exist $C_1,C_2 > 0$ such that
\begin{equation*}
\lambda_{\min}\left(S_k\left(b,g\right)\right)\geq \kappa_6 - C_1\|\delta\|_{\infty} -C_2\Delta^{\max}.
\end{equation*}
\end{lem}

\begin{lem}\label{lem:hoeffding}
Let Assumption \ref{a1new} and \ref{a2} hold. Let $T$ be a deterministic subset of $\left\{1,...,n\right\}$, and define $V_T = \sum_{n\in T} X^n\left(X^n\right)^\top$. Then, for any $z > 0$,
\begin{equation}\label{eq:hoeffding1}
P\left(\|V_T- \mathbb{E}\left(V_T\right)\|_{\text{sp}} > z\right) \leq 2d^2\exp\left(-\frac{z^2}{2d^2 \kappa^4_4\left|T\right|}\right),
\end{equation}
and
\begin{equation}\label{eq:hoeffding2}
P\left(\lambda_{\min}\left(V_T\right) < \frac{1}{2}\kappa_6 K \left|T\right|\right) \leq 2d^2\exp\left(-\frac{\kappa^2_6 K^2\left|T\right|}{8d^2\kappa^4_4}\right).
\end{equation}
\end{lem}

\begin{lem}\label{lem:quadratic}
Let Assumptions \ref{a1new} and \ref{a2} hold. Let $\left\{\pi^n\right\}^{\infty}_{n=1}$ be a sequence of random variables such that each $\pi^n$ takes values in $\left\{1,...,K\right\}$ and is measurable with respect to $\bar{\mathcal{F}}^n$, the sigma-algebra generated by $\pi^1,...,\pi^{n-1}$, $W^1_{\pi^1},...,W^{n-1}_{\pi^{n-1}}$, and $X^1,...,X^n$.

Define $V^n_k = \sum^n_{m=1} X^m \left(X^m\right)^\top 1_{\left\{\pi^n=k\right\}}$. Similarly, let $v^n_k = \sum^n_{m=1} \varepsilon^m_k X^m 1_{\left\{\pi^n=k\right\}}$. The following statements hold:
\begin{enumerate}
\item[1)] For any $n\geq 1$, any $\mu > 0$ and any $0\leq \eta \leq 1$,
\begin{equation}\label{eq:existsm}
P\left(\exists m\leq n: \left(v^m_k\right)^\top\left(V^m_k + \mu I\right)^{-1} v^m_k \geq 2\kappa_5\log\left(\frac{1}{\eta}\right) + \kappa_5 d\log\left(1 + m\frac{\kappa^2_4}{\mu}\right)\right) \leq \eta.
\end{equation}
\item[2)] For any $n\geq 1$ and $\mu > 0$,
\begin{equation}\label{eq:Vandvbound}
\mathbb{E}\left(\left(v^n_k\right)^\top\left(V^n_k + n\mu I\right)^{-1} v^n_k\right) \leq 2\kappa_5\left(\left(1+\kappa^2_4\right)^{\frac{d}{2}}-1\right).
\end{equation}
\end{enumerate}
\end{lem}

We now begin to study the properties of the semi-myopic algorithm. We first establish concentration inequalities on the two types of estimation error (that is, the error of estimating $\beta$ and $g^*$, respectively). The proofs are deferred to the Appendix, but we note that this is essentially the same approach as in Section \ref{sec:rateknown}; in particular, the statement and proof of Lemma \ref{lem:concg} are similar in structure to Lemma \ref{lem:technicalknown2}. The difference is that we now require separate concentration inequalities for two interrelated objects. The result of Lemma \ref{lem:concreg} is used inside the proof of Lemma \ref{lem:concg}.

\begin{lem}\label{lem:concreg}
Let Assumptions \ref{a1new} and \ref{a2} hold, and let $\pi$ be the semi-myopic policy of Section \ref{sec:mainresultsunknown}. There exist constants $C_3,C_4,C_5>0$, which depend only on $d$, $K$ and $\kappa_4,\kappa_5,\kappa_6$, such that
\begin{equation*}
P\left(\max_k \|\Delta^n_k\|_2 > C_3\left(\log n\right)^{-\frac{5}{2}}\right) \leq C_4 \exp\left(-C_5\left(\log n\right)^4\right).
\end{equation*}
\end{lem}

\begin{lem}\label{lem:concg}
Let Assumptions \ref{a1new} and \ref{a2} hold. Suppose that the semi-myopic policy of Section \ref{sec:mainresultsunknown} is used to sample cost functions, while $g^n$ is updated as in Algorithm \ref{fig:sa}. Let $C_1$ be the constant obtained from Lemma \ref{lem:errorrates1}. There exists a constant $C_6>0$ such that
\begin{equation*}
P\left(\|\delta^n\|_2 > \frac{\kappa_6}{4C_1}\right) \leq \frac{C_6}{n^9}.
\end{equation*}
\end{lem}

Finally, we complete the proof. Theorem \ref{thm:errorrates} is the combination of Propositions \ref{lem:expreg} and \ref{lem:expg} below. The proof of Proposition \ref{lem:expreg} is highly technical and deferred to the Appendix. Proposition \ref{lem:expg} is an extension of Theorem \ref{thm:known} incorporating estimation error from regression (and using the result of Proposition \ref{lem:expreg}). As mentioned earlier, Theorem \ref{thm:decision} follows straightforwardly from these results together with Lemma \ref{lem:probineq}.

\begin{prop}\label{lem:expreg}
Let Assumptions \ref{a1new} and \ref{a2} hold, and let $\pi$ be the semi-myopic policy of Section \ref{sec:mainresultsunknown}. There exists a constant $C_7 > 0$ such that
\begin{equation*}
\mathbb{E}\left(\max_j \|\Delta^n_j\|^2_2\right) \leq \frac{C_7}{n}.
\end{equation*}
\end{prop}

\begin{prop}\label{lem:expg}
Let Assumptions \ref{a1new} and \ref{a2} hold. Suppose that the semi-myopic policy of Section \ref{sec:mainresultsunknown} is used to sample cost functions, while $g^n$ is updated as in Algorithm \ref{fig:sa}. There exists some $\bar{\alpha}>0$ such that
\begin{equation*}
\mathbb{E}\left(\|\delta^n\|^2_2\right) \leq \frac{C_8}{n}
\end{equation*}
if $\alpha > \bar{\alpha}$.
\end{prop}

\noindent\textbf{Proof:} Let $\Delta^{\max,n} = \max_k \|\Delta^n_k\|_2$. As in the proof of Theorem \ref{thm:known}, we first write
\begin{eqnarray}
\|\delta^{n+1}\|^2_2 &\leq& \|\delta^n\|^2_2 + \frac{\alpha^2}{\left(n+1\right)^2}\|\zeta^{n+1}-p\|^2_2 - \frac{2\alpha}{n+1}\left(\zeta^{n+1}-p\right)^\top \delta^n\nonumber\\
&\leq & \|\delta^n\|^2_2 + \frac{2\alpha^2}{\left(n+1\right)^2} - \frac{2\alpha}{n+1}\left(\zeta^{n+1}-p^{\hat{\beta}^n,g^n}\right)^\top \delta^n\nonumber\\
&\,& - \frac{2\alpha}{n+1}\left(p^{\hat{\beta}^n,g^n}-p^{\hat{\beta}^n,g^*}\right)^\top \delta^n- \frac{2\alpha}{n+1}\left(p^{\hat{\beta}^n,g^*}-p\right)^\top \delta^n\nonumber\\
&\leq& \|\delta^n\|^2_2 + \frac{2\alpha^2}{\left(n+1\right)^2} -\frac{2\alpha}{n+1}\left(\zeta^{n+1}-p^{\hat{\beta}^n,g^n}\right)^\top \delta^n - \frac{2\alpha D_1}{n+1}\min\left\{\|\delta\|^2_2,\left(\kappa'_1\right)^2\right\}\nonumber\\
&\,& + \frac{2\alpha}{n+1}D_2\left(\Delta^{\max,n}\right)^2 + \frac{2\alpha}{n+1}D_3\|\delta^n\|_2\Delta^{\max,n},\label{eq:deltadecomposition2}
\end{eqnarray}
where (\ref{eq:deltadecomposition2}) is obtained by applying Lemmas \ref{lem:technicalunknown2}-\ref{lem:technicalunknown3} with appropriate $D_1,D_2,D_3 > 0$. Letting $\mathcal{F}^n$ be the sigma-algebra generated by $\pi^1,...,\pi^{n}$, $W^1_{\pi^1},...,W^n_{\pi^{n}}$, and $X^1,...,X^n$, we write
\begin{eqnarray*}
\mathbb{E}\left(\|\delta^{n+1}\|^2_2\mid \mathcal{F}^n\right) &\leq& \|\delta^n\|^2_2 + \frac{2\alpha^2}{\left(n+1\right)^2} - \frac{2\alpha D_1}{n+1}\min\left\{\|\delta\|^2_2,\left(\kappa'_1\right)^2\right\}\\
&\,& + \frac{2\alpha}{n+1}D_2\left(\Delta^{\max,n}\right)^2 + \frac{2\alpha}{n+1}D_3\|\delta^n\|_2\Delta^{\max,n}.
\end{eqnarray*}
Let $D_0 = \frac{\kappa_6}{4C_1}$, where $C_1$ is the constant obtained from Lemma \ref{lem:errorrates1}. On the event $\left\{\|\delta^n\|_2 \leq D_0\right\}$, we obtain
\begin{eqnarray*}
\mathbb{E}\left(\|\delta^{n+1}\|^2_2\mid \mathcal{F}^n\right) &\leq&  \left(1 - \frac{2\alpha D_1}{n+1}\min\left\{1,\frac{\left(\kappa'_1\right)^2}{D^2_0}\right\}\right)\|\delta^n\|^2_2 + \frac{2\alpha^2}{\left(n+1\right)^2}\\
&\,& + \frac{2\alpha}{n+1}D_2\left(\Delta^{\max,n}\right)^2 + \frac{2\alpha}{n+1}D_3\|\delta^n\|_2\Delta^{\max,n}
\end{eqnarray*}
by repeating arguments from the proof of Theorem \ref{thm:known}. Then, in general, we have
\begin{eqnarray}
\mathbb{E}\left(\|\delta^{n+1}\|^2_2\mid \mathcal{F}^n\right) &\leq& \left(1 - \frac{2\alpha D_1}{n+1}\min\left\{1,\frac{\left(\kappa'_1\right)^2}{D^2_0}\right\}\right)\|\delta^n\|^2_2 + \frac{2\alpha^2}{\left(n+1\right)^2}\nonumber\\
&\,& + \frac{2\alpha}{n+1}D_2\left(\Delta^{\max,n}\right)^2 + \frac{2\alpha}{n+1}D_3\|\delta^n\|_2\Delta^{\max,n} + \mathbb{E}\left(\|\delta^{n+1}\|^2_2\cdot 1_{\left\{\delta^n\|_2 > D_0\right\}}\mid \mathcal{F}^n\right)\nonumber\\
&\leq& \left(1 - \frac{2\alpha D_1}{n+1}\min\left\{1,\frac{\left(\kappa'_1\right)^2}{D^2_0}\right\}\right)\|\delta^n\|^2_2 + \frac{2\alpha^2}{\left(n+1\right)^2}\nonumber\\
&\,& + \frac{2\alpha}{n+1}D_2\left(\Delta^{\max,n}\right)^2 + \frac{2\alpha}{n+1}D_3\|\delta^n\|_2\Delta^{\max,n} + \alpha D_4\log\left(n+1\right)1_{\left\{\delta^n\|_2 > D_0\right\}}\label{eq:applytheoldlemma}
\end{eqnarray}
where $D_4>0$ in (\ref{eq:applytheoldlemma}) is a constant obtained from Lemma \ref{lem:technicalknown1}. We now take the unconditional expectation, obtaining
\begin{eqnarray}
\mathbb{E}\left(\|\delta^{n+1}\|^2_2\right) &\leq& \left(1 - \frac{2\alpha D_1}{n+1}\min\left\{1,\frac{\left(\kappa'_1\right)^2}{D^2_0}\right\}\right)\mathbb{E}\left(\|\delta^n\|^2_2\right) + \frac{2\alpha^2}{\left(n+1\right)^2}\nonumber\\
&\,& + \frac{2\alpha}{n+1}D_2\mathbb{E}\left(\left(\Delta^{\max,n}\right)^2\right) + \frac{2\alpha}{n+1}D_3\mathbb{E}\left(\|\delta^n\|_2\Delta^{\max,n}\right) + \alpha D_4\log\left(n+1\right) P\left(\|\delta^n\|_2 > D_0\right)\nonumber\\
&\leq & \left(1 - \frac{2\alpha D_1}{n+1}\min\left\{1,\frac{\left(\kappa'_1\right)^2}{D^2_0}\right\}\right)\mathbb{E}\left(\|\delta^n\|^2_2\right) + \frac{2\alpha^2}{\left(n+1\right)^2}\nonumber\\
&\,& + \frac{2\alpha}{n+1}D_2\mathbb{E}\left(\left(\Delta^{\max,n}\right)^2\right) + \frac{2\alpha}{n+1}D_3\mathbb{E}\left(\|\delta^n\|_2\Delta^{\max,n}\right) + \alpha C_6D_4\frac{\log\left(n+1\right)}{n^9}\label{eq:applyconcg}\\
&\leq & \left(1 - \frac{2\alpha D_1}{n+1}\min\left\{1,\frac{\left(\kappa'_1\right)^2}{D^2_0}\right\}\right)\mathbb{E}\left(\|\delta^n\|^2_2\right) + \frac{2\alpha^2}{\left(n+1\right)^2}\nonumber\\
&\,& + \frac{2\alpha}{n\left(n+1\right)}C_7D_2 + \frac{2\alpha}{\sqrt{n}\left(n+1\right)}\sqrt{C_7}\cdot D_3\sqrt{\mathbb{E}\left(\|\delta^n\|^2_2\right)} + \alpha C_6D_4\frac{\log\left(n+1\right)}{n^9},\label{eq:applyexpreg}
\end{eqnarray}
where (\ref{eq:applyconcg}) uses Lemma \ref{lem:concg}, and (\ref{eq:applyexpreg}) uses Proposition \ref{lem:expreg}. We thus have
\begin{equation*}
\mathbb{E}\left(\|\delta^{n+1}\|^2_2\right) \leq \left(1 - \frac{\alpha D_5}{n+1}\right)\mathbb{E}\left(\|\delta^{n}\|^2_2\right) + \frac{D_6}{\sqrt{n}\left(n+1\right)}\sqrt{\mathbb{E}\left(\|\delta^n\|^2_2\right)} + \frac{D_7}{\left(n+1\right)^2}
\end{equation*}
for suitable $D_5,D_6,D_7>0$. Letting $a_n = n\mathbb{E}\left(\|\delta^n\|^2_2\right)$, we have
\begin{equation*}
a_{n+1} \leq \left(1 - \frac{\alpha D_5-1}{n}\right)a_n + \frac{D_6\sqrt{a_n}}{n} + \frac{D_7}{n}.
\end{equation*}
As in the proof of Theorem \ref{thm:known}, we apply Lemma \ref{lem:weirdineq} and obtain the desired result as long as $\alpha > \frac{1}{D_5}$.\qed

\section{Numerical illustrations}\label{sec:exp}

We illustrate our framework on two numerical examples with linear costs. Section \ref{sec:synthetic} considers synthetic test instances with $X$ generated uniformly on the unit sphere. Section \ref{sec:partitioning} applies our framework to learn facility locations in a geographical partitioning problem (a common application area of semidiscrete optimal transport). Both types of experiments seek to explore the performance of the semi-myopic policy in settings that lie outside the assumptions used in our theoretical analysis.

\subsection{Synthetic linear costs}\label{sec:synthetic}

We consider a set of $1000$ synthetic test problems. In all of these problems, the data $X$ are generated uniformly on the unit sphere in $d=10$ dimensions; since $X$ does not have a density in $\mathbb{R}^d$, Assumption \ref{a1new} does not hold. Each problem uses $K = 5$ different vectors $\beta_k$, also generated uniformly on the unit sphere, and a different $p$ generated uniformly on the $d$-dimensional simplex.

For ease of implementation, we modified the semi-myopic policy to conduct forced exploration probabilistically: that is, in the $n$th iteration, we select $k$ uniformly at random with probability $\frac{\left(\frac{1}{a}\log n\right)^9}{n}$, where we used $a = 4.5$. The rest of the time, we select a cost function according to the myopic policy $\hat{\pi}^n$. We set $\rho^n\equiv 0.001$ for simplicity, and in order to enable use of recursive least squares. At each $n$, we evaluate the policy using the correct selection indicator $1_{\left\{\hat{\pi}^n=\pi^*\left(X^n\right)\right\}}$. By averaging this quantity across $10$ runs, each containing $1000$ iterations and starting from $g^0 \equiv 0$, $\hat{\beta}^0_k \equiv 0$, we obtain a trajectory of the probability $P\left(\hat{\pi}^n = \pi^*\left(X^n\right)\right)$ of correct selection. We then average this trajectory over $1000$ test problems.

The optimal policy $\pi^*$ for each test problem is precomputed via brute force by running (\ref{eq:basicsa}) with the true cost coefficients for a very large number of iterations. As a benchmark, we also implement a policy that knows the true costs (and, therefore, does not need to conduct any exploration or collect any samples), but does not know $g^*$, and runs (\ref{eq:basicsa}) in an online manner. We also calculate the average probability of correct selection for this policy, which allows us to evaluate any loss incurred by having to learn the coefficients.

\begin{figure}[t]
	\centering
	\subfigure[Sensitivity analysis with respect to noise.]{
		\includegraphics[width=0.47\textwidth]{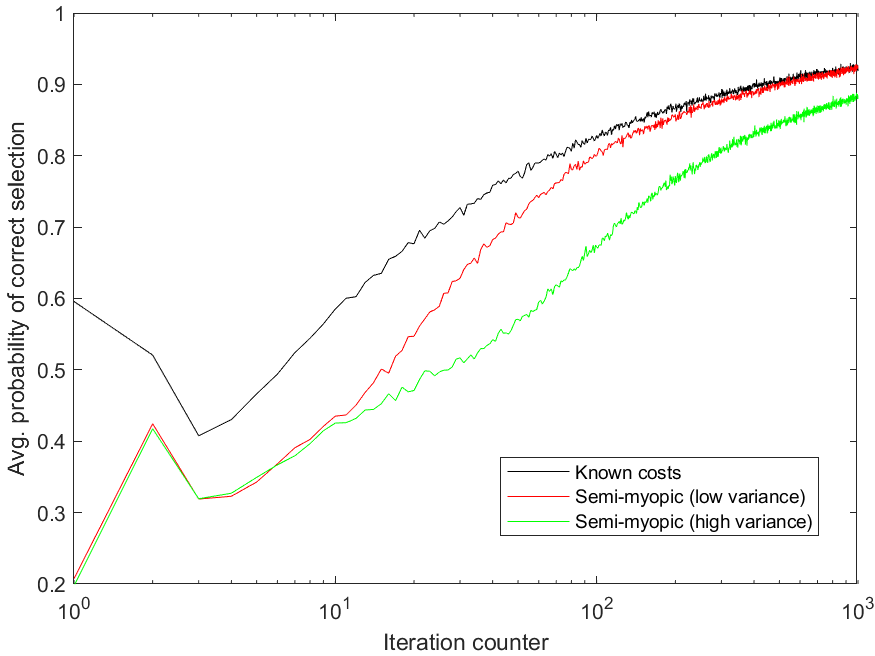}
		\label{fig:normalmain}
	}
	\subfigure[Sensitivity analysis with respect to $\alpha$.]{
		\includegraphics[width=0.47\textwidth]{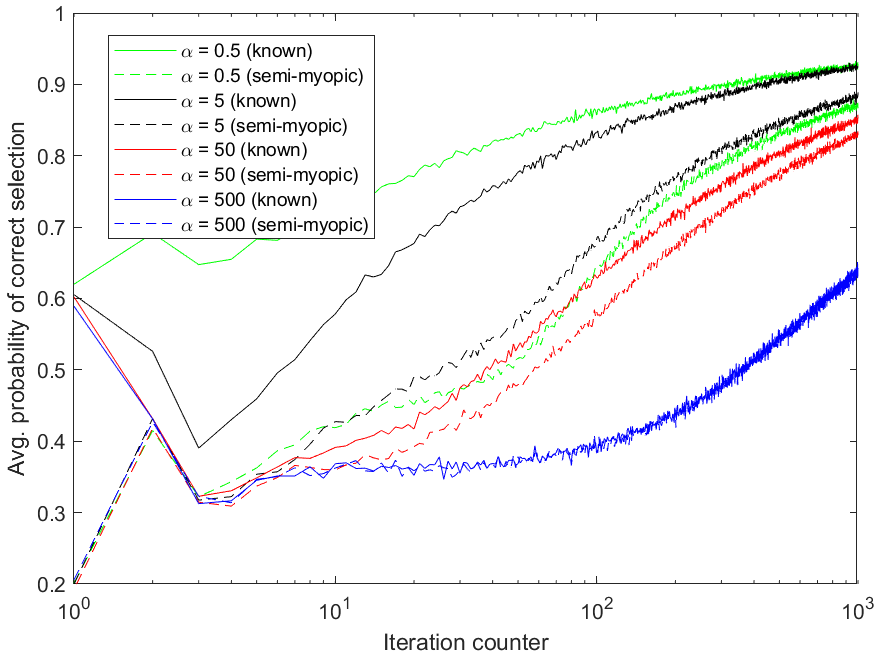}
		\label{fig:normalalpha}
	}
    \caption{Average performance of semi-myopic policy under Gaussian noise.}
\end{figure}

Figure \ref{fig:normalmain} shows average performance for the semi-myopic policy with stepsize parameter $\alpha = 5$ and two levels of Gaussian noise: $Var\left(\varepsilon\right) = 0.02^2$ and $Var\left(\varepsilon\right)=0.2^2$. Since the benchmark policy knows the costs, it is unaffected by the noise variance. We see that, when the variance is low, there is virtually no performance loss caused by unknown costs. When the variance is high, the average PCS of the semi-myopic policy lags behind that of the known-cost policy by $0.039$ (and has narrowed this gap relative to the earlier iterations). Performance also does not appear to be very sensitive to the frequency of forced exploration: we also tested a much higher value $a = 6$, for which the probability of forced exploration is only $0.0036$ by the end of the time horizon, but obtained very similar trajectories. This also indicates that, in an online implementation, we could allow forced exploration to happen quite infrequently while maintaining good performance.

Figure \ref{fig:normalalpha} uses the higher variance level and varies the stepsize parameter $\alpha$ over several orders of magnitude. We observe some sensitivity with respect to this parameter, but, interestingly, this sensitivity is partially smoothed out by the presence of noise: under known costs, the difference between $\alpha = 0.5$ and $\alpha = 50$ is far more pronounced than under unknown costs. By the end of the time horizon, there is virtually no difference between the two best-performing choices, and the third-best is relatively close as well. Only the largest choice $\alpha = 500$ lags behind the others, perhaps reflecting the fact that the multiplicative constant in Theorem \ref{thm:known} is increasing in $\alpha$.

\begin{figure}[t]
	\centering
	\subfigure[Laplace noise.]{
		\includegraphics[width=0.47\textwidth]{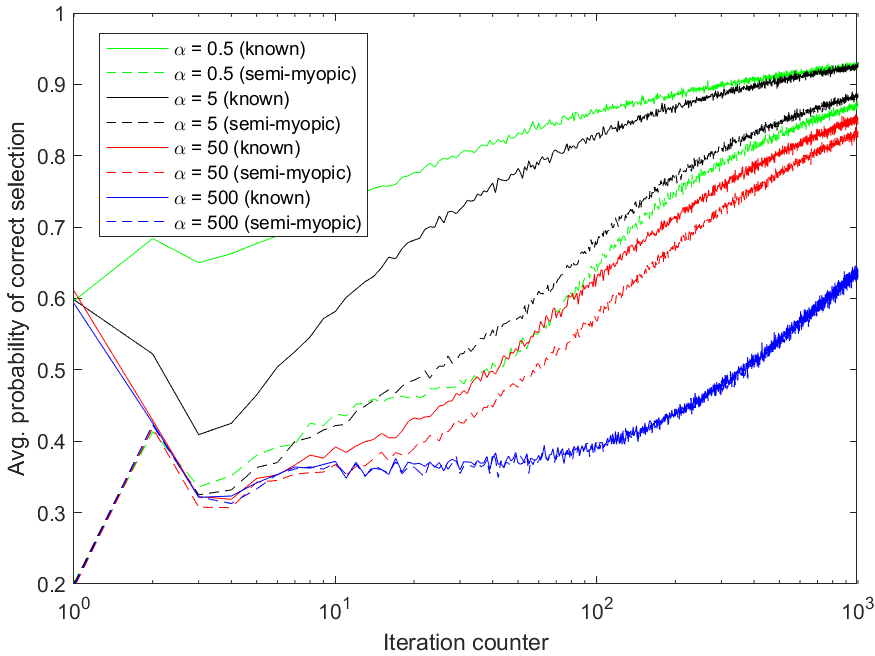}
		\label{fig:laplacealpha}
	}
	\subfigure[Binary observations.]{
		\includegraphics[width=0.47\textwidth]{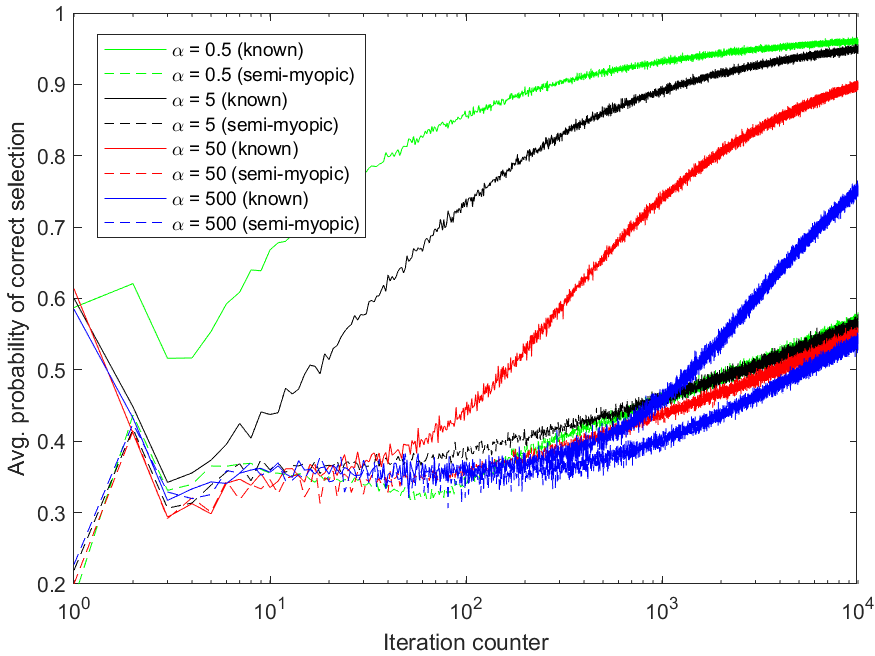}
		\label{fig:logisticalpha}
	}
    \caption{Average performance of semi-myopic policy in non-normal settings.}
\end{figure}

Figure \ref{fig:laplacealpha} repeats the analysis of Figure \ref{fig:normalalpha} for the same experimental setup, with the sole difference that the noise follows a Laplace (rather than Gaussian) distribution with mean zero and $Var\left(\varepsilon\right)=0.2^2$. Thus, this experiment considers a situation where the assumption of sub-Gaussianity is violated (instead, the noise is sub-exponential). However, despite the heavier tails, the results are virtually identical to the normal case, indicating that our approach is fairly robust to violations of this assumption. We also compared noise levels similarly to Figure \ref{fig:normalmain}, but the results were again nearly identical to the Gaussian case and so we omit them here.

Finally, Figure \ref{fig:logisticalpha} uses the same setup for $X$ and $\beta_k$, but now we assume that each observation $W^{n+1}_k$ is generated from a Bernoulli distribution with success probability $\chi\left(\beta^\top_k X^{n+1}\right) = \frac{1}{1+e^{-\beta^\top_k X^{n+1}}}$. In other words, the true cost structure is an instance of logistic regression with coefficients $\beta_k$. However, we do not use logistic regression to estimate the coefficients, due to the high computational cost of estimating a GLM. Instead, we apply recursive least squares directly to the binary observations, treating the error $W^{n+1}_k - \chi\left(\left(\hat{\beta}^n_k\right)^\top X^{n+1}\right)$ as the residual. Thus, the semi-myopic policy works with a partially misspecified statistical model, though \cite{ChRy20} showed that recursive least squares is still consistent in this situation (i.e., we will still learn the true $\beta_k$ values given enough time). Since estimation is more difficult with misspecification, we extend the time horizon to $10^4$ iterations for this experiment.

Because the benchmark policy knows the costs, misspecification is not a concern for it, and its performance only depends on $\alpha$. The semi-myopic policy has much more difficulty in the misspecified setting than in the previous two, but makes consistent improvement throughout the majority of the time horizon. Because cost estimation is now the most difficult aspect of the problem, the performance of the semi-myopic policy is now almost entirely unaffected by $\alpha$.

It is well-known in the stochastic approximation literature that the performance of SA can be sensitive to the stepsize rule. It may be possible to speed up convergence by using other choices of $\alpha_n$, many of which are surveyed in \cite{GePo06}. Other, more sophisticated procedures can be found in \cite{BrCiZe11} or \cite{ScZhLe13}, but every one of these methods involves at least one tunable parameter, so there is no way to avoid the tuning issue entirely. However, it is noteworthy that the presence of noise in the observed costs has the effect of reducing sensitivity to $\alpha$ in all three of the settings considered here.

\subsection{Geographical partitioning}\label{sec:partitioning}

For each $k = 1,...,K$, let $x_k \in \left[0,1\right]^2$ be a fixed location (``facility'') in the unit square. A user appears at some random location $X \in \left[0,1\right]^2$ and is assigned to a facility. The cost incurred by assigning the user to facility $k$ is $c_k\left(X\right) = \|X-x_k\|_2$, the Euclidean distance between the facility and the user. The goal is to choose an assignment rule that minimizes the expected distance subject to the probabilistic targets (which, in this context, require each facility to serve a certain proportion of the population). It is well-known in computational geometry that the optimal policy can be visualized as an additively weighted Voronoi diagram \citep{CaCaDe16} whose weights are the values $g^*_k$. The diagram partitions the unit square into regions, each containing a single facility, to which all users appearing in that region are assigned.

We now suppose that the locations $x_k$ are unknown, but a user appearing at location $X$ can obtain a noisy observation $W_k = \|X-x_k\|_2 + \varepsilon$ of the distance to any one facility of his or her choice. This observation is not linear in $x_k$, but we can still apply a linear regression model using the transformation
\begin{eqnarray}
W^2_k &=& \|X-x_k\|^2_2 + 2\|X-x_k\|_2\varepsilon + \varepsilon^2\nonumber\\
&=& \mathbb{E}\left(\varepsilon^2\right) + \left(X-x_k\right)^\top\left(X-x_k\right) + 2\|X-x_k\|_2\varepsilon + \left(\varepsilon^2-\mathbb{E}\left(\varepsilon^2\right)\right)\nonumber\\
&=& X^\top X + \left(\mathbb{E}\left(\varepsilon^2\right) + x^\top_k x_k\right) - 2X^\top x_k +2\|X-x_k\|_2\varepsilon + \left(\varepsilon^2-\mathbb{E}\left(\varepsilon^2\right)\right)\label{eq:voronoi}.
\end{eqnarray}
The last two terms in (\ref{eq:voronoi}) have zero mean and can be treated as ``residual error.'' Because $X$ is observed, the quantity $X^\top X$ is known. Thus, we can learn $x_k$ using a linear regression model in which $\left[1,-2X^\top\right]^\top$ is the vector of features, and $W^2_k - X^\top X$ is the response variable. The coefficients to be learned are $\beta_k = \left[\mathbb{E}\left(\varepsilon\right)^2 + x^\top_k x_k,x^\top_k\right]^\top$. Clearly, $x_k$ is identifiable in this model.

We considered two instances of this model where $\varepsilon \sim \mathcal{N}\left(0,0.02^2\right)$ and $X$ is uniformly distributed on $\left[0,1\right]^2$. Note that, even with the normality assumption, the residual error in (\ref{eq:voronoi}) is not sub-Gaussian. The instances used different facility locations $x_k$ and targets $p$. The settings of the semi-myopic policy were the same as in Section \ref{sec:synthetic}.

In this discussion, we aim to take advantage of the unique visual interpretation offered by the geographical partitioning setting. Thus, we do not present average trajectories of PCS over a large number of runs, as we did in Section \ref{sec:synthetic}. Rather, we will visualize the \textit{final} partition learned by the semi-myopic policy after a large number $10^6$ of iterations, and compare this to the optimal partition induced by $\pi^*$. Unlike in Section \ref{sec:synthetic}, we do not need to separately present a policy that knows $x_k$ but attempts to learn $g^*$ in an online manner, because the number of iterations is so large that such a policy will have converged to the true optimal partition already.

\begin{figure}[t]
	\centering
	\subfigure[True partition, Instance 1.]{
		\includegraphics[width=0.47\textwidth]{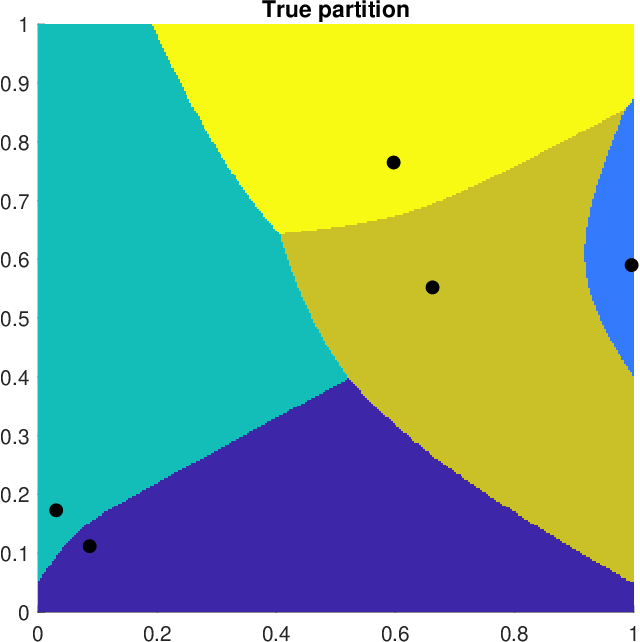}
		\label{fig:d001b}
	}
	\subfigure[Estimated partition, Instance 1.]{
		\includegraphics[width=0.47\textwidth]{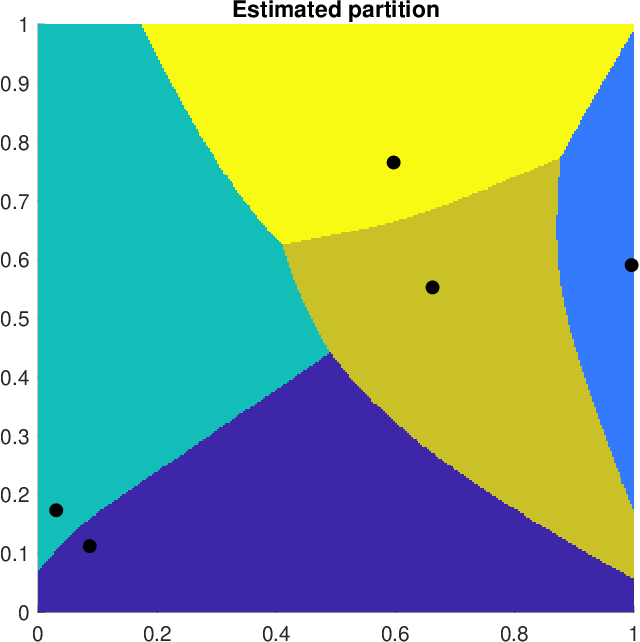}
		\label{fig:d001}
	}
	\subfigure[True partition, Instance 2.]{
		\includegraphics[width=0.47\textwidth]{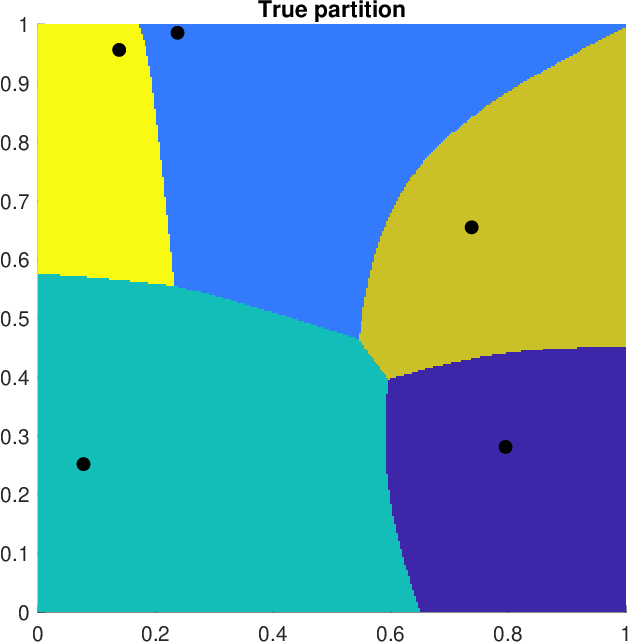}
		\label{fig:s001b}
	}
	\subfigure[Estimated partition, Instance 2.]{
		\includegraphics[width=0.47\textwidth]{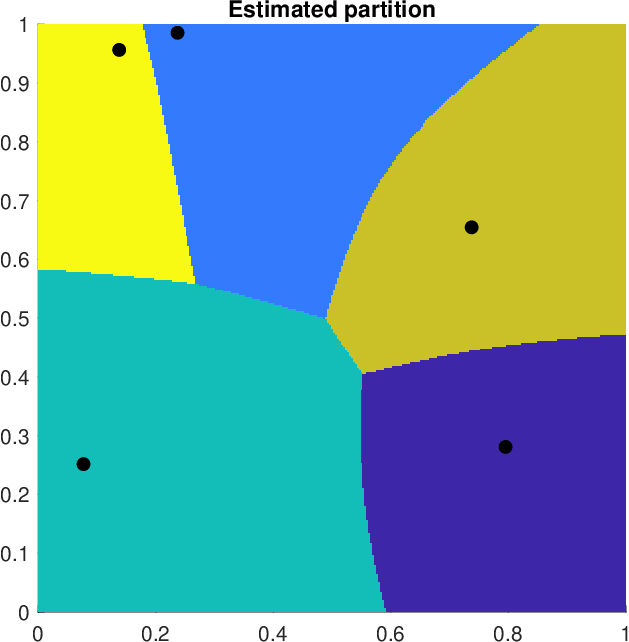}
		\label{fig:s001}
	}
\caption{Optimal and estimated partitions for two instances.}\label{fig:facilities}
\end{figure}

Figure \ref{fig:facilities} presents the true and estimated partitions for both instances, with the facility locations $x_k$ shown as black dots. The difference between locations turns out to exert a very significant impact on performance. In Instance 1, the facility close to the right edge of the unit square is also assigned a small $p_k$ value (specifically, $0.0258$). Thus, only a small portion of observations will lead us to sample this facility, causing us to rely more on forced exploration. Even after $1$ million iterations, the PCS achieved by the semi-myopic policy (i.e., the proportion of users assigned to the correct facility) is only $0.8944$. In Instance 2, this is less of a concern, and the semi-myopic policy achieves a PCS of $0.9542$. The estimated partition is also visibly closer to the true one in Instance 2.

In general, the estimation error of $\beta_k$ is much more of an issue in the partitioning problem than it was in Section \ref{sec:synthetic}. Even after $1$ million iterations, the semi-myopic policy noticeably lags behind SA with known costs (which was used to find the true partition). We found that our policy was effective in learning the regression coefficients, with $\hat{\beta}^n_k$ being within three decimal places of $\beta_k$ by the end of the time horizon. However, by the time this happened, the stepsize in the gradient update had become so small that $g^n$ had fallen behind (though it is still making progress, as PCS continues to increase if we keep running the policy). One reason may be that the regression noise in this problem is not sub-Gaussian, but has heavier tails than it did in Section \ref{sec:synthetic}, so the gradient update is more heavily biased. Even so, this example shows that the semi-myopic policy is able to learn the true partition even when some of the underlying model assumptions are violated.

\section{Conclusion}\label{sec:conc}

We have presented a novel variant of the semidiscrete optimal transport problem in which the costs are unknown, but can be estimated through sequential observations. This process is subject to the usual tradeoff of optimal learning, where sampling one cost function means one less sample available for the others. The problem thus brings together elements of optimal transport, optimal learning, and (through the characterization of the optimal policy) stochastic approximation. We have developed a simple and provably efficient algorithm based on the well-established notion of semi-myopic exploration; however, our analysis couples this structure with stochastic approximation in a novel way. In the process, we are able to resolve several known issues with applying SA to semidiscrete optimal transport, having to do with the non-smoothness of the stochastic gradient and the lack of strong concavity in the objective function.

There are many opportunities for future work. First, it may be interesting to investigate the theoretical performance of other types of learning algorithms, such as Thompson sampling \citep{AgGo13}, in this particular context. Second, it is worth taking a closer look at practical issues such as stepsize selection, which is known to impact the performance of SA procedures. Third, there may be potential for further connections between optimal transport and optimal learning beyond the semidiscrete version studied here. We hope that our work will spark new interest in bridging these fields.

\bibliographystyle{agsm} 
\bibliography{transport} 

\newpage

\section{Appendix: extension to GLMs}

In this section, we extend the results of Sections \ref{sec:mainresultsunknown} and \ref{sec:unknown} to a situation where the cost functions follow a generalized linear model (GLM), rather than linear regression. That is, we now assume $c_k\left(X\right) = \chi\left(\beta^\top_k X\right)$. As before, we collect observations of the form $W^{n+1}_k = c_k\left(X^{n+1}\right) + \upsilon^{n+1}$, where $\upsilon^{n+1}_k$ is an independent noise term with mean zero. Thus, $\chi$ is the ``link function'' \citep{McNe89} that connects the linear form $\beta^\top_k X$ to the expected value of the observation.

To estimate $\hat{\beta}^n_k$, we now replace (\ref{eq:ridge}) with
\begin{equation*}
\hat{\beta}^n_k = \arg\min_{b\in \mathcal{B}} \sum^n_{m=1} L\left(W^m_{\pi^m},b^\top X^m\right)1_{\left\{\pi^m = k\right\}} + \rho^n \|b\|^2_2,
\end{equation*}
where $\rho^n$ is as before, $L$ is a convex loss function (typically the log-likelihood function of the observations), and $\mathcal{B}$ is a known compact set. We suppose that $L$ and $\mathcal{B}$ are chosen in such a way that $\beta_k = \arg\min_{b\in\mathcal{B}} \mathbb{E}\left(L\left(W_k,b^\top X\right)\right)$ for a generic observation $W_k$ and attribute vector $X$.

The semi-myopic policy $\pi^{n+1}$ remains the same as in (\ref{eq:ourpolicy}). However, the plug-in estimator of the optimal decision becomes $\hat{\pi}^{n+1} = \arg\min_j \chi\left(\left(\hat{\beta}^n_j\right)^\top X^{n+1}\right) - g^n_j$. The parameters $g^n$ are updated according to
\begin{equation}\label{eq:saglm}
g^{n+1}_k = g^n_k + \alpha_{n}\left(-1_{\left\{k = \arg\min_j \chi\left(\left(\hat{\beta}^n_j\right)^\top X^{n+1}\right) - g^n_j\right\}}+p_k\right),
\end{equation}
similarly to (\ref{eq:approxsa}), with $\alpha_n = \frac{\alpha}{n+1}$ as before.

The analysis requires some additional assumptions. First, we modify Assumption \ref{a2} as follows. Let $L_2\left(t_1,t_2\right) = \frac{\partial L\left(t_1,t_2\right)}{\partial t_2}$. Essentially, we impose Assumption \ref{a2}(i) on $\varepsilon^n_k = L_2\left(W^n_k,\beta^\top_k X^n\right)$. Assumption \ref{a2}(ii) remains unchanged.

\begin{assum}\label{a2glm}
Suppose that there exist constants $\kappa_5,\kappa_6>0$ such that:
\begin{enumerate}
\item[i)] The random variable $L_2\left(W^n_k,\beta^\top_k X^n\right)$ is sub-Gaussian, i.e.,
\begin{eqnarray*}
\mathbb{E}\left(L_2\left(W^n_k,\beta^\top_k X^n\right)\mid X^n\right) &=& 0,\\
\mathbb{E}\left(\exp\left(t L_2\left(W^n_k,\beta^\top_k X^n\right)\right)\mid X^n\right)&\leq& \exp\left(\frac{1}{2}\kappa_5 t^2\right).
\end{eqnarray*}
\item[ii)] Let $\lambda_{\min}\left(\cdot\right)$ be the function that returns the smallest eigenvalue of the matrix argument. For any $k$,
\begin{equation*}
\lambda_{\min}\left(\mathbb{E}\left(X \left(X\right)^\top1_{\left\{\pi^*\left(X\right)=k\right\}}\right)\right) \geq \kappa_6.
\end{equation*}
\end{enumerate}
\end{assum}

Next, we impose some additional regularity conditions on the GLM, specifically the loss and link functions.

\begin{assum}\label{aextra}
Suppose that there exist constants $\kappa_7,\kappa_{8},\kappa_{9} > 0$ such that:
\begin{enumerate}
\item[i)] $\sup_{b\in\mathcal{B}} \|b\|_2 \leq \kappa_7$.
\item[ii)] For any $k$, $\frac{\partial^2 L\left(t_1,t_2\right)}{\partial t^2_2} \geq \kappa_8$ for any $t_1$ in the support of $W_k$ and any $t_2 \in \left[-\kappa_4\kappa_7,\kappa_4\kappa_7\right]$.
\item[iii)] For any $t_1,t_2\in\mathbb{R}$, $\left|\chi\left(t_1\right)-\chi\left(t_2\right)\right| \leq \kappa_9\left|t_1-t_2\right|$.
\end{enumerate}
\end{assum}

An example of a GLM where Assumptions \ref{a2glm}-\ref{aextra} hold is the logistic regression model with log-likelihood $L\left(t_1,t_2\right) = -t_1t_2 +\log\left(1+\exp\left(t_2\right)\right)$. Then, $\frac{\partial^2 L\left(t_1,t_2\right)}{\partial t^2_2} = \frac{\exp\left(t_2\right)}{\left(1+\exp\left(t_2\right)\right)^2}$, which is bounded away from zero when $t_2$ is in a bounded set. The link function $\chi\left(t\right) = \frac{1}{1+e^{-t}}$ is clearly Lipschitz, and $L_2\left(W^n_k,\beta^\top_k X^n\right) = -W^n_k + \frac{1}{1+e^{-\beta^\top_k X^n}}$ has mean zero and is bounded.

We can now walk through the results in Section \ref{sec:unknown}, adapting each one to handle GLMs as necessary (most results carry over with virtually no changes). The notation is adapted as $F^b\left(g,x\right) = \min_j \chi\left(b^\top_j x\right) - g_j$ and $p^{b,g}_k = P\left(\chi\left(b^\top_k X\right) - g_k = F^b\left(g,X\right)\right)$.

We begin by giving a new version of Lemma \ref{lem:technicalunknown1}. The local strong concavity property is very similar to what was shown before, but with an extra factor $\kappa_9$ included in two places. The proof is nearly identical to the proof of Lemma \ref{lem:technicalunknown1}, with a minor modification explained in a separate section of the Appendix (Section \ref{sec:proofs}).

\begin{lem}\label{lem:technicalunknown1glm}
Let Assumptions \ref{a1new} and \ref{aextra} hold, and take $\delta\in\mathds{R}^K$ satisfying $\delta_1=0$. Fix $b_1,...,b_K$ and let $\Delta_k = b_k-\beta_k$. Define $\Delta^{\max} = \max_k \|\Delta_k\|_2$. The following inequality holds:
\begin{eqnarray*}
&\,& \mathbb{E}\left(F^{b}\left(g^*,X\right)\right)-\mathbb{E}\left(F^{b}\left(g^*+\delta,X\right)\right) - \delta^\top p^{b,g^*}\\
&\geq& \frac{\kappa'_2\kappa'_4}{8K}\left(\sum^K_{k=1} \min\left\{\delta^2_k,\left(\kappa'_1\right)^2\right\}1_{\left\{\min\left\{\left|\delta_k\right|,\kappa'_1\right\} \geq 8\max\left\{1,\frac{\kappa'_3}{\kappa'_2\kappa'_4}\right\}\kappa_4\kappa_9\Delta^{\max}\right\}}\right)1_{\left\{\Delta^{\max}\leq \frac{\kappa'_1}{2\kappa_4\kappa_9}\right\}}.
\end{eqnarray*}
\end{lem}

With the updated definition of $p^{b,g}$, Lemmas \ref{lem:technicalunknown2}-\ref{lem:technicalunknown3} remain unchanged. We omit the proofs, as they are almost completely identical to the original version, with all instances of $b^\top X$ replaced by $\chi\left(b^\top X\right)$, and with $\kappa_4$ replaced by $\kappa_4\kappa_9$ throughout. Essentially, an additional factor $\kappa_9$ is absorbed into the constants $C'$ and $D$.

\begin{lem}\label{lem:technicalunknown2glm}
Suppose that we are in the situation of Lemma \ref{lem:technicalunknown1glm}. Then, we have
\begin{equation*}
\delta^\top\left(p^{b,g^*+\delta}-p^{b,g^*}\right) \geq \max\left\{0,C\min\left\{\|\delta\|^2_2,\kappa^2_1\right\}-C'\left(\Delta^{\max}\right)^2\right\}
\end{equation*}
where $C,C'>0$ are constants.
\end{lem}

\begin{lem}\label{lem:technicalunknown3glm}
Suppose that we are in the situation of Lemma \ref{lem:technicalunknown1glm}. Then,
\begin{equation*}
\|p^{b,g^*}-p\|_2 \leq D\cdot\Delta^{\max},
\end{equation*}
where $D > 0$ is a constant.
\end{lem}

Lemma \ref{lem:probineq} is similarly unchanged. We omit the proof, as it is identical to the original version, with the updated definition of $\pi^{b,g}$ and with $\kappa_4$ replaced by $\kappa_4\kappa_9$.

\begin{lem}\label{lem:probineqglm}
Let Assumptions \ref{a1new} and \ref{aextra} hold. For arbitrary $b,g$, let $\delta = g - g^*$ and $\Delta^{\max} = \max_j \|b_j - \beta_j\|_2$. Define $\pi^{b,g}\left(X\right) = \arg\min_j \chi\left(b^\top X_j\right) - g_j$. Then, there exists $C_0 > 0$ such that
\begin{equation*}
P\left(\pi^{b,g} \neq \pi^{\beta,g^*}\right) \leq C_0 \left(\kappa_4\kappa_9 \Delta^{\max} + \|\delta\|_{\infty}\right).
\end{equation*}
\end{lem}

Lemma \ref{lem:errorrates1} is similarly unchanged. We omit the proof, as it is identical to the original version, with the updated definition of $\pi^{b,g}$ and with Lemma \ref{lem:probineqglm} invoked instead of Lemma \ref{lem:probineq}. These two lemmas differ only by a constant factor in front of $\Delta^{\max}$, which is absorbed into $C_2$.

\begin{lem}\label{lem:errorrates1glm}
Let Assumptions \ref{a1new} and \ref{a2glm}-\ref{aextra} hold. For arbitrary $b,g$, let $\delta = g - g^*$ and $\Delta^{\max} = \max_j \|b_j - \beta_j\|_2$. Define $\pi^{b,g}\left(X\right) = \arg\min_j \chi\left(b^\top X_j\right) - g_j$, and let
\begin{equation}\label{eq:defineSk}
S_k\left(b,g\right) = \mathbb{E}\left(X X^\top 1_{\left\{\pi^{b,g}\left(X\right)=k\right\}}\right).
\end{equation}
Then, there exist $C_1,C_2 > 0$ such that
\begin{equation*}
\lambda_{\min}\left(S_k\left(b,g\right)\right)\geq \kappa_6 - C_1\|\delta\|_{\infty} -C_2\Delta^{\max}.
\end{equation*}
\end{lem}

Lemma \ref{lem:hoeffding} does not require any modifications, as Assumptions \ref{a2}(ii) and \ref{a2glm}(ii) are identical. Lemma \ref{lem:quadraticglm} redefines the residual error according to Assumption \ref{a2glm}(i), but otherwise is unchanged from Lemma \ref{lem:quadratic}.

\begin{lem}\label{lem:hoeffdingglm}
Let Assumptions \ref{a1new} and \ref{a2glm} hold. Let $T$ be a deterministic subset of $\left\{1,...,n\right\}$, and define $V_T = \sum_{n\in T} X^n\left(X^n\right)^\top$. Then, for any $z > 0$, (\ref{eq:hoeffding1})-(\ref{eq:hoeffding2}) hold.
\end{lem}

\begin{lem}\label{lem:quadraticglm}
Let Assumptions \ref{a1new} and \ref{a2glm} hold. Let $\left\{\pi^n\right\}^{\infty}_{n=1}$ be a sequence of random variables such that each $\pi^n$ takes values in $\left\{1,...,K\right\}$ and is measurable with respect to $\bar{\mathcal{F}}^n$, the sigma-algebra generated by $\pi^1,...,\pi^{n-1}$, $W^1_{\pi^1},...,W^{n-1}_{\pi^{n-1}}$, and $X^1,...,X^n$.

Define $V^n_k = \sum^n_{m=1} X^m \left(X^m\right)^\top 1_{\left\{\pi^n=k\right\}}$. Similarly, let $v^n_k = \sum^n_{m=1} \varepsilon^m_k X^m 1_{\left\{\pi^n=k\right\}}$, with $\varepsilon^m_k = L_2\left(W^m_k,\beta^\top_k X^m\right)$. Then, (\ref{eq:existsm}) and (\ref{eq:Vandvbound}) hold.
\end{lem}

Lemma \ref{lem:concreg} requires a new proof, presented in a separate section of the Appendix. We give the statement below.

\begin{lem}\label{lem:concregglm}
Let Assumptions \ref{a1new} and \ref{a2glm}-\ref{aextra} hold, and let $\pi$ be the semi-myopic policy. There exist constants $C_3,C_4,C_5>0$, which depend only on $d$, $K$ and $\kappa_4,...,\kappa_9$, such that
\begin{equation*}
P\left(\max_k \|\Delta^n_k\|_2 > C_3\left(\log n\right)^{-\frac{5}{2}}\right) \leq C_4 \exp\left(-C_5\left(\log n\right)^4\right).
\end{equation*}
\end{lem}

On the other hand, Lemma \ref{lem:concg} remains unchanged. We omit the proof, as it is identical to the original version, with the sole difference that we now refer to Lemmas \ref{lem:technicalunknown2glm}-\ref{lem:technicalunknown3glm} and \ref{lem:concregglm} instead of Lemmas \ref{lem:technicalunknown2}-\ref{lem:technicalunknown3} and \ref{lem:concreg}.

\begin{lem}\label{lem:concgglm}
Let Assumptions \ref{a1new} and \ref{a2glm}-\ref{aextra} hold. Suppose that the semi-myopic policy is used to sample cost functions, while $g^n$ is updated as in (\ref{eq:saglm}). Let $C_1$ be the constant obtained from Lemma \ref{lem:errorrates1glm}. There exists a constant $C_6>0$ such that
\begin{equation*}
P\left(\|\delta^n\|_2 > \frac{\kappa_6}{4C_1}\right) \leq \frac{C_6}{n^9}.
\end{equation*}
\end{lem}

Finally, we reach the main results. Proposition \ref{lem:expregglm} requires a new proof, presented in a separate section of the Appendix. On the other hand, Proposition \ref{lem:expgglm} is virtually unchanged from Proposition \ref{lem:expg}: the sole difference is that all lemmas used in the proof are replaced by their respective counterparts in this section.

\begin{prop}\label{lem:expregglm}
Let Assumptions \ref{a1new} and \ref{a2glm}-\ref{aextra} hold, and let $\pi$ be the semi-myopic policy. There exists a constant $C_7 > 0$ such that
\begin{equation*}
\mathbb{E}\left(\max_j \|\Delta^n_j\|^2_2\right) \leq \frac{C_7}{n}.
\end{equation*}
\end{prop}

\begin{prop}\label{lem:expgglm}
Let Assumptions \ref{a1new} and \ref{a2glm}-\ref{aextra} hold. Suppose that the semi-myopic policy is used to sample cost functions, while $g^n$ is updated as in (\ref{eq:saglm}). There exists some $\bar{\alpha}>0$ such that
\begin{equation*}
\mathbb{E}\left(\|\delta^n\|^2_2\right) \leq \frac{C_8}{n}
\end{equation*}
if $\alpha > \bar{\alpha}$.
\end{prop}

\section{Appendix: additional technical lemmas}

In this section, we present two stand-alone technical results used in various proofs. The first technical lemma is used in the proof of Lemma \ref{lem:technicalunknown1}.

\begin{lem}\label{lem:weirdintegral}
Suppose that $h$ is a non-negative function satisfying $\sup_{z\in\mathbb{R}} h\left(z\right)\leq\rho_1$ and $\int^\rho_0 h\left(z\right) \geq \rho\cdot\rho_2$ for $0 < \rho \leq \rho_3$, where $\rho_1,\rho_2,\rho_3 > 0$. Let $\delta,c,\kappa> 0$ satisfy
\begin{equation}\label{eq:weirdassum}
\min\left\{\delta,\kappa\right\} \geq 4 c \max\left\{1,\frac{\rho_1}{\rho_2}\right\}, \qquad \kappa\leq\rho_3.
\end{equation}
Then,
\begin{equation}\label{eq:weirdintegral}
\int^\kappa_0 \left(\min_{\left|t\right|\leq c} \left|z+t-\delta\right|\cdot 1_{\left\{-\delta < z+t-\delta \leq 0\right\}}\right)h\left(z\right)dz \geq \frac{\rho_2}{8}\cdot\min\left\{\delta^2,\kappa^2\right\}.
\end{equation}
\end{lem}

\noindent\textbf{Proof:} Clearly, if $-c \leq z-\delta \leq c$, the integrand in (\ref{eq:weirdintegral}) equals zero. Hence, we only need to consider $z-\delta > c$ or $z-\delta < -c$. For a lower bound, it is sufficient to consider only one case, so we pick the second, i.e., $z-\delta < -c$. We derive
\begin{eqnarray}
&\,& \int_0^{\min\left\{\delta-c,\kappa\right\}} \left(\min_{\left|t\right|\leq c}\left|z+t-\delta\right|\cdot 1_{\left\{-\delta < z+t-\delta \leq 0\right\}}\right)h\left(z\right)dz\nonumber\\
&=& \int^{\min\left\{\delta-c,\kappa\right\}}_{c} \left(\min_{\left|t\right|\leq c} \left|z+t-\delta\right|\cdot 1_{\left\{-\delta < z+t-\delta \leq 0\right\}}\right)h\left(z\right)dz\label{eq:changelimittoc}\\
&=& \int^{\min\left\{\delta-c,\kappa\right\}}_{c} \left(\min_{\left|t\right|\leq c} \left(\delta-t-z\right)\cdot 1_{\left\{-\delta < z+t-\delta \leq 0\right\}}\right)h\left(z\right)dz\label{eq:getridofabs}\\
&=& \int^{\min\left\{\delta-c,\kappa\right\}}_{c} \left(\delta-c-z\right)h\left(z\right)dz\label{eq:getridofmin}\\
&=& \left.H\left(z\right)\left(\delta-c-z\right)\right|^{\min\left\{\delta-c,\kappa\right\}}_c + \int^{\min\left\{\delta-c,\kappa\right\}}_{c} H\left(z\right)dz,\label{eq:intbyparts}
\end{eqnarray}
where $H$ is the antiderivative of $h$ with $H\left(0\right) = 0$. In this derivation, (\ref{eq:changelimittoc}) follows because, if $0\leq z< c$, the integrand in (\ref{eq:weirdintegral}) can be made to equal zero by taking $t=c$. Equation (\ref{eq:getridofabs}) follows by taking the positive part of the absolute value, (\ref{eq:getridofmin}) sets $t=-c$, and (\ref{eq:intbyparts}) applies integration by parts.

For any $0 < \rho \leq \rho_3$, we have
\begin{equation}\label{eq:Hbound}
H\left(\rho\right) = H\left(\rho\right) - H\left(0\right) = \int^\rho_0 h\left(z\right) dz \geq \rho\cdot\rho_2.
\end{equation}
To continue bounding (\ref{eq:intbyparts}), let us first consider the case $\delta - c < \kappa$. If this holds, then
\begin{equation*}
\left.H\left(z\right)\left(\delta-c-z\right)\right|^{\min\left\{\delta-c,\kappa\right\}}_c = \left.H\left(z\right)\left(\delta-c-z\right)\right|^{\delta-c}_c = -H\left(c\right)\left(\delta-2c\right),
\end{equation*}
and
\begin{eqnarray}
\int^{\min\left\{\delta-c,\kappa\right\}}_{c} H\left(z\right)dz &=& \int^{\delta-c}_{c} H\left(z\right)dz\nonumber\\
&\geq& \rho_2 \int^{\delta - c}_c z dz\label{eq:uselowerboundh}\\
&=& \frac{\rho_2}{2}\delta\left(\delta-2c\right)\nonumber
\end{eqnarray}
where (\ref{eq:uselowerboundh}) follows by (\ref{eq:Hbound}). Thus, we can continue (\ref{eq:intbyparts}) as
\begin{eqnarray}
&\,& \int_0^{\min\left\{\delta-c,\kappa\right\}} \left(\min_{\left|t\right|\leq c}\left|z+t-\delta\right|\cdot 1_{\left\{-\delta < z+t-\delta \leq 0\right\}}\right)h\left(z\right)dz\nonumber\\
&\geq & \left(\delta-2c\right)\left(\frac{\rho_2}{2}\delta - H\left(c\right)\right)\nonumber\\
&=&\left(\delta-2c\right)\left(\frac{\rho_2}{2}\delta - \left(H\left(c\right)-H\left(0\right)\right)\right)\nonumber\\
&=&\left(\delta-2c\right)\left(\frac{\rho_2}{2}\delta - \int^c_0 h\left(z\right)dz\right)\nonumber\\
&\geq& \left(\delta-2c\right)\left(\frac{\rho_2}{2}\delta - \rho_1 c\right)\label{eq:usedensitybound}\\
&\geq& \frac{\delta}{2}\cdot\frac{\delta\rho_2}{4}\label{eq:manipulatemaxassum}\\
&=& \frac{\delta^2\rho_2}{8},\label{eq:weirddeltacase}
\end{eqnarray}
where (\ref{eq:usedensitybound}) uses the boundedness of $h$, and (\ref{eq:manipulatemaxassum}) follows from (\ref{eq:weirdassum}).

It remains to consider the case $\delta - c \geq \kappa$. If this holds, then
\begin{equation*}
\left.H\left(z\right)\left(\delta-c-z\right)\right|^{\min\left\{\delta-c,\kappa\right\}}_c = \left.H\left(z\right)\left(\delta-c-z\right)\right|^{\kappa}_c = H\left(\kappa\right)\left(\delta-c-\kappa\right) - H\left(c\right)\left(\delta-2c\right),
\end{equation*}
and
\begin{eqnarray*}
\int^{\min\left\{\delta-c,\kappa\right\}}_{c} H\left(z\right)dz &=& \int^{\kappa}_{c} H\left(z\right)dz\\
&\geq & \int^{\kappa}_c zdz\\
&=& \frac{\rho_2}{2}\left(\kappa^2-c^2\right).
\end{eqnarray*}
Continuing (\ref{eq:intbyparts}), we have
\begin{eqnarray}
&\,& \int_0^{\min\left\{\delta-c,\kappa\right\}} \left(\min_{\left|t\right|\leq c}\left|z+t-\delta\right|\cdot 1_{\left\{-\delta < z+t-\delta \leq 0\right\}}\right)h\left(z\right)dz\nonumber\\
&\geq & H\left(\kappa\right)\left(\delta-c-\kappa\right) - H\left(c\right)\left(\delta-2c\right) + \frac{\rho_2}{2}\left(\kappa^2-c^2\right)\nonumber\\
&=& \left(H\left(\kappa\right)-H\left(c\right)\right)\left(\delta-c-\kappa\right)-H\left(c\right)\left(\kappa-c\right)+\frac{\rho_2}{2}\left(\kappa^2-c^2\right)\nonumber\\
&\geq &-H\left(c\right)\left(\kappa-c\right)+\frac{\rho_2}{2}\left(\kappa^2-c^2\right)\label{eq:removekappac}\\
&=&\left(\kappa-c\right)\left(\frac{\rho_2}{2}\left(\kappa+c\right)-H\left(c\right)\right)\nonumber\\
&=&\left(\kappa-c\right)\left(\frac{\rho_2}{2}\left(\kappa+c\right)-\int^c_0 h\left(z\right)dz\right)\nonumber\\
&\geq& \left(\kappa-c\right)\left(\frac{\rho_2}{2}\left(\kappa+c\right)-\rho_1c\right)\nonumber\\
&\geq& \frac{3\kappa}{4}\left(\frac{\rho_2}{2}\kappa - \rho_1c\right)\nonumber\\
&\geq & \frac{3\kappa}{4}\cdot\frac{\kappa\rho_2}{4}\label{eq:manipulatemaxassum2}\\
&=& \frac{3\kappa^2\rho_2}{16}\nonumber\\
&\geq& \frac{\kappa^2\rho_2}{8},\label{eq:weirdkappacase}
\end{eqnarray}
where (\ref{eq:removekappac}) uses $\kappa\geq c$ and $\delta-c\geq\kappa$, and (\ref{eq:manipulatemaxassum2}) uses (\ref{eq:weirdassum}). Combining (\ref{eq:weirddeltacase}) and (\ref{eq:weirdkappacase}) yields the desired result.\qed

The second technical lemma is used in the proof of Theorems \ref{thm:known} and \ref{thm:errorrates}.

\begin{lem}\label{lem:weirdineq}
Let $\left\{a_n\right\}^{\infty}_{n=0}$ be a non-negative sequence. Suppose that $C_0,C_1 > 0$ and $C_2,C_3\geq 0$ are fixed constants, and that, for $n\geq C_0$ we have
\begin{equation}\label{eq:weirdineq}
a_{n+1} \leq \left(1-C_1\frac{1}{n}\right) a_n + \frac{C_2}{n} + \frac{C_3\sqrt{a_n}}{n}.
\end{equation}
Then,
\begin{equation*}
\sup_{n\geq C_0} a_n \leq \max\left\{a_{C_0},\left(C^{-2}_1+C^{-1}_1\right)\left(C^2_3+2C_1C_2\right)\right\}.
\end{equation*}
\end{lem}

\noindent\textbf{Proof:} Define
\begin{equation*}
D_1 = \left(\frac{C_3 + \sqrt{C^2_3 + 4C_1 C_2}}{2C_1}\right)^2, \qquad D_2 = D_1 + C_2 + C_3\sqrt{D_1}.
\end{equation*}
It is straightforward to verify that $-C_1D_1 + C_2 +C_3\sqrt{D_1} = 0$ by checking the root of the quadratic equation $-C_1x^2 + C_3x + C_2 = 0$. Hence,
\begin{eqnarray}
D_2 &=& \left(1+C_1\right)D_1\nonumber\\
&\leq & \left(1+C_1\right)\frac{2C^2_3 + 2\left(C^2_3 + 4C_1C_2\right)}{4C^2_1}\label{eq:elementary}\\
&=& \left(C^{-2}_1 + C^{-1}_1\right)\left(C^2_3 + 2C_1C_2\right),\nonumber
\end{eqnarray}
where (\ref{eq:elementary}) follows by the elementary inequality $\left(x+y\right)^2 \leq 2x^2+2y^2$ for $x,y\in\mathbb{R}$.

We now show by induction that $a_n \leq \max\left\{D_2,a_{C_0}\right\}$ for all $n \geq C_0$. The inequality obviously holds for $n = C_0$. Suppose that it holds for $n$. We consider two cases: $a_n \geq D_1$ and $a_n < D_1$. In the first case, we observe that
\begin{equation*}
a_{n+1}-a_n \leq -\frac{C_1a_n}{n} + \frac{C_2}{n} + \frac{C_3\sqrt{a_n}}{n} \leq 0
\end{equation*}
where the second inequality is due to the non-negativity of $a_n$ together with the assumption that $a_n \geq D_1$. In the second case, we observe that
\begin{equation*}
a_{n+1} < \left(1-\frac{C_1}{n}\right)D_1 + \frac{C_2}{n} + \frac{C_3\sqrt{D_1}}{n} \leq D_2
\end{equation*}
where the first inequality is due to (\ref{eq:weirdineq}) and the assumption that $a_n < D_1$, and the second inequality follows by $n \geq 1$. This completes the proof.\qed

\section{Appendix: proofs}\label{sec:proofs}

Below, we provide complete proofs for all results that were stated in the text.

\subsection{Proof of Proposition \ref{prop:ipa}}


By assumption, $P\left(c_j\left(X\right)-g_j = c_k\left(X\right)-g_k\right) = 0$ for any $j\neq k$, implying that $\arg\min_j c_j\left(X\right) - g_j$ is unique with probability $1$. Fix any $h \in \mathbb{R}^K$ and a sequence $\left\{t_n\right\}$ with each $t_n \in \left(0,1\right)$ and $\lim_{n\rightarrow\infty} t_n = 0$. Define
\begin{equation*}
q_n\left(\omega\right) = \frac{F\left(g + t_n h, X\left(\omega\right)\right) - F\left(g,X\left(\omega\right)\right)}{t_n}.
\end{equation*}
Let $A$ be the set of all $\omega$ such that $\arg\min_j\left(c_j\left(X\left(\omega\right)\right) - g_j\right)$ is unique. For $\omega\in A$, let $j^*\left(\omega\right)$ be the unique argmin. Define
\begin{equation*}
q\left(\omega\right) = \left\{
\begin{array}{c l}
- h_{j^*\left(\omega\right)} & \omega\in A,\\
0 & \omega\notin A.
\end{array}
\right.
\end{equation*}
Consider $\omega\in A$. For sufficiently large $n$, $j^*\left(\omega\right) = \arg\min_j\left(c_j\left(X\left(\omega\right)\right) - g_j - t_nh_j\right)$, and thus $q_n\left(\omega\right) = q\left(\omega\right)$. Since $P\left(A\right) = 1$, we have $q_n \rightarrow q$ almost surely.

Observe that
\begin{equation*}
\left|F\left(g+t_n h,X\right) - F\left(g,X\right)\right| \leq t_n \max_j\left|h_j\right|,
\end{equation*}
whence $\left|q_n\right| \leq \max_j \left|h_j\right|$. We can then apply the dominated convergence theorem to obtain
\begin{eqnarray*}
\lim_{n\rightarrow\infty} \frac{\mathbb{E}\left(F\left(g + t_n h, X\right)\right) - \mathbb{E}\left(F\left(g,X\right)\right)}{t_n} &=& \mathbb{E}\left(q\right)\\
&=& - \sum^K_{k=1} h_k \cdot P\left(k = \arg\min_j \left(c_j\left(X\right) - g_j\right)\right),
\end{eqnarray*}
which completes the proof.

\subsection{Proof of Lemma \ref{a1}}

Let $\varphi$ denote the joint density of $Z$, and denote by $\varphi_{j,k}$ the density of $\left(Z_j,Z_k\right)$ for $j < k$. That is,
\begin{equation}\label{eq:a1-1}
\varphi_{j,k}\left(z_j,z_k\right) = \iint \varphi\left(u_1,...,u_{j-1},z_j,u_{j+1},...,u_{k-1},z_k,u_{k+1},...,u_K\right)du_1 ... du_K.
\end{equation}
By the usual calculation, the density of the pairwise difference $Z_j - Z_k$ is given by
\begin{equation}\label{eq:varrho}
\varrho\left(a\right) = \int^{\infty}_{-\infty} \varphi_{j,k}\left(a+z,z\right)dz.
\end{equation}
By Assumption \ref{a1new}(ii), we have $Z\in\left[-\kappa_4,\kappa_4\right]^K$ and so we may restrict the range of integration in (\ref{eq:varrho}) to a bounded interval. Specifically, because $z_j,z_k$ must both lie in $\left[-\kappa_4,\kappa_4\right]$, we may let $S_a = \left[-\kappa_4-a,\kappa_4-a\right]\cap\left[-\kappa_4,\kappa_4\right]$ and write
\begin{eqnarray*}
\varrho\left(a\right) &=& \int_{S_a} \varphi_{j,k}\left(a+z,z\right)dz\\ 
&\leq& \kappa_3 \cdot Leb\left(S_a\right)\cdot\left(2\kappa_4\right)^{K-2}\\
&\leq& \kappa_3 \left(2\kappa_4\right)^{K-1},
\end{eqnarray*}
where the first inequality uses Assumption \ref{a1new}(i).


Next, since $\varphi$ is bounded below by $\kappa_2$ on $\left[-\kappa_1,\kappa_1\right]^K$, we can restrict the integral in (\ref{eq:a1-1}) to $\left[-\kappa_1,\kappa_1\right]^{K-2}$, whence
\begin{equation*}
\varphi_{j,k}\left(z_j,z_k\right) \geq \kappa_2\left(2\kappa_1\right)^{K-2}, \quad \left(z_j,z_k\right)\in\left[-\kappa_1,\kappa_1\right]^2.
\end{equation*}
Therefore, for any $a$ such that $\left|a\right|\leq \frac{1}{2}\kappa_1$, we have
\begin{equation*}
\varrho\left(a\right) \geq \int^{\frac{1}{2}\kappa_1}_{-\frac{1}{2}\kappa_1} \varphi_{j,k}\left(a+z,z\right)dz \geq 2^{K-2}\kappa_2\kappa_1^{K-1}.
\end{equation*}
Thus, statement i) of the lemma is established with $\kappa'_1 = \frac{1}{2}\kappa_1$, $\kappa'_2 = 2^{K-2}\kappa_1^{K-1}\kappa_2$, and $\kappa'_3 = \kappa_3 \left(2\kappa_4\right)^{K-1}$.

We now verify statement ii) of the lemma. Take $0 < \kappa < \kappa'_1$. For $z = \left(z_1,...,z_K\right)$, define a function
\begin{equation*}
f\left(z\right) = 1_{\left\{\min_{\ell\neq j}z_\ell - z_k \geq \kappa'_1, \; 0 \leq z_j - z_k \leq \kappa\right\}}.
\end{equation*}
We then write
\begin{eqnarray}
P\left(\min_{\ell\neq j} Z_{\ell,k} \geq \kappa'_1 \mid 0 \leq Z_{j,k} \leq \kappa\right) &=& \frac{P\left(\min_{\ell\neq j} Z_{\ell,k} \geq \kappa'_1, \; 0 \leq Z_{j,k} \leq \kappa\right)}{P\left(0\leq Z_{j,k}\leq \kappa\right)}\nonumber\\
&\geq & \frac{\int_{\left[-\kappa_1,\kappa_1\right]^K}f\left(z\right)\varphi\left(z\right)dz}{P\left(0\leq Z_{j,k}\leq \kappa\right)}.\label{eq:a1-2}
\end{eqnarray}
Let $U_1,...,U_K$ be i.i.d. uniform on $\left[-\kappa_1,\kappa_1\right]$. We have
\begin{equation*}
P\left(\min_{\ell\neq j} U_\ell - U_k \geq \kappa'_1, \; 0\leq U_j - U_k\leq \kappa\right) = \frac{1}{\left(2\kappa_1\right)^K}\int_{\left[-\kappa_1,\kappa_1\right]^K}f\left(z\right)dz.
\end{equation*}
Since $\varphi$ is bounded below by $\kappa_2$ on $\left[-\kappa_1,\kappa_1\right]^K$, we have
\begin{eqnarray}
\int_{\left[-\kappa_1,\kappa_1\right]^K}f\left(z\right)\varphi\left(z\right)dz &\geq& \kappa_2 \int_{\left[-\kappa_1,\kappa_1\right]^K} f\left(z\right)dz\nonumber\\
&=& \left(2\kappa_1\right)^K\kappa_2 P\left(\min_{\ell\neq j} U_\ell - U_k \geq \kappa'_1, \; 0\leq U_j - U_k\leq \kappa\right).\label{eq:connecttoU}
\end{eqnarray}
We may now focus on the i.i.d. uniform case. First, observe that
\begin{equation*}
P\left(\min_{\ell\neq j} U_\ell - U_k \geq \kappa'_1, \; 0\leq U_j - U_k\leq \kappa\right) = \mathbb{E}\left( P\left(\min_{\ell\neq j} U_\ell \geq U_k + \kappa'_1\mid U_j,U_k\right) 1_{\left\{0\leq U_j - U_k\leq \kappa\right\}}\right).
\end{equation*}
The conditional distribution, given $U_j,U_k$, of $U_\ell$ for any $\ell\neq j,k$ is still uniform on $\left[-\kappa_1,\kappa_1\right]$. Furthermore, since $U_k\in\left[-\kappa_1,\kappa_1\right]$, we have $U_k + \kappa'_1 > -\kappa_1$. Thus, we may explicitly compute
\begin{equation*}
P\left(\min_{\ell\neq j} U_\ell \geq U_k + \kappa'_1\mid U_j,U_k\right) = 1_{\left\{U_k+\kappa'_1\leq\kappa_1\right\}}\frac{\left(\kappa_1 - \left(U_k+\kappa'_1\right)\right)^{K-2}}{\left(2\kappa_1\right)^{K-2}}.
\end{equation*}
Define $\kappa_0 = \int^{\kappa_1 -\kappa'_1}_{-\kappa_1} \left(\kappa_1 - \left(u+\kappa'_1\right)\right)^{K-2}du$. Then, we derive
\begin{eqnarray*}
&\,& P\left(\min_{\ell\neq j} U_\ell - U_k \geq \kappa'_1, \; 0\leq U_j - U_k\leq \kappa\right)\\
&=& \mathbb{E}\left(1_{\left\{U_k+\kappa'_1\leq\kappa_1\right\}}1_{\left\{0\leq U_j - U_k\leq \kappa\right\}}\frac{\left(\kappa_1 - \left(U_k+\kappa'_1\right)\right)^{K-2}}{\left(2\kappa_1\right)^{K-2}}\right)\\
&=& \int^{\kappa_1}_{-\kappa_1}\int^{\kappa_1}_{-\kappa_1} 1_{\left\{u+\kappa'_1\leq\kappa_1\right\}}1_{\left\{0\leq v - u\leq \kappa\right\}}\frac{\left(\kappa_1 - \left(u+\kappa'_1\right)\right)^{K-2}}{\left(2\kappa_1\right)^{K}}du\,dv\\
&=& \int^{\kappa_1}_{-\kappa_1} 1_{\left\{u+\kappa'_1\leq\kappa_1\right\}}\frac{\left(\kappa_1 - \left(u+\kappa'_1\right)\right)^{K-2}}{\left(2\kappa_1\right)^{K}}\left(\int^{\kappa_1}_{-\kappa_1} 1_{\left\{0\leq v - u\leq \kappa\right\}} dv\right)du\\
&=& \int^{\kappa_1-\kappa'_1}_{-\kappa_1} \frac{\left(\kappa_1 - \left(u+\kappa'_1\right)\right)^{K-2}}{\left(2\kappa_1\right)^{K}}\left(\min\left\{\kappa_1,\kappa+u\right\}-\max\left\{-\kappa_1,u\right\}\right)du\\
&=& \kappa \int^{\kappa_1-\kappa'_1}_{-\kappa_1} \frac{\left(\kappa_1 - \left(u+\kappa'_1\right)\right)^{K-2}}{\left(2\kappa_1\right)^{K}}du\\
&=& \kappa\cdot \kappa_0\frac{1}{\left(2\kappa_1\right)^K},
\end{eqnarray*}
where the last line follows from the fact that $\kappa+u\leq\kappa_1$ when $u \leq \kappa_1-\kappa'_1$. Combining this derivation with (\ref{eq:connecttoU}) yields
\begin{equation*}
\int_{\left[-\kappa_1,\kappa_1\right]^K}f\left(z\right)\varphi\left(z\right)dz \geq \kappa\cdot\kappa_0\cdot \kappa_2.
\end{equation*}
By (\ref{eq:a1-2}), we have
\begin{equation*}
P\left(\min_{\ell\neq j} Z_{\ell,k} \geq \kappa'_1 \mid 0 \leq Z_{j,k} \leq \kappa\right) \geq \frac{\kappa\cdot \kappa_0\cdot \kappa_2}{P\left(0\leq Z_{j,k}\leq \kappa\right)}.
\end{equation*}
Recall that $Z_{j,k}$ has a density $\varrho$ which is bounded above by $\kappa'_3$. Thus,
\begin{equation*}
\frac{1}{\kappa}P\left(0\leq Z_{j,k}\leq \kappa\right) \leq \kappa'_3,
\end{equation*}
whence
\begin{equation*}
P\left(\min_{\ell\neq j} Z_{\ell,k} \geq \kappa'_1 \mid 0 \leq Z_{j,k} \leq \kappa\right) \geq \frac{\kappa_0\kappa_2}{\kappa'_3}.
\end{equation*}
Thus, statement ii) of the lemma is established with $\kappa'_4 = \frac{\kappa_0\kappa_2}{\kappa'_3}$.

\subsection{Proof of Lemma \ref{lem:technicalknown1}}

We observe that $\|\zeta^{n+1}-p\|^2_2 \leq 2$ w.p. $1$. Then,
\begin{eqnarray*}
\|\delta^{n}\|_2 - \|\delta^0\|_2 &=& \sum^{n-1}_{m=0} \|\delta^{m+1}\|_2 - \|\delta_m\|_2\\
&\leq & \sum^{n-1}_{m=0} \frac{\alpha}{m+1}\|\zeta^{m+1}-p\|_2\\
&\leq & \sqrt{2}\cdot \alpha \sum^{n-1}_{m=0}\frac{1}{m+1}\\
&\leq& \sqrt{2}\cdot\alpha\left(1 + \log n\right)\\
&\leq & 2\sqrt{2}\cdot \alpha\log\left(n+1\right),
\end{eqnarray*}
which completes the proof.

\subsection{Proof of Lemma \ref{lem:technicalknown2}}

Fix arbitrary constants $M>1$ and $C_0>0$ such that $C_0 + 2\sqrt{2}\alpha \log\left(M\right) \leq 2$. Fix $n$ and define $N = \lceil\exp\left(\left(\log n\right)^{\frac{1}{3}}\right)\rceil$. Let $p^n = p^{g^*+\delta^n}$. Using (\ref{eq:basicdelta}), we derive
\begin{eqnarray}
\|\delta^{n+1}\|^2_2 &\leq & \|\delta^n\|^2_2 + \frac{\alpha^2}{\left(n+1\right)^2}\|\zeta^{n+1}-p\|^2_2 - \frac{2\alpha}{n+1}\left(\zeta^{n+1}-p\right)^\top \delta^n\nonumber\\
&\leq & \|\delta^n\|^2_2 + \frac{2\alpha^2}{\left(n+1\right)^2} - \frac{2\alpha}{n+1}\left(\zeta^{n+1}-p\right)^\top \delta^n\nonumber\\
&=& \|\delta^n\|^2_2 + \frac{2\alpha^2}{\left(n+1\right)^2} - \frac{2\alpha}{n+1}\left(\zeta^{n+1}-p^n\right)^\top \delta^n - \frac{2\alpha}{n+1}\left(p^n-p\right)^\top \delta^n\nonumber\\
&\leq& \|\delta^n\|^2_2 + \frac{2\alpha^2}{\left(n+1\right)^2} -\frac{2\alpha}{n+1}\left(\zeta^{n+1}-p^n\right)^\top \delta^n - \frac{2\alpha C_1}{n+1}\min\left\{\|\delta^n\|^2_2,\left(\kappa'_1\right)^2\right\},\label{eq:applyl2}
\end{eqnarray}
where (\ref{eq:applyl2}) uses Lemma \ref{lem:technical2} with an appropriate $C_1 > 0$.

Due to Lemma \ref{lem:technicalknown1}, we may write $\|\delta^n\|_2 \leq \alpha C_2\log\left(n+1\right)$ for some sufficiently large constant $C_2 > 0$. It is also easy to show that
\begin{equation*}
\min\left\{\|\delta^n\|^2_2,\left(\kappa'_1\right)^2\right\} \geq \min\left\{\left(\kappa'_1\right)^2 C^{-2}_0,1\right\}\cdot\min\left\{\|\delta^n\|^2_2,C^2_0\right\}.
\end{equation*}
Combining this with (\ref{eq:applyl2}), we have
\begin{equation*}
\|\delta^{n+1}\|^2_2 \leq \|\delta^n\|^2_2 + \frac{2\alpha^2}{\left(n+1\right)^2} -\frac{2\alpha}{n+1}\left(\zeta^{n+1}-p^n\right)^\top \delta^n - \frac{2\alpha C_3}{n+1}\min\left\{\|\delta^n\|^2_2,C^2_0\right\}
\end{equation*}
for some suitable $C_3 > 0$.

Pick any $n' \in \left\{1,...,n\right\}$ and define the event
\begin{equation*}
E_{n'} = \left\{\|\delta^{n-n'}\|^2_2 < C^2_0\right\} \cap \left\{\min_{n-n'+1\leq m \leq n} \|\delta^m\|^2_2 \geq C^2_0\right\},
\end{equation*}
with an additional special case
\begin{equation*}
E_{n+1} = \left\{\min_{0\leq m \leq n} \|\delta^m\|^2_2 \geq C^2_0\right\}.
\end{equation*}
We will study the behavior of $\|\delta^n\|_2$ on $E_{n'}$ for three possible cases representing different ranges of $n-n'+1$. Specifically, we consider 1) $n-n'+1 > \frac{n}{M}$, 2) $N \leq n -n'+1 \leq \frac{n}{M}$, and 3) $n-n'+1 < N$.

\textit{Case 1}: $n-n'+1 > \frac{n}{M}$. By repeating the arguments in the proof of Lemma \ref{lem:technicalknown1}, we obtain
\begin{equation}\label{eq:repeatknown1steps}
\|\delta^n\|_2 \leq \|\delta^{n-n'}\|_2 + 2\sqrt{2}\cdot\alpha\log\left(\frac{n}{n-n'+1}\right).
\end{equation}
On the event $E_{n'}$, (\ref{eq:repeatknown1steps}) combined with the assumption $n-n'+1 > \frac{n}{M}$ yields
\begin{equation*}
\|\delta^n\|_2 < C_0 + 2\sqrt{2}\cdot \alpha\log M.
\end{equation*}

\textit{Case 2}: $N \leq n -n'+1 \leq \frac{n}{M}$. On the event $E_{n'}$, we have
\begin{eqnarray}
\|\delta^n\|^2_2 \hspace{-0.1in} &\leq& \hspace{-0.1in} \|\delta^{n-n'}\|^2_2 + \sum^{n-1}_{m=n-n'} \frac{2\alpha^2}{\left(m+1\right)^2} - \frac{2\alpha}{m+1}\left(\zeta^{m+1}-p^m\right)^\top \delta^m - \frac{2\alpha C_3}{m+1}\min\left\{\|\delta^m\|^2_2,C^2_0\right\}\nonumber\\
\hspace{-0.1in} &\leq & \hspace{-0.1in} \|\delta^{n-n'}\|^2_2 + 2\alpha^2 \sum^{n-1}_{m=n-n'} \frac{1}{\left(m+1\right)^2} -2\alpha\sum^{n-1}_{m=n-n'} \frac{1}{m+1}\left(\zeta^{m+1}-p^m\right)^\top \delta^m\nonumber\\
\hspace{-0.1in} &\,& \hspace{-0.1in} - 2\alpha C^2_0 C_3\sum^{n-1}_{m=n-n'} \frac{1}{m+1}\label{eq:usingE}\\
\hspace{-0.1in} &\leq& \hspace{-0.1in} C^2_0 + \frac{4\alpha^2}{n-n'+1} -2\alpha\sum^{n-1}_{m=n-n'} \frac{1}{m+1}\left(\zeta^{m+1}-p^m\right)^\top \delta^m -2\alpha C^2_0 C_3\log\left(\frac{n}{n-n'+1}\right),\label{eq:intapprox}
\end{eqnarray}
where (\ref{eq:usingE}) uses the definition of $E_{n'}$, while (\ref{eq:intapprox}) uses integral bounds on partial sums. At the same time, on the event $E_{n'}$, we have $\|\delta^n\|^2_2 > C^2_0$, meaning that
\begin{equation}\label{eq:longpositive}
\frac{4\alpha}{n-n'+1} -2\sum^{n-1}_{m=n-n'} \frac{1}{m+1}\left(\zeta^{m+1}-p^m\right)^\top \delta^m -2 C^2_0 C_3\log\left(\frac{n}{n-n'+1}\right) \geq 0.
\end{equation}

We will now bound $P\left(E_{n'}\right)$ for Case 2. If the condition $N \leq n -n'+1 \leq \frac{n}{M}$ of Case 2 holds and $n$ is sufficiently large, we have
\begin{equation}\label{eq:midrange}
C^2_0 C_3\log\left(\frac{n}{n-n'+1}\right) \geq \frac{4\alpha}{n-n'+1}.
\end{equation}
Together, (\ref{eq:longpositive}) and (\ref{eq:midrange}) imply
\begin{equation*}
C^2_0 C_3\log\left(\frac{n}{n-n'+1}\right) \leq 2\left|\sum^{n-1}_{m=n-n'} \frac{1}{m+1}\left(\zeta^{m+1}-p^m\right)^\top \delta^m\right|,
\end{equation*}
whence
\begin{equation}\label{eq:boundPE1}
P\left(E_{n'}\right) \leq P\left(\left|\sum^{n-1}_{m=n-n'} \frac{1}{m+1}\left(\zeta^{m+1}-p^m\right)^\top \delta^m\right| \geq \frac{1}{2}C^2_0 C_3\log\left(\frac{n}{n-n'+1}\right)\right).
\end{equation}
For any $z > 0$, Azuma's inequality yields
\begin{eqnarray}
P\left(\left|\sum^{n-1}_{m=n-n'} \frac{1}{m+1}\left(\zeta^{m+1}-p^m\right)^\top \delta^m\right| \geq z\right) &\leq & 2\exp\left(-\frac{z^2}{4C^2_2 \sum^{n-1}_{m=n-n'} \left(\frac{\log \left(m+1\right)}{m+1}\right)^2}\right)\label{eq:applyC2bound}\\
&\leq & 2\exp\left(-\frac{z^2}{4C^2_2\sum^{\infty}_{m=n-n'} \left(\frac{\log \left(m+1\right)}{m+1}\right)^2}\right)\nonumber\\
&\leq & 2\exp\left(-\frac{z^2\left(n-n'+1\right)}{4C^2_2 C_4\left(\log\left(n-n'+1\right)\right)^2}\right),\label{eq:sumapproxagain}
\end{eqnarray}
where (\ref{eq:applyC2bound}) is due to the inequality
\begin{equation*}
\left|\frac{1}{m+1}\left(\zeta^{m+1}-p^m\right)^\top \delta^m\right| \leq \sqrt{2}\cdot C_2\frac{\log \left(m+1\right)}{m+1},
\end{equation*}
and (\ref{eq:sumapproxagain}) follows by
\begin{equation*}
\sum^{\infty}_{m=n-n'} \left(\frac{\log \left(m+1\right)}{m+1}\right)^2 \leq C_4 \frac{\left(\log\left(n-n'+1\right)\right)^2}{n-n'+1}
\end{equation*}
for some suitable constant $C_4 > 0$. Applying this result to (\ref{eq:boundPE1}) yields
\begin{equation}\label{eq:boundPE2}
P\left(E_{n'}\right) \leq 2\exp\left(-\frac{C^4_0 C^2_3\left(n-n'+1\right)\left(\log\left(\frac{n}{n-n'+1}\right)\right)^2}{16C^2_2 C_4\left(\log\left(n-n'+1\right)\right)^2}\right).
\end{equation}

The remainder of the proof of Case 2 focuses on showing the bound
\begin{equation}\label{eq:onethirdbound}
\min_{N \leq n -n'+1 \leq \frac{n}{M}} \frac{\left(n-n'+1\right)\left(\log\left(\frac{n}{n-n'+1}\right)\right)^2}{\left(\log\left(n-n'+1\right)\right)^2} \geq \exp\left(\left(\log n\right)^\frac{1}{3}\right).
\end{equation}
Since $\exp\left(z^{\frac{1}{3}}\right)\gg z$, combining (\ref{eq:boundPE2}) and (\ref{eq:onethirdbound}) yields
\begin{equation}\label{eq:boundPE3}
\sum_{N\leq n-n'+1\leq \frac{n}{M}} P\left(E_{n'}\right) \leq 2n \exp\left(-10\log n\right) \leq \frac{2}{n^9},
\end{equation}
which will be used later. To show (\ref{eq:onethirdbound}), we first observe that
\begin{equation*}
\frac{\left(n-n'+1\right)\left(\log\left(\frac{n}{n-n'+1}\right)\right)^2}{\left(\log\left(n-n'+1\right)\right)^2} = \exp\left(h\left(n-n'+1\right)\right)
\end{equation*}
where
\begin{equation*}
h\left(z\right) = \log z + 2\log\log\left(\frac{n}{z}\right)-2\log\log z, \quad z\in\left[N,\frac{n}{M}\right].
\end{equation*}
We then calculate
\begin{equation*}
h'\left(z\right) = \frac{1}{z}\left(1 - \frac{2}{\log n - \log z}-\frac{2}{\log z}\right).
\end{equation*}
Thus, $h'\left(z\right)\geq 0$ is equivalent to $\left(\log n - \log z\right)\left(\log z\right) \geq 2 \log n$, which can be rewritten as
\begin{equation*}
\left(\log z\right)^2 - \left(\log n\right)\left(\log z\right) + 2\log n \leq 0.
\end{equation*}
This holds if and only if $r^{n,1} \leq \log z \leq r^{n,2}$, where
\begin{eqnarray*}
r^{n,1} &=& \frac{\log n - \sqrt{\left(\log n\right)^2 - 8 \log n}}{2},\\
r^{n,2} &=& \frac{\log n + \sqrt{\left(\log n\right)^2 - 8 \log n}}{2}.
\end{eqnarray*}
It can be seen that $r^{n,1}\rightarrow 2$ by multiplying and dividing the expression by $\log n + \sqrt{\left(\log n\right)^2 - 8 \log n}$. On the other hand, $r^{n,2} = o\left(1\right) - 2 + \log n$. Since we are considering $z \geq N$, plugging in the definition of $N$ yields $\log z \geq \left(\log n\right)^{\frac{1}{3}}$, whence $\log z \geq r^{n,1}$ for large $n$. Therefore, for large $n$, $h$ is increasing on $\left[N,\exp\left(r^{n,2}\right)\right]$, and, if $\exp\left(r^{n,2}\right)\leq \frac{n}{M}$, $h$ is decreasing on $\left[\exp\left(r^{n,2}\right),\frac{n}{M}\right]$. In any event, we have
\begin{equation*}
\min_{N\leq z\leq \frac{n}{M}} h\left(z\right) \geq \min\left\{h\left(N\right),h\left(\frac{n}{M}\right)\right\} \geq \left(\log n\right)^{\frac{1}{3}}
\end{equation*}
for sufficiently large $n$, thus establishing (\ref{eq:onethirdbound}).

\textit{Case 3}: $n -n'+1 < N$. On the event $E_{n'}$, we have $\min_{N\leq m \leq n} \|\delta^m\|_2 > C_0$. By repeating the derivation of (\ref{eq:intapprox}), we obtain
\begin{equation}\label{eq:intapprox2}
\|\delta^n\|^2_2 \leq \|\delta^N\|^2_2 + \frac{4\alpha^2}{N+1} - 2\alpha\sum^{n-1}_{m=N} \frac{1}{m+1}\left(\zeta^{m+1}-p^m\right)^\top \delta^m -2\alpha C^2_0 C_3\log\left(\frac{n}{N+1}\right).
\end{equation}
Recall that $\|\delta^N\|^2_2 \leq \alpha^2 C^2_2\left(\log\left(N+1\right)\right)^2$. Plugging in the definition of $N$, we obtain $\|\delta^N\|^2_2 \lesssim \left(\log n\right)^{\frac{2}{3}}$. We also have
\begin{equation*}
\log\left(\frac{n}{N+1}\right) \geq \log n - \left(\log n\right)^{\frac{1}{3}},
\end{equation*}
whence, for large $n$, we have
\begin{equation*}
\|\delta^N\|^2_2 + \frac{4\alpha^2}{N+1} \leq \alpha C^2_0 C_3\log\left(\frac{n}{N+1}\right).
\end{equation*}
Consequently, (\ref{eq:intapprox2}) becomes
\begin{eqnarray}
\|\delta^n\|^2_2 &\leq& - 2\alpha\sum^{n-1}_{m=N} \frac{1}{m+1}\left(\zeta^{m+1}-p^m\right)^\top \delta^m - \alpha C^2_0 C_3\log\left(\frac{n}{N+1}\right)\nonumber\\
&\leq& - 2\alpha\sum^{n-1}_{m=N} \frac{1}{m+1}\left(\zeta^{m+1}-p^m\right)^\top \delta^m - \frac{1}{2}\alpha C^2_0 C_3\log n.\label{eq:intapprox3}
\end{eqnarray}
On $E_{n'}$, we have $\|\delta^n\|^2_2 \geq C^2_0$, whence (\ref{eq:intapprox3}) yields
\begin{equation*}
-\sum^{n-1}_{m=N} \frac{1}{m+1}\left(\zeta^{m+1}-p^m\right)^\top \delta^m \geq \frac{1}{4}C^2_0 C_3\log n + \frac{C_0}{2\alpha}.
\end{equation*}
Applying (\ref{eq:sumapproxagain}) yields
\begin{eqnarray*}
P\left(E_{n'}\right) &\leq& 2\exp\left(-\frac{\left(\frac{1}{4}C^2_0 C_3 \log n + \frac{C_0}{2\alpha}\right)^2 \left(N+1\right)}{4C^2_2C_4\left(\log \left(N+1\right)\right)^2}\right)\\
&\leq & \frac{2}{n^{10}},
\end{eqnarray*}
with the second inequality due to $\frac{N\left(\log n\right)^2}{\left(\log N\right)^2}\gg \log n$. Consequently, for large $n$, we have
\begin{equation*}
\sum_{n-n'+1< N} P\left(E_{n'}\right) \leq 2N \exp\left(-10\log n\right) \leq \frac{2}{n^9},
\end{equation*}
the same bound as in (\ref{eq:boundPE3}). Note that the same argument applies to the special case $n' = n+1$.

\textit{Putting Cases 1-3 together}. We derive
\begin{eqnarray}
&\,& P\left(\|\delta^n\|_2 \geq C_0 + 2\sqrt{2}\cdot\alpha\log M\right)\nonumber\\
&=& P\left(\|\delta^n\|_2 \geq C_0 + 2\sqrt{2}\cdot\alpha\log M, \; \min_{1\leq m\leq n-1} \|\delta^m\|_2 \geq C_0\right)\nonumber\\
&\,& +P\left(\|\delta^n\|_2 \geq C_0 + 2\sqrt{2}\cdot\alpha\log M, \; \min_{1\leq m\leq n-1} \|\delta^m\|_2 < C_0\right)\nonumber\\
&\leq& P\left(E_{n+1}\right) + P\left(\left\{\|\delta^n\|_2 \geq C_0 + 2\sqrt{2}\cdot\alpha\log M\right\}\cap\left(\bigcup_{n-n'+1\in\left\{1,...,n\right\}} E_{n'}\right)\right)\nonumber\\
&\leq& P\left(E_{n+1}\right) + \sum_{1\leq n-n'+1\leq n} P\left(\left\{\|\delta^n\|_2 \geq C_0 + 2\sqrt{2}\cdot\alpha\log M\right\}\cap E_{n'}\right)\nonumber\\
&=& P\left(E_{n+1}\right) + \sum_{1\leq n-n'+1 \leq \frac{n}{M}} P\left(E_{n'}\right)\label{eq:applycase1}\\
&=& \sum_{n-n'+1 < N} P\left(E_{n'}\right) + \sum_{N \leq n-n'+1 \leq \frac{n}{M}} P\left(E_{n'}\right)\nonumber\\
&\leq& \frac{4}{n^9},\label{eq:applycases23}
\end{eqnarray}
where (\ref{eq:applycase1}) is due to our analysis of Case 1, and (\ref{eq:applycases23}) is the result of the other two cases. By construction, $C_0 + 2\sqrt{2}\cdot\alpha\log M\leq 2$. The coefficient in (\ref{eq:applycases23}) can be increased to make the bound hold for all $n$. The desired result follows.

\subsection{Proof of Lemma \ref{lem:technicalunknown1}}

To avoid notational clutter, we write $g^*$ as simply $g$ in this proof, since no other $g$ will be considered. Note, however, that we do distinguish between $b$ and $\beta$.

Initially, we proceed exactly as in the proof of Lemma \ref{lem:technical1}. By making the same arguments (omitted to avoid redundancy), we derive
\begin{equation*}
F^b\left(g,X\right) - F^b\left(g+\delta,X\right) = T_1 + T_2,
\end{equation*}
where $T_1$ satisfies $\mathbb{E}\left(T_1\right) = \delta^\top p^{g,b}$, and
\begin{eqnarray}
\hspace{-0.1in} &\,& \hspace{-0.1in} T_2\nonumber\\
\hspace{-0.1in} &=& \hspace{-0.1in} \sum^K_{k=1} \left(b_k^\top X -g_k-\delta_k\right)\left(1_{\left\{b_k^\top X-g_k = F^b\left(g,X\right)\right\}}-1_{\left\{b_k^\top X-g_k-\delta_k = F^b\left(g+\delta,X\right)\right\}}\right)\nonumber\\
\hspace{-0.1in} &\geq & \hspace{-0.1in} \frac{1}{K} \sum^K_{k=1}\sum^K_{j=1}\left[\left(b^\top_k X-g_k-\delta_k\right)-\left(b^\top_j X-g_j-\delta_j\right)\right]\cdot 1_{\left\{b^\top_k X-g_k = F\left(g,X\right),b^\top_k X - g_k-\delta_k > b^\top_j X-g_j-\delta_j\right\}}.\label{eq:T2-4}
\end{eqnarray}
As in the proof of Lemma \ref{lem:technical1}, our goal is to bound the expectation of each term in (\ref{eq:T2-4}).

For notational convenience, we let $Z^{\beta}_{j,k} = \beta^\top_j X-g_j-\left(\beta^\top_k X-g_k\right)$ for any $j,k$. As in the proof of Lemma \ref{lem:technical1}, let $\bar{\delta}_{j,k} = \delta_j-\delta_k$. We also introduce the new notation
\begin{equation*}
\bar{\Delta}_{j,k} = \left(b_j-\beta_j\right)^\top X - \left(b_k-\beta_k\right)^\top X,
\end{equation*}
and note that
\begin{equation*}
\left|\bar{\Delta}_{j,k}\right| \leq \|X\|_2 \cdot\left(\|\Delta_j\|_2 + \|\Delta_k\|_2\right) \leq 2\kappa_4 \Delta^{\max}.
\end{equation*}
In addition,
\begin{equation*}
b^\top_j X - g_j - \left(b^\top_k X - g_k\right) = Z^{\beta}_{j,k} + \bar{\Delta}_{j,k}.
\end{equation*}

In the following, we may assume, without loss of generality, that $\bar{\delta}_{j,k} > 0$ and $\Delta^{\max}$ satisfies $\Delta^{\max}\leq\frac{\kappa'_1}{2\kappa_4}$. The bound becomes trivial if $\Delta^{\max}>\frac{\kappa'_1}{2\kappa_4}$ due to the concavity of $F^b$. Furthermore, for any $j\neq k$, either $\bar{\delta}_{j,k}$ or $\bar{\delta}_{k,j}$ will be non-negative, and a later step in the analysis allows us to only consider one of these.

With these assumptions, we write
\begin{eqnarray*}
&\,& \left[\left(b^\top_k X-g_k-\delta_k\right)-\left(b^\top_j X-g_j-\delta_j\right)\right]\cdot 1_{\left\{b^\top_k X-g_k = F\left(g,X\right),b^\top_k X - g_k-\delta_k > b^\top_j X-g_j-\delta_j\right\}}\\
&=& \left(\bar{\delta}_{j,k}-Z^\beta_{j,k}-\bar{\Delta}_{j,k}\right)\cdot 1_{\left\{\min_{\ell \neq j}Z^\beta_{\ell,k}+\bar{\Delta}_{\ell,k}\geq 0,\; 0\leq Z^\beta_{j,k}+\bar{\Delta}_{j,k}<\bar{\delta}_{j,k}\right\}}\\
&\geq & \min_{\left|t\right|\leq 2\kappa_4\Delta^{\max}}\left(\bar{\delta}_{j,k}-Z^\beta_{j,k}-t\right)\cdot 1_{\left\{\min_{\ell \neq j}Z^\beta_{\ell,k}+\bar{\Delta}_{\ell,k}\geq 0,\; 0\leq Z^\beta_{j,k}+t<\bar{\delta}_{j,k}\right\}}\\
&\geq & \min_{\left|t\right|\leq 2\kappa_4\Delta^{\max}}\left(\bar{\delta}_{j,k}-Z^\beta_{j,k}-t\right)\cdot 1_{\left\{\min_{\ell \neq j}Z^\beta_{\ell,k}+\min_{\ell \neq j}\bar{\Delta}_{\ell,k}\geq 0,\; 0\leq Z^\beta_{j,k}+t<\bar{\delta}_{j,k}\right\}}\\
&\geq & \min_{\left|t\right|\leq 2\kappa_4\Delta^{\max}}\left(\bar{\delta}_{j,k}-Z^\beta_{j,k}-t\right)\cdot 1_{\left\{\min_{\ell \neq j}Z^\beta_{\ell,k}-2\kappa_4\Delta^{\max}\geq 0,\; 0\leq Z^\beta_{j,k}+t<\bar{\delta}_{j,k}\right\}}\\
&\geq & \min_{\left|t\right|\leq 2\kappa_4\Delta^{\max}}\left(\bar{\delta}_{j,k}-Z^\beta_{j,k}-t\right)\cdot 1_{\left\{\min_{\ell \neq j}Z^\beta_{\ell,k}-\kappa'_1\geq 0,\; 0\leq Z^\beta_{j,k}+t<\bar{\delta}_{j,k}\right\}},
\end{eqnarray*}
where the last line is due to the assumption $\Delta^{\max} \leq \frac{\kappa'_1}{2\kappa_4}$. We now take expectations to obtain
\begin{eqnarray}
\hspace{-0.1in} &\,& \hspace{-0.1in} \mathbb{E}\left\{\left[\left(b^\top_k X-g_k-\delta_k\right)-\left(b^\top_j X-g_j-\delta_j\right)\right]\cdot 1_{\left\{b^\top_k X-g_k = F\left(g,X\right),b^\top_k X - g_k-\delta_k > b^\top_j X-g_j-\delta_j\right\}}\right\}\nonumber\\
\hspace{-0.1in} &\geq & \hspace{-0.1in} \mathbb{E}\left\{\min_{\left|t\right|\leq 2\kappa_4\Delta^{\max}}\left(\bar{\delta}_{j,k}-Z^\beta_{j,k}-t\right)\cdot 1_{\left\{\min_{\ell \neq j}Z^\beta_{\ell,k}-\kappa'_1\geq 0,\; 0\leq Z^\beta_{j,k}+t<\bar{\delta}_{j,k}\right\}}\right\}\nonumber\\
\hspace{-0.1in} &=& \hspace{-0.1in} \mathbb{E}\left\{P\left(\min_{\ell \neq j}Z^\beta_{\ell,k}\geq \kappa'_1\mid Z^\beta_{j,k} \right)\min_{\left|t\right|\leq 2\kappa_4\Delta^{\max}}\left(\bar{\delta}_{j,k}-Z^\beta_{j,k}-t\right)\cdot 1_{\left\{0\leq Z^\beta_{j,k}+t<\bar{\delta}_{j,k}\right\}}\right\}\nonumber\\
\hspace{-0.1in} &\geq& \hspace{-0.1in} \int^{\kappa'_1}_0 \min_{\left|t\right|\leq 2\kappa_4\Delta^{\max}}\left(\bar{\delta}_{j,k}-z-t\right)\cdot 1_{\left\{0\leq z+t<\bar{\delta}_{j,k}\right\}}P\left(\min_{\ell \neq j}Z^\beta_{\ell,k}\geq \kappa'_1\mid Z_{j,k}=z \right)P\left(Z_{j,k}\in dz\right)\nonumber\\
\hspace{-0.1in} &=& \hspace{-0.1in} \int^{\kappa'_1}_0 \min_{\left|t\right|\leq 2\kappa_4\Delta^{\max}}\left(\bar{\delta}_{j,k}-z-t\right)\cdot 1_{\left\{-\bar{\delta}_{j,k} \leq z+t-\bar{\delta}_{j,k}<0\right\}}P\left(\min_{\ell \neq j}Z^\beta_{\ell,k}\geq \kappa'_1\mid Z_{j,k}=z \right)P\left(Z_{j,k}\in dz\right)\label{eq:againcondint}
\end{eqnarray}
By repeating arguments from the proof of Lemma \ref{lem:technical1}, we obtain
\begin{equation*}
\int^\kappa_0 P\left(\min_{\ell \neq j}Z_{\ell,k}\geq \kappa'_1 \mid Z_{j,k}=z\right)P\left(Z_{j,k}\in dz\right) \geq \kappa\cdot \kappa'_2\kappa'_4
\end{equation*}
for any $0 < \kappa < \kappa'_1$. We can then apply Lemma \ref{lem:weirdintegral} and continue (\ref{eq:againcondint}) as
\begin{eqnarray*}
&\,& \mathbb{E}\left\{\left[\left(b^\top_k X-g_k-\delta_k\right)-\left(b^\top_j X-g_j-\delta_j\right)\right]\cdot 1_{\left\{b^\top_k X-g_k = F\left(g,X\right),b^\top_k X - g_k-\delta_k > b^\top_j X-g_j-\delta_j\right\}}\right\}\\
&\geq & \frac{\kappa'_2\kappa'_4}{8}\min\left\{\bar{\delta}^2_{j,k},\left(\kappa'_1\right)^2\right\}\cdot 1_{\left\{\min\left\{\bar{\delta}_{j,k},\kappa'_1\right\}\geq 8 \max\left\{1,\frac{\kappa'_3}{\kappa'_2\kappa'_4}\right\}\kappa_4\Delta^{\max}\right\}}1_{\left\{\Delta^{\max}\leq \frac{\kappa'_1}{2\kappa_4}\right\}}
\end{eqnarray*}
We then repeat the arguments at the end of the proof of Lemma \ref{lem:technical1} (starting from (\ref{eq:T2-3}) onwards) to obtain the final bound.

\subsection{Proof of Lemma \ref{lem:technicalunknown2}}

To avoid notational clutter, we write $g^*$ as simply $g$ in this proof, since no other $g$ will be considered. Note, however, that we do distinguish between $b$ and $\beta$.

As in the proof of Lemma \ref{lem:technical2}, we apply Lemma \ref{lem:technicalunknown1} and obtain
\begin{equation}\label{eq:applylemmaunknown1}
\delta^\top\left(p^{b,g+\delta}-p^{b,g}\right) \geq \frac{\kappa'_2\kappa'_4}{8K}\left(\sum^K_{k=1} \min\left\{\delta^2_k,\left(\kappa'_1\right)^2\right\}1_{\left\{\min\left\{\left|\delta_k\right|,\kappa'_1\right\} \geq 8\max\left\{1,\frac{\kappa'_3}{\kappa'_2\kappa'_4}\right\}\kappa_4\Delta^{\max}\right\}}\right)1_{\left\{\Delta^{\max}\leq \frac{\kappa'_1}{2\kappa_4}\right\}}.
\end{equation}
Let $C_0 = 8\max\left\{1,\frac{\kappa'_3}{\kappa'_2\kappa'_4}\right\}\kappa_4$. We wish to show that, if $\kappa'_1 \geq C_0\Delta^{\max}$, then
\begin{equation}\label{eq:tobeshownlemma2}
\min\left\{\delta^2_k,\left(\kappa'_1\right)^2\right\} 1_{\left\{\min\left\{\left|\delta_k\right|,\kappa'_1\right\} \geq C_0\Delta^{\max}\right\}} \geq \min\left\{\delta^2_k,\left(\kappa'_1\right)^2\right\}-C^2_0\left(\Delta^{\max}\right)^2.
\end{equation}
To show (\ref{eq:tobeshownlemma2}), we consider two cases. First, if $\min\left\{\left|\delta_k\right|,\kappa'_1\right\} < C_0\Delta^{\max}$, then the left-hand side of (\ref{eq:tobeshownlemma2}) is zero and the right-hand side is negative. If $\min\left\{\left|\delta_k\right|,\kappa'_1\right\} \geq C_0\Delta^{\max}$, then the left-hand side becomes $\min\left\{\delta^2_k,\left(\kappa'_1\right)^2\right\}$ and the right-hand side becomes $\min\left\{\delta^2_k,\left(\kappa'_1\right)^2\right\}-C^2_0\left(\Delta^{\max}\right)^2$.


Combining (\ref{eq:applylemmaunknown1}) and (\ref{eq:tobeshownlemma2}) yields
\begin{eqnarray}
\delta^\top\left(p^{b,g+\delta}-p^{b,g}\right) &\geq& \frac{\kappa'_2\kappa'_4}{8K}\left(\sum^K_{k=1} \min\left\{\delta^2_k,\left(\kappa'_1\right)^2\right\}-C^2_0\left(\Delta^{\max}\right)^2\right)1_{\left\{\Delta^{\max}\leq \frac{\kappa'_1}{2\kappa_4}\right\}}1_{\left\{\kappa'_1\geq C_0\Delta^{\max}\right\}}\nonumber\\
&=& \frac{\kappa'_2\kappa'_4}{8K}\left(\sum^K_{k=1} \min\left\{\delta^2_k,\left(\kappa'_1\right)^2\right\}-C^2_0\left(\Delta^{\max}\right)^2\right)1_{\left\{\Delta^{\max}\leq C_1\right\}}\label{eq:reformulateC1}\\
&\geq & \frac{\kappa'_2\kappa'_4}{8K}\left(\min\left\{\|\delta\|^2_2,\left(\kappa'_1\right)^2\right\}-C^2_0\left(\Delta^{\max}\right)^2\right)1_{\left\{\Delta^{\max}\leq C_1\right\}},\label{eq:repeattechnical2}
\end{eqnarray}
where $C_1>0$ in (\ref{eq:reformulateC1}) is a suitable constant obtained by combining the two conditions on $\Delta^{\max}$, and (\ref{eq:repeattechnical2}) repeats arguments in the proof of Lemma \ref{lem:technical2}. Thus, we have shown the desired result
\begin{equation}\label{eq:intermsofCandCprime}
\delta^\top\left(p^{b,g+\delta}-p^{b,g}\right) \geq C\left(\min\left\{\|\delta\|^2_2,\left(\kappa'_1\right)^2\right\}-C'\left(\Delta^{\max}\right)^2\right)
\end{equation}
in the case $\Delta^{\max}\leq C_1$. To handle the other case $\Delta^{\max}> C_1$, we write
\begin{equation}\label{eq:otherDeltabound}
C\left(\min\left\{\|\delta\|^2_2,\left(\kappa'_1\right)^2\right\}-C'\left(\Delta^{\max}\right)^2\right) \leq C\left(\kappa'_1\right)^2-C'\cdot C_1.
\end{equation}
We can make $C'$ large enough for the right-hand side of (\ref{eq:otherDeltabound}) to be negative. Since
\begin{equation*}
\delta^\top\left(p^{b,g+\delta}-p^{b,g}\right)\geq 0
\end{equation*}
by concavity of $F^b$, this means that (\ref{eq:intermsofCandCprime}) always holds. This completes the proof.

\subsection{Proof of Lemma \ref{lem:errorrates1}}

We first write
\begin{eqnarray*}
\lambda_{\min}\left(S_k\left(b,g\right)\right) &=& \lambda_{\min}\left(S_k\left(\beta,g^*\right) + S_k\left(b,g\right) - S_k\left(\beta,g^*\right)\right)\\
&\geq & \lambda_{\min}\left(S_k\left(\beta,g^*\right)\right) - \|S_k\left(b,g\right) - S_k\left(\beta,g^*\right)\|_{\text{sp}}\\
&\geq & \kappa_6 - \|S_k\left(b,g\right) - S_k\left(\beta,g^*\right)\|_{\text{sp}}.
\end{eqnarray*}
For any vector $a \in \mathbb{R}^d$, we have
\begin{eqnarray}
\left|a^\top \left(S_k\left(b,g\right) - S_k\left(\beta,g^*\right)\right) a\right| &=& \left|\mathbb{E}\left(\left(a^\top X\right)^2\left(1_{\left\{\pi^{b,g}=k\right\}}-1_{\left\{\pi^{\beta,g^*}=k\right\}}\right)\right)\right|\nonumber\\
&\leq & \|a\|^2_2 \kappa^2_4 \mathbb{E}\left|1_{\left\{\pi^{b,g}=k\right\}}-1_{\left\{\pi^{\beta,g^*}=k\right\}}\right|\label{eq:expabsprob}\\
&\leq & \|a\|^2_2 \kappa^2_4 P\left(\pi^{b,g} \neq \pi^{\beta,g^*}\right)\nonumber\\
&\leq & \|a\|^2_2 \kappa^2_4 C_0 \left(\kappa_4 \Delta^{\max} + \|\delta\|_{\infty}\right),\label{eq:useprobineq}
\end{eqnarray}
where (\ref{eq:expabsprob}) is due to Assumption \ref{a1new}(ii), and (\ref{eq:useprobineq}) follows by Lemma \ref{lem:probineq}. The desired bound follows.

\subsection{Proof of Lemma \ref{lem:hoeffding}}

Recall that, for any square matrix $A$, $\|A\|^2_{\text{sp}} \leq \tr\left(A^\top A\right)$. Therefore,
\begin{equation*}
\|V_T-\mathbb{E}\left(V_T\right)\|^2_{\text{sp}} \leq \sum_{\ell_1,\ell_2} \left(\left(V_T\right)_{\ell_1,\ell_2} - \left(\mathbb{E}\left(V_T\right)\right)_{\ell_1,\ell_2}\right)^2.
\end{equation*}
By Hoeffding's inequality,
\begin{equation*}
P\left(\left|\left(V_T\right)_{\ell_1,\ell_2} - \left(\mathbb{E}\left(V_T\right)\right)_{\ell_1,\ell_2}\right|>z\right) \leq 2\exp\left(-\frac{z^2}{2\kappa^4_4\left|T\right|}\right).
\end{equation*}
Thus,
\begin{eqnarray*}
P\left(\|V_T-\mathbb{E}\left(V_T\right)\|^2_{\text{sp}} > d^2 z^2\right) &\leq& P\left(\sum_{\ell_1,\ell_2} \left(\left(V_T\right)_{\ell_1,\ell_2} - \left(\mathbb{E}\left(V_T\right)\right)_{\ell_1,\ell_2}\right)^2 > d^2z^2\right)\\
&\leq& P\left(\bigcup_{\ell_1,\ell_2} \left\{\left(\left(V_T\right)_{\ell_1,\ell_2} - \left(\mathbb{E}\left(V_T\right)\right)_{\ell_1,\ell_2}\right)^2 > z^2\right\}\right)\\
&\leq& \sum_{\ell_1,\ell_2} P\left(\left(\left(V_T\right)_{\ell_1,\ell_2} - \left(\mathbb{E}\left(V_T\right)\right)_{\ell_1,\ell_2}\right)^2 > z^2\right)\\
&\leq& 2d^2\exp\left(-\frac{z^2}{2 \kappa^4_4\left|T\right|}\right),
\end{eqnarray*}
and (\ref{eq:hoeffding1}) follows by a change of variables.

To prove (\ref{eq:hoeffding2}), we observe that
\begin{equation*}
\lambda_{\min}\left(\sum^K_{k=1} \mathbb{E}\left(X X^\top 1_{\left\{\pi^*\left(X\right)=k\right\}}\right)\right) \geq \sum^K_{k=1} \lambda_{\min}\left(\mathbb{E}\left(X X^\top 1_{\left\{\pi^*\left(X\right)=k\right\}}\right)\right) \geq \kappa_6 K,
\end{equation*}
where the last inequality is due to Assumption \ref{a2}(ii). We then write
\begin{eqnarray*}
\lambda_{\min}\left(V_T\right) &\geq & \lambda_{\min}\left(\mathbb{E}\left(V_T\right)\right) - \|V_T - \mathbb{E}\left(V_T\right)\|_{\text{sp}}\\
&\geq & \kappa_6 K\left|T\right| - \|V_T - \mathbb{E}\left(V_T\right)\|_{\text{sp}}.
\end{eqnarray*}
Therefore,
\begin{eqnarray*}
P\left(\lambda_{\min}\left(V_T\right) < \frac{1}{2}\kappa_6 K \left|T\right|\right) &\leq & P\left(\|V_T - \mathbb{E}\left(V_T\right)\|_{\text{sp}} > \frac{1}{2}\kappa_6 K \left|T\right|\right)\\
&\leq& 2d^2\exp\left(-\frac{\kappa^2_6 K^2\left|T\right|}{8d^2\kappa^4_4}\right),
\end{eqnarray*}
with the last line following by (\ref{eq:hoeffding1}). This completes the proof.

\subsection{Proof of Lemma \ref{lem:quadratic}}

Fix $\mu$ and let
\begin{eqnarray*}
M^n_k\left(z\right) &=& \exp\left(z^\top v^n_k - \frac{1}{2}\kappa_5 z^\top V^n_k z\right)\\
\bar{M}^n_k &=& \int_{\mathbb{R}^d} M^n_k\left(z\right)h\left(z\right)dz,
\end{eqnarray*}
where $h$ is the density of the $\mathcal{N}\left(0,\frac{1}{\kappa_5\mu}I\right)$ distribution. Also let $H^n_k = \kappa_5\left(V^n_k + \mu\cdot I\right)$. We first bound $\bar{M}^n_k$ by deriving
\begin{eqnarray}
\bar{M}^n_k &=& \left(2\pi\right)^{-\frac{d}{2}}\left(\kappa_5\mu\right)^{\frac{d}{2}}\int_{\mathbb{R}^d} \exp\left(z^\top v^n_k - \frac{1}{2}\kappa_5 z^\top V^n_k z - \frac{1}{2}\kappa_5\mu\|z\|^2_2\right)dz\nonumber\\
&=& \left(2\pi\right)^{-\frac{d}{2}}\left(\kappa_5\mu\right)^{\frac{d}{2}} \int_{\mathbb{R}^d} \exp\left(z^\top v^n_k - z^\top H^n_k z\right)dz\nonumber\\
&=& \left(2\pi\right)^{-\frac{d}{2}}\left(\kappa_5\mu\right)^{\frac{d}{2}} \int_{\mathbb{R}^d} \exp\left(-\frac{1}{2}\left(z-\left(H^n_k\right)^{-1}v^n_k\right)^\top H^n_k\left(z-\left(H^n_k\right)^{-1}v^n_k\right) + \frac{1}{2}\left(v^n_k\right)^\top \left(H^n_k\right)^{-1} v^n_k\right)dz\nonumber\\
&=& \left(2\pi\right)^{-\frac{d}{2}}\left(\kappa_5\mu\right)^{\frac{d}{2}} \exp\left(\frac{1}{2}\left(v^n_k\right)^\top \left(H^n_k\right)^{-1} v^n_k\right)\cdot\left(2\pi\right)^{\frac{d}{2}}\left(\det H^n_k\right)^{-\frac{1}{2}}\nonumber\\
&=& \mu^{\frac{d}{2}} \exp\left(\frac{1}{2}\left(v^n_k\right)^\top \left(H^n_k\right)^{-1} v^n_k\right)\cdot\left(\det \left(V^n_k + \mu\cdot I\right)\right)^{-\frac{1}{2}}\nonumber\\
&\geq & \exp\left(\frac{1}{2}\left(v^n_k\right)^\top \left(H^n_k\right)^{-1} v^n_k\right)\left(\frac{\mu}{\mu+n\kappa^2_4}\right)^{\frac{d}{2}}\label{eq:meanofM1}\\
&=& \exp\left(\frac{1}{2}\left(v^n_k\right)^\top \left(H^n_k\right)^{-1} v^n_k\right)\left(1+n\frac{\kappa^2_4}{\mu}\right)^{-\frac{d}{2}},\label{eq:meanofM2}
\end{eqnarray}
where (\ref{eq:meanofM1}) is due to
\begin{equation*}
\det\left(V^n_k + \mu\cdot I\right) \leq \prod^d_{\ell=1} \left(\mu + \lambda_{\max}\left(V^n_k\right)\right) \leq \left(\mu + n\kappa^2_4\right)^d,
\end{equation*}
using Assumption \ref{a1new}(ii).

At the same time, we observe that
\begin{eqnarray*}
&\,& \mathbb{E}\left(M^{n+1}_k\left(z\right)\mid\bar{\mathcal{F}}^{n}\right)\\
&=& M^n_k\left(z\right)\cdot\mathbb{E}\left(\exp\left(z^\top X^{n+1}\varepsilon^{n+1}_k 1_{\left\{\pi^{n+1}=k\right\}} - \frac{1}{2}\kappa_5\left(z^\top X^{n+1}\right)^21_{\left\{\pi^{n+1}=k\right\}}\right)\mid\bar{\mathcal{F}}^{n}\right)\\
&=& M^n_k\left(z\right) \cdot \mathbb{E}\left\{\mathbb{E}\left[\exp\left(z^\top X^{n+1}\varepsilon^{n+1}_k 1_{\left\{\pi^{n+1}=k\right\}}\right)\mid\bar{\mathcal{F}}^{n+1}\right]\exp\left(- \frac{1}{2}\kappa_5\left(z^\top X^{n+1}\right)^21_{\left\{\pi^{n+1}=k\right\}}\right)\mid\bar{\mathcal{F}}^n\right\}\\
&\leq & M^n_k\left(z\right),
\end{eqnarray*}
where the last line is obtained by applying Assumption \ref{a2}(i). Thus, for any $z$, $\left\{M^n_k\left(z\right)\right\}^{\infty}_{n=1}$ is a supermartingale. It follows that $\bar{M}^n_k$ is also a supermartingale. By Theorem 1 in Sec. 7.3 of \cite{Sh19}, we have
\begin{equation*}
P\left(\max_{m\leq n} \bar{M}^n_k \geq \frac{1}{\mu}\right) \leq \mu\mathbb{E}\left(\bar{M}^0_k\right) = \mu.
\end{equation*}
From (\ref{eq:meanofM2}), we conclude (\ref{eq:existsm}).

To show (\ref{eq:Vandvbound}), we take $\mu = n$ and observe, from (\ref{eq:meanofM2}), that
\begin{equation*}
\mathbb{E}\left(\exp\left(\frac{1}{2\kappa_5} \left(v^n_k\right)^\top\left(V^n_k + n\cdot I\right)^{-1} v^n_k\right)\right)\left(1+\kappa^2_4\right)^{-\frac{d}{2}} \leq \mathbb{E}\left(\bar{M}^n_k\right).
\end{equation*}
Since $\mathbb{E}\left(\bar{M}^n_k\right) \leq 1$, the desired result follows.

\subsection{Proof of Lemma \ref{lem:concreg}}

We use the notation $V^n_k = \sum^n_{m=1} X^m \left(X^m\right)^\top 1_{\left\{\pi^n=k\right\}}$ and $v^n_k = \sum^n_{m=1} \varepsilon^m_k X^m 1_{\left\{\pi^n=k\right\}}$ from Lemma \ref{lem:quadratic}. Using this notation, and recalling (\ref{eq:ridge}), we can write
\begin{equation*}
\beta^n_k = \left(\rho^n\cdot I + V^n_k\right)^{-1}\left(V^n_k\beta_k + v^n_k\right).
\end{equation*}
Then,
\begin{equation*}
\Delta^n_k = -\rho^n\left(\rho^n\cdot I + V^n_k\right)^{-1}\beta_k + \left(\rho^n\cdot I + V^n_k\right)^{-1}v^n_k,
\end{equation*}
and
\begin{eqnarray}
\|\Delta^n_k\|_2 &\leq & \frac{\rho^n}{\rho^n+\lambda_{\min}\left(V^n_k\right)}\|\beta_k\|_2 + \frac{1}{\sqrt{\rho^n + \lambda_{\min}\left(V^n_k\right)}}\|\left(\rho^n\cdot I + V^n_k\right)^{-\frac{1}{2}}v^n_k\|_2\nonumber\\
&\leq & \frac{\rho^n}{\rho^n+\lambda_{\min}\left(V^n_k\right)}\|\beta_k\|_2 + \frac{1}{\sqrt{\rho^n + \lambda_{\min}\left(V^n_k\right)}}\|\left(\rho^n\cdot I + V^n_k\right)^{-\frac{1}{2}}\left(I+V^n_k\right)^{\frac{1}{2}}\|_{\text{sp}}\|\left(I+V^n_k\right)^{\frac{1}{2}}v^n_k\|_2\nonumber\\
&\leq & \frac{\rho^n}{\rho^n+\lambda_{\min}\left(V^n_k\right)}\|\beta_k\|_2 + \frac{1}{\sqrt{\rho^n + \lambda_{\min}\left(V^n_k\right)}}\|\left(I+V^n_k\right)^{\frac{1}{2}}v^n_k\|_2,\label{eq:Deltalowerbound}
\end{eqnarray}
where (\ref{eq:Deltalowerbound}) follows because $\rho^n\geq 1$.

Recall that $\mathcal{T}_k$ is the set of time periods in which the policy is forced to explore the $k$th cost function. Then, $\left|\mathcal{T}_k\right|\geq D_1\left(\log n\right)^9$ for some $D_1>0$. Letting $V^n_{\mathcal{T}_k} =\sum_{m\in\mathcal{T}_k,m\leq n} X^m \left(X^m\right)^\top$, we obviously have $\lambda_{\min}\left(V^n_k\right) \geq \lambda_{\min}\left(V^n_{\mathcal{T}_k}\right)$. Then, (\ref{eq:Deltalowerbound}) is continued as
\begin{equation}\label{eq:Deltalowerbound2}
\|\Delta^n_k\|_2 \leq \frac{\rho^n}{\rho^n+\lambda_{\min}\left(V^n_{\mathcal{T}_k}\right)}\|\beta_k\|_2 + \frac{1}{\sqrt{\rho^n + \lambda_{\min}\left(V^n_{\mathcal{T}_k}\right)}}\|\left(I+V^n_k\right)^{\frac{1}{2}}v^n_k\|_2.
\end{equation}

By Lemma \ref{lem:quadratic}, we have
\begin{equation*}
P\left(\|\left(I+V^n_k\right)^{\frac{1}{2}}v^n_k\|_2 \leq \sqrt{2\kappa_5 \log\left(\frac{1}{\eta}\right) + \kappa_5 d \log\left(1+n\kappa^2_4\right)}\right) \geq 1-\eta
\end{equation*}
for any $0 < \eta < 1$. Combining this with (\ref{eq:Deltalowerbound2}) yields
\begin{equation}\label{eq:problambdamin}
P\left(\|\Delta^n_k\|_2 > \frac{\rho^n}{\rho^n+\lambda_{\min}\left(V^n_{\mathcal{T}_k}\right)}\|\beta_k\|_2 + \sqrt{\frac{2\kappa_5\log\left(\frac{1}{\eta}\right) + \kappa_5 d\log\left(1+n\kappa^2_4\right)}{\rho^n+\lambda_{\min}\left(V^n_{\mathcal{T}_k}\right)}}\right)\leq \eta.
\end{equation}
By Lemma \ref{lem:hoeffding}, we have
\begin{equation*}
P\left(\lambda_{\min}\left(V^n_{\mathcal{T}_k}\right) < \frac{\kappa_6}{2}D_1 K\left(\log n\right)^9\right) \leq 2d^2\exp\left(-\frac{\kappa^2_6 D_1 K^2\left(\log n\right)^9}{8d^2\kappa^4_4}\right).
\end{equation*}
Combining this with (\ref{eq:problambdamin}) yields
\begin{eqnarray*}
&\,& P\left(\|\Delta^n_k\|_2 > \frac{\rho^n}{\rho^n+\frac{\kappa_6}{2}D_1 K\left(\log n\right)^9}\|\beta_k\|_2 + \sqrt{\frac{2\kappa_5\log\left(\frac{1}{\eta}\right) + \kappa_5 d\log\left(1+n\kappa^2_4\right)}{\rho^n+\frac{\kappa_6}{2}D_1 K\left(\log n\right)^9}}\right)\\
&\leq& 2d^2\exp\left(-\frac{\kappa^2_6 D_1 K^2\left(\log n\right)^9}{8d^2\kappa^4_4}\right) + \eta.
\end{eqnarray*}
Recalling that $\rho^n = 1 + \left(\log n\right)^3$, and taking $\eta = \exp\left(-\left(\log n\right)^4\right)$, we have the desired result (taking the constants to be sufficiently large to make it hold for all $n$).

\subsection{Proof of Lemma \ref{lem:concg}}

The overall structure of this proof is very similar to that of Lemma \ref{lem:technicalknown2}. We will omit those details that are identical, referring to arguments from the proof of Lemma \ref{lem:technicalknown2} as necessary. As in that proof, we begin by fixing $M > 1$ and $D_0 > 0$ such that $D_0 + 2\sqrt{2}\alpha \log\left(M\right)\leq \frac{\kappa_6}{4C_1}$. We let $N = \lceil \exp\left(\left(\log n\right)^{\frac{1}{3}}\right)\rceil$. Again, we use the recursion
\begin{equation*}
\delta^{n+1}=\delta^n-\frac{\alpha}{n+1}\left(\zeta^{n+1}-p\right),
\end{equation*}
where $\zeta^{n+1}_k = 1_{\left\{k=\hat{\pi}^{n+1}\right\}}$. Let $\Delta^{\max,n} = \max_k \|\Delta^n_k\|_2$.

Similarly to (\ref{eq:applyl2}) in the proof of Lemma \ref{lem:technicalknown2}, we have
\begin{eqnarray}
\|\delta^{n+1}\|^2_2\hspace{-0.1in} &\leq & \hspace{-0.1in}\|\delta^n\|^2_2 + \frac{\alpha^2}{\left(n+1\right)^2}\|\zeta^{n+1}-p\|^2_2 - \frac{2\alpha}{n+1}\left(\zeta^{n+1}-p\right)^\top \delta^n\nonumber\\
\hspace{-0.1in}&\leq & \hspace{-0.1in}\|\delta^n\|^2_2 + \frac{2\alpha^2}{\left(n+1\right)^2} - \frac{2\alpha}{n+1}\left(\zeta^{n+1}-p^{\hat{\beta}^n,g^n}\right)^\top \delta^n\nonumber\\
\hspace{-0.1in}&\,&\hspace{-0.1in} - \frac{2\alpha}{n+1}\left(p^{\hat{\beta}^n,g^n}-p^{\hat{\beta}^n,g^*}\right)^\top \delta^n- \frac{2\alpha}{n+1}\left(p^{\hat{\beta}^n,g^*}-p\right)^\top \delta^n\nonumber\\
\hspace{-0.1in}&\leq&\hspace{-0.1in} \|\delta^n\|^2_2 + \frac{2\alpha^2}{\left(n+1\right)^2} -\frac{2\alpha}{n+1}\left(\zeta^{n+1}-p^{\hat{\beta}^n,g^n}\right)^\top \delta^n - \frac{2\alpha D_1}{n+1}\min\left\{\|\delta\|^2_2,\kappa^2_1\right\}\nonumber\\
\hspace{-0.1in}&\,& \hspace{-0.1in}+\frac{2\alpha D_2}{n+1}\left(\Delta^{\max,n}\right)^2+\frac{2\alpha D_3}{n+1}\|\delta^n\|_2\Delta^{\max,n},\label{eq:applyl2-2}
\end{eqnarray}
where (\ref{eq:applyl2-2}) is obtained by applying Lemmas \ref{lem:technicalunknown2}-\ref{lem:technicalunknown3} with appropriate $D_1,D_2,D_3>0$.

Due to Lemma \ref{lem:technicalknown1}, we may write $\|\delta^n\|_2 \leq \alpha D_4\log\left(n+1\right)$ for some suitably large constant $D_4>0$. By repeating the arguments immediately after (\ref{eq:applyl2}) in the proof of Lemma \ref{lem:technicalknown2}, we have
\begin{eqnarray}
\|\delta^{n+1}\|^2_2 &\leq& \|\delta^n\|^2_2 + \frac{2\alpha^2}{\left(n+1\right)^2} -\frac{2\alpha}{n+1}\left(\zeta^{n+1}-p^{\hat{\beta}^n,g^n}\right)^\top \delta^n - \frac{2\alpha D_5}{n+1}\min\left\{\|\delta^n\|^2_2,D^2_0\right\}\nonumber\\
&\,& + \frac{2\alpha D_2}{n+1}\left(\Delta^{\max,n}\right)^2+\frac{2\alpha D_3}{n+1}\|\delta^n\|_2\Delta^{\max,n}\label{eq:recursionextraterms}
\end{eqnarray}
for some suitable $D_5>0$.

Pick any $n' \in \left\{1,...,n\right\}$ and define two events
\begin{eqnarray*}
E_{n'} &=& \left\{\|\delta^{n-n'}\|^2_2 < D^2_0\right\} \cap \left\{\min_{n-n'+1\leq m \leq n} \|\delta^m\|^2_2 \geq D^2_0\right\},\\
F_{n'} &=& \bigcap^n_{m={n-n'+1}}\left\{\Delta^{\max,n} \leq C_3\left(\log n\right)^{-\frac{5}{2}}\right\},
\end{eqnarray*}
where $C_3$ is the constant from Lemma \ref{lem:concreg}. We also define an additional special case
\begin{equation*}
E_{n+1} = \left\{\min_{0\leq m \leq n} \|\delta^m\|^2_2 \geq D^2_0\right\}.
\end{equation*}
Similarly to the proof of Lemma \ref{lem:technicalknown2}, we will study the behavior of $\|\delta^n\|_2$ on $E_{n'}\cap F_{n'}$ for three possible cases representing different ranges of $n-n'+1$, namely, 1) $n-n'+1 > \frac{n}{M}$, 2) $N \leq n -n'+1 \leq \frac{n}{M}$, and 3) $n-n'+1 < N$.

\textit{Case 1}: $n-n'+1 > \frac{n}{M}$. This case proceeds identically to the proof of Lemma \ref{lem:technicalknown2}, and yields
\begin{equation*}
\|\delta^n\|_2 < D_0 + 2\sqrt{2}\cdot \alpha\log M
\end{equation*}
on the event $E_{n'}\cap F_{n'}$.

\textit{Case 2}: $N \leq n -n'+1 \leq \frac{n}{M}$. On the event $E_{n'}\cap F_{n'}$, we continue (\ref{eq:recursionextraterms}) as
\begin{eqnarray}
\|\delta^{n+1}\|^2_2 &\leq& \|\delta^n\|^2_2 + \frac{2\alpha^2}{\left(n+1\right)^2} -\frac{2\alpha}{n+1}\left(\zeta^{n+1}-p^{\hat{\beta}^n,g^n}\right)^\top \delta^n - \frac{2\alpha D_5}{n+1}\min\left\{\|\delta^n\|^2_2,D^2_0\right\}\nonumber\\
&\,& + \frac{2\alpha C^2_4 D_2}{n+1}\left(\log n\right)^{-5}+\frac{2\alpha C_3D_3D_4}{n+1}\left(\log n\right)^{-\frac{3}{2}}.\label{eq:recursionextraterms2}
\end{eqnarray}
The last two terms in (\ref{eq:recursionextraterms2}) are obtained from the definition of $F_{n'}$ and from the bound on $\|\delta^n\|_2$.

Starting with (\ref{eq:recursionextraterms2}), we now repeat the derivation of (\ref{eq:intapprox}) in the proof of Lemma \ref{lem:technicalknown2} and obtain
\begin{eqnarray}
\|\delta^n\|^2_2 &\leq& D^2_0 + \frac{4\alpha^2}{n-n'+1} -2\alpha\sum^{n-1}_{m=n-n'} \frac{1}{m+1}\left(\zeta^{m+1}-p^m\right)^\top \delta^m -2\alpha D^2_0 D_5\log\left(\frac{n}{n-n'+1}\right)\nonumber\\
&\,& + C^2_4 D_2\alpha\left(\log\left(n-n'+1\right)\right)^{-4} + 8C_3D_3D_4\alpha^2\left(\log\left(n-n'+1\right)\right)^{-\frac{1}{2}}.\label{eq:newintapprox}
\end{eqnarray}
By repeating the arguments after the derivation of (\ref{eq:intapprox}) in the proof of Lemma \ref{lem:technicalknown2}, we obtain
\begin{equation}\label{eq:midrange2}
D^2_0 D_5\log\left(\frac{n}{n-n'+1}\right) \geq \frac{4\alpha}{n-n'+1} + C^2_4 D_2\left(\log\left(n-n'+1\right)\right)^{-4} + 8C_3D_3D_4\alpha\left(\log\left(n-n'+1\right)\right)^{-\frac{1}{2}}
\end{equation}
for sufficiently large $n$, by analogy with (\ref{eq:midrange}). In other words, the additional terms on the right-hand side of (\ref{eq:midrange2}), which are new to this proof, can be ignored. Therefore, the remainder of Case 2 proceeds exactly as in the proof of Lemma \ref{lem:technicalknown2}, and we obtain
\begin{equation*}
\sum_{N\leq n-n'+1\leq \frac{n}{M}} P\left(E_{n'}\cap F_{n'}\right) \leq \frac{2}{n^9},
\end{equation*}
by analogy with (\ref{eq:boundPE3}).

\textit{Case 3}: $n -n'+1 < N$. On the event $E_{n'}\cap F_{n-N+1}$, we have $\min_{N\leq m \leq n} \|\delta^m\|_2 > D_0$. By repeating the derivation of (\ref{eq:newintapprox}), we obtain
\begin{eqnarray*}
\|\delta^n\|^2_2 &\leq & \|\delta^N\|^2_2 + \frac{4\alpha^2}{N+1} -2\alpha\sum^{n-1}_{m=N} \frac{1}{m+1}\left(\zeta^{m+1}-p^m\right)^\top \delta^m -2\alpha D^2_0 D_5\log\left(\frac{n}{N+1}\right)\\
&\,& + 2 C^2_4 D_2\alpha\left(\log N\right)^{-4} + 16 C_3D_3D_4\alpha^2\left(\log N\right)^{-\frac{1}{2}}.
\end{eqnarray*}
Repeating the arguments in the proof of Lemma \ref{lem:technicalknown2}, starting with eq. (\ref{eq:intapprox2}), we obtain
\begin{equation}\label{eq:ignorethenewterms}
\|\delta^N\|^2_2 + \frac{4\alpha^2}{N+1} + 2 C^2_4 D_2\alpha\left(\log N\right)^{-4} + 16 C_3D_3D_4\alpha^2\left(\log N\right)^{-\frac{1}{2}} \leq \alpha D^2_0 D_5\log\left(\frac{n}{N+1}\right)
\end{equation}
for sufficiently large $n$. Again, this essentially allows us to ignore the additional terms on the left-hand side of (\ref{eq:ignorethenewterms}) that are new to this proof. Therefore, the remainder of Case 3 proceeds exactly as in the proof of Lemma \ref{lem:technicalknown2}, and we obtain $P\left(E_{n'}\cap F_N\right) \leq \frac{2}{n^{10}}$ and
\begin{equation*}
\sum_{n-n'+1< N} P\left(E_{n'}\cap F_{n-N+1}\right) \leq \frac{2}{n^9}.
\end{equation*}

\textit{Putting Cases 1-3 together}. As in the proof of Lemma \ref{lem:technicalknown2}, we write
\begin{eqnarray}
&\,& P\left(\|\delta^n\|_2 \geq D_0 + 2\sqrt{2}\cdot\alpha\log M\right)\nonumber\\
&\leq& P\left(E_{n+1}\right) + \sum_{1\leq n-n'+1\leq n} P\left(\left\{\|\delta^n\|_2 \geq D_0 + 2\sqrt{2}\cdot\alpha\log M\right\}\cap E_{n'}\right)\nonumber\\
&=& P\left(E_{n+1}\right) + \sum_{1\leq n-n'+1 \leq \frac{n}{M}} P\left(E_{n'}\right)\label{eq:applycase1again}\\
&\leq& P\left(E_{n+1}\cap F_{n-N+1}\right) + P\left(F^c_{n-N+1}\right) + \sum_{1\leq n-n'+1< N} P\left(E_{n'}\right) + \sum_{N\leq n-n'+1\leq\frac{n}{M}} P\left(E_{n'}\right)\nonumber\\
&\leq & P\left(E_{n+1}\cap F_{n-N+1}\right) + \left(N+1\right)P\left(F^c_{n-N+1}\right)\nonumber\\
&\,& + \sum_{1\leq n-n'+1< N} P\left(E_{n'}\cap F_{n-N+1}\right) + \sum_{N\leq n-n'+1\leq\frac{n}{M}} P\left(E_{n'}\cap F_{n'}\right) + \sum_{N\leq n-n'+1\leq\frac{n}{M}} P\left(F^c_{n'}\right)\nonumber\\
&\leq& \frac{4}{n^9} + \left(N+1\right)P\left(F^c_{n-N+1}\right) + \sum_{N\leq n-n'+1\leq\frac{n}{M}} P\left(F^c_{n'}\right),\label{eq:applycases23again}
\end{eqnarray}
where (\ref{eq:applycase1again}) follows from the analysis of Case 1, and (\ref{eq:applycases23again}) is the result of the other two cases. Applying Lemma \ref{lem:concreg}, we obtain
\begin{equation*}
P\left(F^c_{n'}\right) \leq \sum^n_{m=n-n'+1} C_4\exp\left(-C_5 \left(\log n\right)^4\right) \leq C_4\left(n'\right)\exp\left(-C_5 \left(\log n-n'+1\right)^4\right),
\end{equation*}
whence
\begin{eqnarray*}
\sum^{n-N+1}_{n'=1} P\left(F^c_{n'}\right) &\leq& C_4 n^2 \exp\left(-C_5\left(\log N\right)^4\right)\\
&\leq& C_4 n^2\exp\left(-C_5\left(\log n\right)^{\frac{4}{3}}\right)\\
&\leq& \frac{1}{n^{9}}.
\end{eqnarray*}
Using similar reasoning, we also obtain $\left(N+1\right)P\left(F^c_{n-N+1}\right) \leq \frac{1}{n^9}$, whence
\begin{equation}\label{eq:concdeltafinal}
P\left(\|\delta^n\|_2 \geq D_0 + 2\sqrt{2}\cdot\alpha\log M\right) \leq \frac{6}{n^9}.
\end{equation}
By construction, $D_0 + 2\sqrt{2}\alpha \log\left(M\right)\leq \frac{\kappa_6}{4C_1}$. The coefficient in (\ref{eq:concdeltafinal}) can be increased to make the bound hold for all $n$. The desired result follows.

\subsection{Proof of Proposition \ref{lem:expreg}}

As in Lemma \ref{lem:quadratic}, define $V^n_k = \sum^n_{m=1} X^m \left(X^m\right)^\top 1_{\left\{\pi^n=k\right\}}$. Also let
\begin{equation*}
S^n_k = \mathbb{E}\left(X^n \left(X^n\right)^\top 1_{\left\{\pi^n\left(X^n\right)=k\right\}}\mid\mathcal{F}^{n-1}\right).
\end{equation*}
where $\mathcal{F}^{n-1}$ is the sigma-algebra generated by $\pi^1,...,\pi^{n-1}$, $W^1_{\pi^1},...,W^{n-1}_{\pi^{n-1}}$, and $X^1,...,X^{n-1}$. By repeating the same arguments as in the proof of Lemma \ref{lem:hoeffding} (replacing Hoeffding's inequality with Azuma's inequality), we can obtain
\begin{equation}\label{eq:anotherregconc}
P\left(\|V^n_k - \sum^n_{m=1} S^m_k\|^2_{\text{sp}} > d^2 z^2\right) \leq 2d^2\exp\left(-\frac{z^2}{2\kappa^4_4 n}\right)
\end{equation}
for any $z > 0$.

Recall that $\mathcal{T}_k$ is the set of time periods in which the policy is forced to explore the $k$th cost function. For any $m\in\left\{1,...,n\right\}\cap \mathcal{T}^c_k$, we have $S^m_k = S_k\left(\hat{\beta}^n,g^n\right)$ where $S_k$ was defined in eq. (\ref{eq:defineSk}) of Lemma \ref{lem:errorrates1}. By the result of Lemma \ref{lem:errorrates1},
\begin{equation*}
\lambda_{\min}\left(S_k\left(\hat{\beta}^n,g^n\right)\right) \geq \kappa_6 - C_1\|\delta^n\|_{\infty} - C_2 \Delta^{\max,n},
\end{equation*}
where $\Delta^{\max,n} = \max_k \|\Delta^n_k\|_2$. Applying Lemmas \ref{lem:concreg} and \ref{lem:concg}, we may obtain
\begin{equation}\label{eq:lambdaminbound}
P\left(\lambda_{\min}\left(S^m_k\right)<\frac{1}{2}\kappa_6\right) \leq \frac{D_1}{n^9}
\end{equation}
for $m\in\left\{1,...,n\right\}\cap \mathcal{T}^c_k$ and some suitable $D_1 > 0$. Since $\left|\left\{1,...,n\right\}\cap\mathcal{T}_k\right|$ is of order $\mathcal{O}\left(\left(\log n\right)^9\right)$, we have
\begin{equation}\label{eq:cardineq}
\frac{2}{3}n - \left|\left\{1,...,n\right\}\cap\mathcal{T}_k\right| \geq \frac{2}{5}n
\end{equation}
for large $n$.

Then, for large $n$, we may derive
\begin{eqnarray}
&\,& P\left(\lambda_{\min}\left(V^n_k\right) < \frac{\kappa_6}{10}n\right)\nonumber\\
&\leq & P\left(\sum^n_{m=1} \lambda_{\min}\left(S^m_k\right) - \|V^n_k - \sum^n_{m=1}S^m_k\|_{\text{sp}} < \frac{\kappa_6}{10}n\right)\nonumber\\
&\leq & P\left(-\|V^n_k - \sum^n_{m=1}S^m_k\|_{\text{sp}} < -\frac{\kappa_6}{10}n\right)+P\left(\sum^n_{m=1} \lambda_{\min}\left(S^m_k\right) \leq \frac{\kappa_6}{5}n\right)\nonumber\\
&\leq& 2d^2\exp\left(-\frac{\kappa^2_6 n}{200\kappa^4_4 d^2}\right) + P\left(\sum_{m\in\left\{\frac{n}{3},...,n\right\}\cap\mathcal{T}_k} \lambda_{\min}\left(S^m_k\right) \leq \frac{\kappa_6}{5}n\right)\label{eq:regineq1}\\
&\leq& 2d^2\exp\left(-\frac{\kappa^2_6 n}{200\kappa^4_4 d^2}\right) + \sum_{m\in\left\{\frac{n}{3},...,n\right\}\cap\mathcal{T}_k} P\left(\lambda_{\min}\left(S^m_k\right) \leq \frac{\kappa_6n}{5\left(\frac{2}{3}n-\left|\left\{1,...,n\right\}\cap\mathcal{T}_k\right|\right)}\right)\nonumber\\
&\leq& 2d^2\exp\left(-\frac{\kappa^2_6 n}{200\kappa^4_4 d^2}\right) + \sum_{m\in\left\{\frac{n}{3},...,n\right\}\cap\mathcal{T}_k} P\left(\lambda_{\min}\left(S^m_k\right) \leq \frac{1}{2}\kappa_6\right)\label{eq:applycard}\\
&\leq& 2d^2\exp\left(-\frac{\kappa^2_6 n}{200\kappa^4_4 d^2}\right) + D_1\sum^n_{m=\frac{n}{3}} \frac{1}{m^9}.\label{eq:applyprevlemmas}
\end{eqnarray}
In this derivation, (\ref{eq:regineq1}) is due to (\ref{eq:anotherregconc}), while (\ref{eq:applycard}) applies (\ref{eq:cardineq}), and (\ref{eq:applyprevlemmas}) follows by (\ref{eq:lambdaminbound}). Thus, we arrive at the bound
\begin{equation}\label{eq:lambdaminbound2}
P\left(\lambda_{\min}\left(V^n_k\right) < \frac{\kappa_6}{10}n\right) \leq \frac{D_2}{n^8}
\end{equation}
for some suitable $D_2>0$.

From the proof of Lemma \ref{lem:concreg}, we recall that
\begin{equation*}
\Delta^n_k = -\rho^n\left(\rho^n\cdot I + V^n_k\right)^{-1}\beta_k + \left(\rho^n\cdot I + V^n_k\right)^{-1}v^n_k,
\end{equation*}
where $v^n_k = \sum^n_{m=1} \varepsilon^m_k X^m 1_{\left\{\pi^m=k\right\}}$. Therefore,
\begin{equation*}
\|\Delta^n_k\|_2 \leq \frac{\rho^n}{\rho^n+\lambda_{\min}\left(V^n_k\right)}\|\beta_k\|_2 + \|\left(\rho^n\cdot I + V^n_k\right)^{-1}v^n_k\|_2.
\end{equation*}
Taking expectations, we have
\begin{equation}\label{eq:concregdecomp}
\mathbb{E}\left(\|\Delta^n_k\|^2_2\right) \leq 2\|\beta_k\|^2_2 \mathbb{E}\left[\left(\frac{\rho^n}{\rho^n+\lambda_{\min}\left(V^n_k\right)}\right)^2\right] + 2\mathbb{E}\left(\|\left(\rho^n\cdot I + V^n_k\right)^{-1}v^n_k\|^2_2\right).
\end{equation}
Applying (\ref{eq:lambdaminbound2}) and the fact that $\rho^n = 1 + \left(\log n\right)^3$, we obtain
\begin{eqnarray*}
\mathbb{E}\left[\left(\frac{\rho^n}{\rho^n+\lambda_{\min}\left(V^n_k\right)}\right)^2\right] &\leq& \left(\frac{\rho^n}{\rho^n + \frac{\kappa_6}{10}n}\right)^2 + P\left(\lambda_{\min}\left(V^n_k\right) < \frac{\kappa_6}{10}n\right)\\
&\leq & \frac{D_3}{n}
\end{eqnarray*}
for some suitable $D_3>0$. The second term on the right-hand side of (\ref{eq:concregdecomp}) is handled similarly. We let $E = \left\{\lambda_{\min}\left(V^n_k\right) \geq \frac{\kappa_6}{10}n\right\}$ and write
\begin{eqnarray}
&\,& \mathbb{E}\left(\|\left(\rho^n\cdot I + V^n_k\right)^{-1}v^n_k\|^2_2\right)\nonumber\\
&\leq & \mathbb{E}\left(\|\left(\rho^n\cdot I + V^n_k\right)^{-1}v^n_k\|^2_2\cdot 1_E\right) + \mathbb{E}\left(\frac{\|\left(\rho^n\cdot I + V^n_k\right)^{-\frac{1}{2}}v^n_k\|^2_2}{\rho^n + \lambda_{\min}\left(V^n_k\right)}\cdot 1_{E^c}\right)\nonumber\\
&\leq & \mathbb{E}\left(\|\left(n\cdot I + V^n_k\right)^{\frac{1}{2}}\left(\rho^n\cdot I + V^n_k\right)^{-1}\|^2_{\text{sp}}\cdot\|\left(n\cdot I + V^n_k\right)^{-\frac{1}{2}}v^n_k\|^2_2\cdot 1_E\right)\nonumber\\
&\,& + \mathbb{E}\left(\frac{\|\left(\rho^n\cdot I + V^n_k\right)^{-\frac{1}{2}}v^n_k\|^2_2}{\rho^n + \lambda_{\min}\left(V^n_k\right)}\cdot 1_{E^c}\right)\nonumber\\
&\lesssim& \mathbb{E}\left(\frac{n+\lambda_{\min}\left(V^n_k\right)}{\left(\rho^n+\lambda_{\min}\left(V^n_k\right)\right)^2}\|\left(n\cdot I + V^n_k\right)^{-\frac{1}{2}}v^n_k\|^2_2\cdot 1_E\right)\nonumber\\
&\,& + \mathbb{E}\left(\frac{\|\left(\rho^n\cdot I + V^n_k\right)^{-\frac{1}{2}}v^n_k\|^2_2}{\rho^n + \lambda_{\min}\left(V^n_k\right)}\cdot 1_{E^c}\right)\nonumber\\
&\lesssim& \frac{1}{n}\mathbb{E}\left(\|\left(n\cdot I + V^n_k\right)^{-\frac{1}{2}}v^n_k\|^2_2\right) + \mathbb{E}\left(\frac{\|\left(\rho^n\cdot I + V^n_k\right)^{-\frac{1}{2}}v^n_k\|^2_2}{\rho^n}\cdot 1_{E^c}\right)\nonumber\\
&\lesssim & \frac{1}{n} + \mathbb{E}\left(\frac{\|\left(\rho^n\cdot I + V^n_k\right)^{-\frac{1}{2}}v^n_k\|^2_2}{\rho^n}\cdot 1_{E^c}\right),\label{eq:applysecondclaim}
\end{eqnarray}
where (\ref{eq:applysecondclaim}) follows by the second claim of Lemma \ref{lem:quadratic}.

We now use the first claim of Lemma \ref{lem:quadratic}. We take $z=\log\left(\frac{1}{\eta}\right)$ and let $z \geq \sqrt{\rho^n}$. For sufficiently large $n$,
\begin{equation}\label{eq:applyfirstclaim}
P\left(\|\left(\rho^n\cdot I + V^n_k\right)^{-\frac{1}{2}}v^n_k\|^2_2 \geq z\right) \leq \exp\left(-\frac{z}{4\kappa_5}\right).
\end{equation}
Therefore, when $n$ is large, we may derive
\begin{eqnarray}
&\,& \mathbb{E}\left(\|\left(\rho^n\cdot I + V^n_k\right)^{-\frac{1}{2}}v^n_k\|^2_2\cdot 1_{E^c}\right)\nonumber\\
&=& \int^{\infty}_0 P\left(\left\{\|\left(\rho^n\cdot I + V^n_k\right)^{-\frac{1}{2}}v^n_k\|^2_2 > z\right\}\cap E^c\right)dz\nonumber\\
&\leq& \int^{\infty}_0 \min\left\{P\left(\|\left(\rho^n\cdot I + V^n_k\right)^{-\frac{1}{2}}v^n_k\|^2_2 > z\right),P\left(E^c\right)\right\}dz\nonumber\\
&\leq & \int^{\infty}_0 \min\left\{P\left(\|\left(\rho^n\cdot I + V^n_k\right)^{-\frac{1}{2}}v^n_k\|^2_2 > z\right),\frac{D_2}{n^8}\right\}dz\label{eq:applylambdaminbound2}\\
&=& \int^{\sqrt{\rho^n}}_0 \min\left\{P\left(\|\left(\rho^n\cdot I + V^n_k\right)^{-\frac{1}{2}}v^n_k\|^2_2 > z\right),\frac{D_2}{n^8}\right\}dz\nonumber\\
&\,& + \int^{\infty}_{\sqrt{\rho^n}} \min\left\{P\left(\|\left(\rho^n\cdot I + V^n_k\right)^{-\frac{1}{2}}v^n_k\|^2_2 > z\right),\frac{D_2}{n^8}\right\}dz\nonumber\\
&\leq & \frac{\sqrt{\rho^n}}{n^8}D_2 + \int^{\infty}_{\sqrt{\rho^n}} \min\left\{P\left(\|\left(\rho^n\cdot I + V^n_k\right)^{-\frac{1}{2}}v^n_k\|^2_2 > z\right),\frac{D_2}{n^8}\right\}dz\nonumber\\
&\leq & \frac{\sqrt{\rho^n}}{n^8}D_2 + \int^{\infty}_{\sqrt{\rho^n}} P\left(\|\left(\rho^n\cdot I + V^n_k\right)^{-\frac{1}{2}}v^n_k\|^2_2 > z\right)dz\nonumber\\
&\leq & \frac{\sqrt{\rho^n}}{n^8}D_2 + \int^{\infty}_{\sqrt{\rho^n}} \exp\left(-\frac{z}{4\kappa_5}\right)dz\label{eq:applyfirstclaim2}\\
&=& \frac{\sqrt{\rho^n}}{n^8}D_2 + 4\kappa_5\exp\left(-\frac{\sqrt{\rho^n}}{4\kappa_5}\right).\nonumber
\end{eqnarray}
In this derivation, (\ref{eq:applylambdaminbound2}) is due to (\ref{eq:lambdaminbound2}), and (\ref{eq:applyfirstclaim2}) follows by (\ref{eq:applyfirstclaim}). Returning to (\ref{eq:applysecondclaim}), we have shown
\begin{eqnarray*}
\mathbb{E}\left(\|\left(\rho^n\cdot I + V^n_k\right)^{-1}v^n_k\|^2_2\right) &\lesssim& \frac{1}{n} + \frac{\sqrt{\rho^n}}{n^8}D_2 + 4\kappa_5\exp\left(-\frac{\sqrt{\rho^n}}{4\kappa_5}\right)\\
&\lesssim& \frac{1}{n}.
\end{eqnarray*}
Plugging this bound into (\ref{eq:concregdecomp}) completes the proof.

\subsection{Proof of Lemma \ref{lem:technicalunknown1glm}}

We proceed exactly as in the proof of Lemma \ref{lem:technicalunknown1}, replacing all expressions of the form $b^\top_k X$ and $\beta^\top_k X$ with $\chi\left(b^\top_k X\right)$ and $\chi\left(\beta^\top_k X\right)$. As before, we let
\begin{equation*}
\bar{\Delta}_{j,k} = \left[\chi\left(b^\top_j X\right) - \chi\left(\beta^\top_j X\right)\right] - \left[\chi\left(b^\top_k X\right) - \chi\left(\beta^\top_k X\right)\right],
\end{equation*}
and observe that the Lipschitz property of $\chi$ (Assumption \ref{aextra}(iii)), together with the arguments made in the proof of Lemma \ref{lem:technicalunknown1}, yields
\begin{equation*}
\left|\bar{\Delta}_{j,k}\right| \leq 2\kappa_4 \kappa_9\Delta^{\max}.
\end{equation*}
Then, the arguments are the same as in the proof of Lemma \ref{lem:technicalunknown1}.

\subsection{Proof of Lemma \ref{lem:concregglm}}

Let $\Delta^n_k = \hat{\beta}^n_k - \beta_k$ and $L_{22}\left(t_1,t_2\right) = \frac{\partial^2 L\left(t_1,t_2\right)}{\partial t^2_2}$. By Taylor's theorem, we have
\begin{eqnarray}
&\,& L\left(W^m_k,\left(\hat{\beta}^n_k\right)^\top X^m\right) - L\left(W^m_k,\beta^\top X^m\right)\nonumber\\
&=& L_2\left(W^m_k,\beta^\top X^m\right)\cdot\left(\Delta^n_k\right)^\top X^m + \frac{1}{2}L_{22}\left(W^m_k,\beta^\top_k X^m + \eta_m\left(\Delta^n_k\right)^\top X^m\right)\cdot\left(\left(\Delta^n_k\right)^\top X^m\right)^2,\label{eq:concregglm1}
\end{eqnarray}
where $\eta_m \in \left[0,1\right]$, for any $m\leq n$. Observe that both $\beta^\top_k X^m$ and $\left(\hat{\beta}^n_k\right)^\top X^m$ are in $\left[-\kappa_4\kappa_7,\kappa_4,\kappa_7\right]$, following from Assumptions \ref{a1new}(ii) and \ref{aextra}(i). Therefore, the value of the convex combination $\beta^\top_k X^m + \eta_m\left(\Delta^n_k\right)^\top X^m$ is also in $\left[-\kappa_4\kappa_7,\kappa_4,\kappa_7\right]$. By Assumption \ref{aextra}(ii), it follows that
\begin{equation*}
L_{22}\left(W^m_k,\beta^\top_k X^m + \eta_m\left(\Delta^n_k\right)^\top X^m\right)\geq \kappa_8.
\end{equation*}
Therefore, (\ref{eq:concregglm1}) implies
\begin{eqnarray}
&\,& L\left(W^m_k,\left(\hat{\beta}^n_k\right)^\top X^m\right) - L\left(W^m_k,\beta^\top X^m\right)\nonumber\\
&\geq& L_2\left(W^m_k,\beta^\top X^m\right)\cdot\left(\Delta^n_k\right)^\top X^m + \frac{1}{2}\kappa_8\left(\left(\Delta^n_k\right)^\top X^m\right)^2.\label{eq:concregglm2}
\end{eqnarray}
By the definition of $\hat{\beta}^n_k$, we have
\begin{equation}\label{eq:concregglm3}
\sum^n_{m=1} L\left(W^m_k,\left(\hat{\beta}^n_k\right)^\top X^m\right)1_{\left\{\pi^m=k\right\}} + \rho^n\|\hat{\beta}^n_k\|^2_2 \leq \sum^n_{m=1} L\left(W^m_k,\beta_k^\top X^m\right)1_{\left\{\pi^m=k\right\}} + \rho^n\|\beta_k\|^2_2.
\end{equation}
Together, (\ref{eq:concregglm2})-(\ref{eq:concregglm3}) imply that
\begin{equation}
\rho^n\|\beta_k\|^2_2 -\rho^n\|\hat{\beta}^n_k\|^2_2 \geq \left(v^n_k\right)^\top\Delta^n_k + \frac{1}{2}\kappa_8\left(\Delta^n_k\right)^\top V^n_k\Delta^n_k,\label{eq:concregglm4}
\end{equation}
where
\begin{equation}\label{eq:vVglm}
v^n_k = \sum^n_{m=1} L_2\left(W^m_k,\beta^\top_k X^m\right)\cdot X^m\cdot 1_{\left\{\pi^m=k\right\}}, \qquad V^n_k = \sum^n_{m=1} X^m\left(X^m\right)^\top 1_{\left\{\pi^m=k\right\}}.
\end{equation}
Observing that
\begin{eqnarray*}
\|\beta_k\|^2_2 -\|\hat{\beta}^n_k\|^2_2 &=& \|\beta_k\|^2_2 -\|\beta_k+\Delta^n_k\|^2_2\\
&=& -\|\Delta^n_k\|^2_2 - 2\beta^\top_k\Delta^n_k\\
&\leq & -\|\Delta^n_k\|^2_2 + 2\|\beta_k\|_2\|\Delta^n_k\|_2\\
&\leq & -\|\Delta^n_k\|^2_2 + 2\kappa_7\|\Delta^n_k\|_2,
\end{eqnarray*}
where the last line is due to Assumption \ref{aextra}(i), (\ref{eq:concregglm4}) becomes
\begin{equation}
2\kappa_7 \rho^n\|\Delta^n_k\|_2 \geq \left(v^n_k\right)^\top\Delta^n_k + \frac{1}{2}\kappa_8\left(\Delta^n_k\right)^\top \left(V^n_k+\mu^n I\right)\Delta^n_k,\label{eq:concregglm5}
\end{equation}
with $\mu^n = 2\rho^n\frac{\kappa_7}{\kappa_8}$. Define
\begin{equation}\label{eq:DeltaHglm}
\tilde{\Delta}^n_k = \left(V^n_k + \mu^n I\right)^{\frac{1}{2}}\Delta^n_k, \qquad H^n_k = \left(V^n_k+\mu^n I\right)^{-\frac{1}{2}} v^n_k.
\end{equation}
Then, (\ref{eq:concregglm5}) is rewritten as
\begin{equation}\label{eq:concregglm6}
\frac{1}{2}\kappa_8\|\tilde{\Delta}^n_k\|^2_2 + \left(H^n_k\right)^\top\tilde{\Delta}^n_k \leq 2\kappa_7\rho_n\|\Delta^n_k\|_2 \leq 2\kappa^2_7\rho_n,
\end{equation}
where the last inequality is due to Assumption \ref{aextra}(i). Using $\left|\left(H^n_k\right)^\top \tilde{\Delta}^n_k\right| \leq \|H^n_k\|_2\|\tilde{\Delta}^n_k\|_2$ and rearranging terms, (\ref{eq:concregglm6}) becomes
\begin{equation}\label{eq:concregquadratic}
\frac{1}{2}\kappa_8\|\tilde{\Delta}^n_k\|^2_2 - \|H^n_k\|_2\|\tilde{\Delta}^n_k\|_2 - 4\kappa^2_7 \rho_n \leq 0,
\end{equation}
which is a quadratic inequality in $\|\tilde{\Delta}^n_k\|_2$. Applying the quadratic formula, we have
\begin{equation*}
\|\tilde{\Delta}^n_k\|_2 \leq \frac{\|H^n_k\|_2 + \sqrt{\|H^n_k\|^2_2 + 8\kappa^2_7 \rho_n}}{\kappa_8} \leq \frac{2\|H^n_k\|_2 + 3\kappa_7\sqrt{\rho_n}}{\kappa_8}.
\end{equation*}
Because
\begin{equation*}
\|\tilde{\Delta}^n_k\|_2 \geq \|\Delta^n_k\|_2 \sqrt{\lambda_{\min}\left(V^n_k + \mu^n I\right)} \geq \|\Delta^n_k\|_2\sqrt{\lambda_{\min}\left(V^n_k\right)},
\end{equation*}
it follows that
\begin{eqnarray*}
\|\Delta^n_k\|_2 &\leq & \frac{2\|H^n_k\|_2 + 3\kappa_7\sqrt{\rho_n}}{\kappa_8\sqrt{\lambda_{\min}\left(V^n_k\right)}}\\
&\leq & \frac{2\|H^n_k\|_2 + 3\kappa_7\sqrt{\rho_n}}{\kappa_8\sqrt{\lambda_{\min}\left(V^n_{\mathcal{T}_k}\right)}},
\end{eqnarray*}
where $V^n_{\mathcal{T}_k} =\sum_{m\in\mathcal{T}_k,m\leq n} X^m\left(X^m\right)^\top 1_{\left\{\pi^m=k\right\}}$ and the last line follows because $\lambda_{\min}\left(V^n_k\right) \geq \lambda_{\min}\left(V^n_{\mathcal{T}_k}\right)$.

The remaining arguments proceed as in the last part of the proof of Lemma \ref{lem:concreg}. We apply Lemmas \ref{lem:hoeffdingglm}-\ref{lem:quadraticglm} to obtain
\begin{eqnarray*}
&\,& P\left(\|\Delta^n_k\|_2 > \frac{2\sqrt{2\kappa_5 \left(\log n\right)^4 + \kappa_5 d\log\left(1 + \frac{\kappa^2_4}{\mu^n}n\right)} + 3\kappa_7\sqrt{\rho_n}}{\kappa_8\sqrt{\frac{\kappa_6}{2}D_1 K \left(\log n\right)^9}}\right)\\
&\leq& 2\exp\left(-\frac{\kappa^2_6 D_1 K^2\left(\log n\right)^9}{8 d^2\kappa^4_4}\right) + \exp\left(-\left(\log n\right)^4\right),
\end{eqnarray*}
which completes the proof.

\subsection{Proof of Proposition \ref{lem:expregglm}}

Recalling the proof of Lemma \ref{lem:concregglm}, define $v^n_k$, $V^n_k$ as in (\ref{eq:vVglm}) and $\tilde{\Delta}^n_k$, $H^n_k$ as in (\ref{eq:DeltaHglm}), with $\mu^n = 2\rho^n\frac{\kappa_7}{\kappa_8}$.

Repeating the arguments in the beginning of the proof of Proposition \ref{lem:expreg}, we obtain (\ref{eq:lambdaminbound2}) for some suitable $D_2 > 0$. By Lemma \ref{lem:quadraticglm}, we have
\begin{equation}\label{eq:expregD3bound}
\mathbb{E}\left(\|H^n_k\|^2_2\right) \leq D_3
\end{equation}
for some suitably large $D_3 > 0$. Recalling (\ref{eq:concregquadratic}) from the proof of Lemma \ref{lem:concregglm}, we obtain
\begin{eqnarray*}
\frac{1}{2}\kappa_8\mathbb{E}\left(\|\tilde{\Delta}^n_k\|^2_2\right) &\leq & 4\kappa^2_7 \rho_n + \mathbb{E}\left(\|H^n_k\|_2\|\tilde{\Delta^n_k}\|_2\right)\\
&\leq & 4\kappa^2_7 \rho_n + \sqrt{\mathbb{E}\left(\|H^n_k\|^2_2\right)}\sqrt{\mathbb{E}\left(\|\tilde{\Delta}^n_k\|^2_2\right)}\\
&\leq & 4\kappa^2_7 \rho_n + D^{\frac{1}{2}}_3\sqrt{\mathbb{E}\left(\|\tilde{\Delta}^n_k\|^2_2\right)},
\end{eqnarray*}
which is a quadratic inequality in $\sqrt{\mathbb{E}\left(\|\tilde{\Delta}^n_k\|^2_2\right)}$. Applying the quadratic formula, we have
\begin{eqnarray*}
\sqrt{\mathbb{E}\left(\|\tilde{\Delta}^n_k\|^2_2\right)} &\leq & \frac{D^{\frac{1}{2}}_4 + \sqrt{D_4 + 8\kappa^2_7\kappa_8\rho^n}}{\kappa_8}\\
&<& \kappa^{-1}_8\left(2D^{\frac{1}{2}}_4 + 3\kappa_7\sqrt{\kappa_8\rho^n}\right).
\end{eqnarray*}
Recalling that $\rho^n = 1 + \left(\log n\right)^3$, we can take $D_4 > 0$ large enough such that
\begin{equation}\label{eq:tildeDeltalogbound}
\mathbb{E}\left(\|\tilde{\Delta}^n_k\|^2_2\right) \leq D_4\left(\log n\right)^3.
\end{equation}

Observing that
\begin{equation}\label{eq:expregglm1}
\|\tilde{\Delta}^n_k\|_2 = \|\left(V^n_k + \mu^n I\right)^{\frac{1}{2}}\Delta^n_k\|_2 \geq \|\Delta^n_k\|_2\sqrt{\lambda_{\min}\left(V^n_k\right) + \mu^n},
\end{equation}
we derive
\begin{eqnarray}
\mathbb{E}\left(\|\Delta^n_k\|^2_2\right) &\leq & \mathbb{E}\left(\|\Delta^n_k\|^2_2 \cdot 1_{\left\{\lambda_{\min}\left(V^n_k\right) < \frac{\kappa_6}{10}n\right\}}\right)+ \mathbb{E}\left(\|\Delta^n_k\|^2_2 \cdot 1_{\left\{\lambda_{\min}\left(V^n_k\right) \geq \frac{\kappa_6}{10}n\right\}}\right)\nonumber\\
&\leq & 4\kappa^2_7 P\left(\lambda_{\min}\left(V^n_k\right)<\frac{\kappa_6}{10}n\right) + \mathbb{E}\left(\|\Delta^n_k\|^2_2 \cdot 1_{\left\{\lambda_{\min}\left(V^n_k\right) \geq \frac{\kappa_6}{10}n\right\}}\right)\label{eq:expregglm2}\\
&\leq & 4\kappa^2_7 \frac{D_2}{n^8} + \mathbb{E}\left(\frac{\|\tilde{\Delta}^n_k\|^2_2\cdot 1_{\left\{\lambda_{\min}\left(V^n_k\right) \geq \frac{\kappa_6}{10}n\right\}}}{\lambda_{\min}\left(V^n_k\right) + \mu^n}\right)\label{eq:expregglm3}\\
&\leq & 4\kappa^2_7 \frac{D_2}{n^8} + \mathbb{E}\left(\frac{\|\tilde{\Delta}^n_k\|^2_2}{\frac{\kappa_6}{10}n}\right)\nonumber\\
&\leq& \frac{D_5}{n}\left(\log n\right)^3,\label{eq:expregglm3-2}
\end{eqnarray}
where (\ref{eq:expregglm2}) is due to Assumption \ref{aextra}(i), (\ref{eq:expregglm3}) follows by (\ref{eq:lambdaminbound2}) and (\ref{eq:expregglm1}), while (\ref{eq:expregglm3-2}) follows from (\ref{eq:tildeDeltalogbound}) and by taking $D_5 > 0$ to be sufficiently large.

Recalling (\ref{eq:concregglm6}) from the proof of Lemma \ref{lem:concregglm}, we have
\begin{equation*}
\frac{1}{2}\kappa_8\|\tilde{\Delta}^n_k\|^2_2 \leq \|H^n_k\|_2\|\tilde{\Delta}^n_k\|_2 + 2\kappa_7\rho^n\|\Delta^n_k\|_2.
\end{equation*}

Taking expectations, we obtain
\begin{eqnarray}
\frac{1}{2}\kappa_8\mathbb{E}\left(\|\tilde{\Delta}^n_k\|^2_2\right) &\leq & \sqrt{\mathbb{E}\left(\|H^n_k\|^2_2\right)}\sqrt{\mathbb{E}\left(\|\tilde{\Delta}^n_k\|^2_2\right)} + 2\kappa_7\rho^n\mathbb{E}\left(\|\Delta^n_k\|_2\right)\nonumber\\
&\leq & D^{\frac{1}{2}}_3\sqrt{\mathbb{E}\left(\|\tilde{\Delta}^n_k\|^2_2\right)} + 2\kappa_7\rho^n D^{\frac{1}{2}}_5 n^{-\frac{1}{2}}\left(\log n\right)^{\frac{3}{2}}\label{eq:expregglm4}\\
&=& D^{\frac{1}{2}}_3\sqrt{\mathbb{E}\left(\|\tilde{\Delta}^n_k\|^2_2\right)} + o\left(1\right),\label{eq:expregglm5}
\end{eqnarray}
where (\ref{eq:expregglm4}) follows from (\ref{eq:expregD3bound}) and (\ref{eq:expregglm3-2}), and (\ref{eq:expregglm5}) follows from the definition of $\rho^n$. Again, (\ref{eq:expregglm5}) is a quadratic inequality in $\sqrt{\mathbb{E}\left(\|\tilde{\Delta}^n_k\|^2_2\right)}$. Solving it yields
\begin{equation*}
\sqrt{\mathbb{E}\left(\|\tilde{\Delta}^n_k\|^2_2\right)} \leq \frac{D^{\frac{1}{2}}_3 + \sqrt{D_3 + o\left(1\right)}}{\kappa_8} \leq D_6
\end{equation*}
for some suitably large $D_6 > 0$. Then, repeating the arguments of (\ref{eq:expregglm1})-(\ref{eq:expregglm3-2}), we obtain
\begin{equation*}
\mathbb{E}\left(\|\Delta^n_k\|^2_2\right) \leq 4\kappa^2_7 \frac{D_2}{n^8} + \frac{D_6}{\frac{\kappa_6}{10}n},
\end{equation*}
which completes the proof.

\end{document}

%% file: notation_ams.tex
\def \qed{\hfill $\Box$}

\def \notin{{\not \in}}

\def \hbar{\bar h}

\def \v1 {\mathbb{F}}



%% file: macro_ams.tex
\newcommand{\doublespace}{\addtolength{\baselineskip}{.5\baselineskip}}
\newcounter{stepno}

\newtheorem{thm}{Theorem}[section]

\newtheorem{lem}{Lemma}[section]

\newtheorem{corol}{Corollary}[section]

\newtheorem{prop}{Proposition}[section]

\newcommand{\bnn}{\\ \\$\begin{array}{rcll}}
\newcommand{\enn}{\end{array}$\\ \\}
\newcommand{\ba}{\begin{eqnarray}}
\newcommand{\ea}{\end{eqnarray}}
\newcommand{\bas}{\begin{eqnarray*}}
\newcommand{\eas}{\end{eqnarray*}}
\newcommand{\bdm}{\begin{displaymath}}
\newcommand{\edm}{\end{displaymath}}
\newcommand{\be}{\begin{equation}}
\newcommand{\ee}{\end{equation}}
\newcommand{\bn}{\begin{eqnarray}}
\newcommand{\en}{\end{eqnarray}}
\newcommand{\bns}{\begin{eqnarray*}}
\newcommand{\ens}{\end{eqnarray*}}

\newcommand{\defvarbegin}{\begin{quotation}\vspace{-15pt}\begin{tabbing}}
\newcommand{\defvarend}  {\end{tabbing}\vspace{-10pt}\end{quotation}}

\newcommand{\bnarray}{\begin{equation}\begin{array}{rcll}}
\newcommand{\enarray}{\end{array}\end{equation}}

\newcommand{\barr}{\begin{array}}
\newcommand{\earr}{\end{array}}

\newcounter{cnum}

\newcommand{\beginalg}{\setcounter{stepno}{1}
                \begin{list}{\bf Step~\arabic{stepno}}
                         {\usecounter{stepno}\settowidth{\labelwidth}{\bf Step~9m}
                \addtolength{\leftmargin}{2\parindent}}
                }
\newcommand{\eg}{\end{list}}

\def \define{\begin{quote}\begin{itemize}}
\def \enddefine{\end{itemize}\end{quote}}

\newlength{\boxedparwidth} 
  {\begin{center} \begin{tabular}{|@{\hspace{.15in}}c@{\hspace{.15in}}|}
                 \hline \\ \begin{minipage}[t]{\boxedparwidth}
                 \setlength{\parindent}{0.0in}}%
   {\end{minipage} \\ \\ \hline \end{tabular} \end{center}}
\newcounter{chapterctr}
\newcounter{example}[chapterctr]
\newcounter{ctr}